\documentclass{article}

\usepackage[reqno,centertags,intlimits]{amsmath}
\usepackage{amsfonts}
\usepackage{cite,epsfig,graphics,ifthen}

\usepackage[A4]{QRP}

\begin{document}

  
  \title{Sum-Over-Histories Quantization of Relativistic Particle}
  
  \author{Pavel Krtou\v{s}\\
    {\small Institute of Theoretical Physics,}\\
    {\small Faculty of Mathematics and Physics, Charles University,}\\
    {\small V Hole\v{s}ovi\v{c}k\'{a}ch 2, 180 00 Praha 8, Czech Republic}
    }
  
  \date{gr-qc/0011077}
  
  \maketitle

\begin{abstract}
Sum-over-histories quantization of particle-like
theory in curved space is discussed. It is reviewed 
that the propagator satisfies the Schr\"odinger 
equation respective wave equation with 
a Laplace-like operator. 
The exact dependence of the operator on the choice 
of measure is shown.

Next, modifications needed for a manifold with 
a boundary are introduced, and the exact form of 
the equation for the propagator is derived. 
It is shown that the Laplace-like operator 
contains some distributional 
terms localized on the boundary. These terms induce 
proper boundary conditions for the propagator. 
This choice of boundary conditions is explained
as a consequence of a measurement of particles 
on the boundary.

The interaction with sources inside of the domain 
and sources on the boundary is also discussed. 

\end{abstract}


\section*{Introduction}

The main goal of this work is to investigate a quantization 
of a relativistic particle using the sum-over-histories 
approach. We have the following motivations to do such a study.

The usual approach how to quantize a relativistic 
particle is the scalar field theory. 
Yet, it is a quantization of a completely different system 
--- a quantization of a continuous field on spacetime. 
It is true that we can identify some states of such system 
as \emph{particle states} --- states with some properties 
of particles. But is there any other way to make a connection 
to the particle theory? Is it possible to quantize 
a classical relativistic particle, and does it give 
predictions equivalent to predictions of 
quantum scalar field theory?

There exists a candidate for the direct quantization 
of a particle theory --- quantization using 
the sum-over-histories approach. The classical explanation 
of this approach for usual nonrelativistic physics can 
be found in \cite{Feynman:lectures} and more technically 
in \cite{FeynmanHibbs:book,Schulman:book,Simon:book}. 
A nice non-technical overview for a relativistic theory can be found 
in \cite{Feynman:book1985}. Beside these classical introductions, 
this approach has received considerable attention in recent years 
(see for example \cite{Hartle:LH1992}). The new development 
has led to a generalization of this method called 
\defterm{generalized quantum mechanics} (see 
\cite{HartleGellMann:1990a,Hartle:LH1992,Hartle:1993a}).
 
In this approach the transition amplitudes associated 
with the chosen criteria are computed by summing over 
amplitudes of all possible histories 
which meet the criteria. It is known that some of 
these amplitudes computed for a relativistic particle 
lead in the special cases 
(e.g., in flat spacetime) to quantities which can 
also be obtained from scalar field theory. 
The goal of our work is to investigate this correspondence 
in more detail.

There are more reasons for studying 
relativistic particle theory. One of the attempts to understand 
the quantization of the gravitational field coupled to matter 
is to reduce the full gravitation theory to a system with 
a finite number of degrees of freedom and try to quantize 
this simplified system. These reduced theories are called 
minisuperspace models. It is well known that such reduced 
system is essentially equivalent to a particle theory in 
a Lorentzian space with (usually) a complicated potential. 
A common method for the quantization of minisuperspace 
models is the sum-over-histories approach.

Key feature of our investigation is that we study 
the particle theory on a bounded domain of spacetime,
and that we pay an attention to the exact form of
boundary conditions. The usual approach is a bit generous 
on this question --- the theory is usually formulated 
on the whole spacetime with not always clearly formulated 
special behavior at infinities. In the flat spacetime such 
an approach is justifiable because there exists 
a preferred behavior at infinities, but in a general 
curved spacetime we have to be more careful. 
The question of boundary conditions is usually completely 
ignored in definitions of the path integral. 
We try to formulate the theory in a more careful way 
and identify its boundary-condition dependence.

The plan of our work is the following. In the first part
of the paper we review sum-over-histories approach to
the quantization of the particle in curved spacetime without
boundary. Equations for key amplitudes (the propagator and
Feynmann Green function) are derived and their exact
dependence on the definition of the path integral is
shown.

It the second part we investigate the theory on a bounded 
domain. On the bounded domain we have to modify our definition 
of the path integral. The new definition leads to a
modification of the equations for physical amplitudes.
Some new distributional terms localized on the boundary
appear in these equation and we find that they
specify the exact form of boundary conditions for
the amplitudes. It is shown that the exact form of
the boundary conditions depends on the details
of the definition of the path integral.

Finally we discuss shortly a physical meaning of the
boundary conditions. We argue that boundary conditions
can serve as a phenomenological description of 
apparatuses measuring particles on the boundary of the domain.

We will continue in this discussion in the following paper
\cite{Krtous:QRPBC}, where a relationship of different
boundary conditions will be investigated and
a connection with a particle definition in the quantum scalar 
field theory will be shown.

Sections 1 and 2 contain the main line of arguments,
details of computations can be found in the appendix 
\ref{apx:AEHK}. The appendix \ref{apx:GdTh} contains a general 
overview of the geodesic theory necessary for the computations, 
including not common theory for reflected geodesics.

Let us note that our study can accommodate a wider range of theories.
The theory will be parameterized by two signature factors.
One of them (the factor $\stsign$) characterizes a signature of
a target space metric (the metric in which a particle lives),
and another (the factor $\isign$) describes whether 
the theory is physical or Euclidian. For real $\stsign$
and imaginary $\isign$ we obtain non-relativistic particle
in curved space, for $\stsign$ and $\isign$ both imaginary
we obtain relativistic particle in curved spacetime
and for real signature factors we get mathematically better
behaving, but non-physical Euclidian version of the theory.
Usually most quantities are well defined in the Euclidian 
sector and the definition for physical signature are
obtained by an analytical continuation in the signature factors.



\section{Particle in a curved space without boundary}
\label{sc:SOH-partwobnd}

\subsection{Space of histories, action and amplitudes}

In this section we shortly formulate 
the sum-over-histories approach
to a quantization of a particle-like theory.
A similar calculation have been done for example 
in \cite{Parker:1979,DeWitt:1975}. We present 
our derivation here because we will generalize 
our line of reasoning in then next section 
to the case when a particle
is moving in a domain with a boundary.

As usual in the sum-over-histories approach, 
the theory is characterized by a space of histories 
and an action. An elementary history in our case is 
a \defterm{trajectory} --- an imbedding of 
a $1$-dimensional manifold $\imfld$ (called the 
\defterm{inner space}) to a $\stdim$-dimensional 
\defterm{spacetime manifold} $\stmfld$ --- and an 
\defterm{inner space metric} $\imtrc$ on 
the inner manifold $\imfld$.
$\stmfld$ is equipped with a spacetime
metric $\stmtrc$ and scalar potential $\stpot$.  

In the Euclidian version of the theory $\imtrc$ 
is positive definite; in the physical version 
it is negative definite. In the Euclidian version
we allow the spacetime metric to be also positive 
definite\note{nt:Conventions}.

The whole theory is invariant under diffeomorphisms 
of the inner manifold $\imfld$. As usual 
(e.g. \cite{Polyakov:book}), we factorize
over this symmetry. If we fix a coordinate 
$\eta : \imfld \rightarrow \langle 0, 1 \rangle$
on the inner manifold $\imfld$ we can characterize
a class of equivalent histories using a pair 
$[\traj{x}, \tau]$, where $\traj{x}$ is a map
$\traj{x}\,:\, \langle0,1\rangle \quad\rightarrow\quad \stmfld$
and $\tau$ is a total \defterm{inner time} or a 
total \defterm{inner length}
of $\imfld$ measured using the inner metric $\imtrc$.

The Euclidian action\note{nt:WhyEuclidian}
in these variables has the form
\begin{equation}\label{particleaction}
  I(\tau,\traj{x}) =
  \frac12 \int_{\langle0,1\rangle} \left( \frac1{\isign\tau} 
  \tder{\traj{x}}^{\absidx{\alpha}}\tder{\traj{x}}^{\absidx{\beta}}
  \stmtrc_{\absidx{\alpha\beta}}(\traj{x}) +
  \isign \tau \stpot(\traj{x}) \right) \dvol\eta\period
\end{equation}
Here $\isign$ is a constant signature 
factor distinguishing Euclidian and physical versions 
of the theory. In the former case $\isign = 2$, in the later
it is multiplied by $i$.

In the sum-over-histories approach to quantum theory 
we can define an amplitude $\ampl{A}(\hist{H})$
for any set of histories $\hist{H}$ by \vague{summing} 
over amplitudes of elementary histories in the set. 
The quantum amplitude is not directly a physical 
measurable quantity. We need an additional notion of 
\defterm{distinguishable} or \defterm{decoherent} histories 
to give a probabilistic interpretation to the square of 
amplitudes. We expect that this notion has the same 
symmetry as the action and a measure on histories. 
This means that we will be always interested in amplitudes 
of sets of histories which are invariant under the action 
of the diffeomorphism group. For such sets we can factorize 
the  path integral and eliminate the reference 
to the diffeomorphism (e.g. \cite{Polyakov:book}).
In the factorized integral we are summing only over variables 
$[\traj{x}, \tau]$:
\begin{equation}
  \ampl{A}(\hist{H}) = \int_{[\tau, \traj{x}] \in \hist{H}}
  \measure{M}_{red}(\tau,\traj{x})
  \;\exp\bigl( -I(\tau,\traj{x}) \bigr)\comma
\end{equation}
with a reduced, renormalized measure $\measure{M}_{red}$.

\subsection{Propagator}

It is useful to compute an amplitude 
$\frac1{\stsign}\hkrnl(\tau,x_\fix|x_\iix)$ 
-- called the \defterm{propagator} or \defterm{heat kernel}
-- for the set of histories restricted only by positions 
of end points of the trajectory $x_\fix$ and $x_\iix$ 
in the spacetime $\stmfld$ and by fixing an inner time 
to a particular value $\tau$:
\begin{equation}\label{hkrnlampl}
  \frac1{\stsign} \hkrnl(\tau,x_\fix|x_\iix) =
  \int_{\traj{x}\in\trajset(x_\fix|x_\iix)} 
  \fmsr(\tau, x_\fix|x_\iix)[\traj{x}] \;
  \exp\bigl( -I(\tau,\traj{x}) \bigr) \commae
\end{equation}
where $\trajset(x_\fix|x_\iix)$ is a set of trajectories 
$\traj{x}:\langle0,1\rangle \rightarrow \stmfld$ with 
$\traj{x}(1) = x_\fix$ and $\traj{x}(0) = x_\iix$ and 
$\stsign$ is a constant factor\note{nt:stsignAndProp} governing 
signature of the spacetime metric $\stmtrc$.

Because the set of histories 
$[\tau,\trajset(x_\fix|x_\iix)] = \{\tau\} \times 
\trajset(x_\fix|x_\iix)$ 
is a lower dimensional subset of the space of all histories, 
$\hkrnl(\tau,x_\fix|x_\iix)$ is essentially 
an amplitude \vague{density} on the space 
$\realn^+\times\stmfld\times\stmfld$ of values 
$[\tau, x_\fix, x_\iix]$. Therefore we have to expect that 
the restriction $\fmsr(\tau, x_\fix|x_\iix)$ of the measure 
$\measure{M}_{red}$ to the space 
$[\tau,\trajset(x_\fix|x_\iix)]$, which we call the 
\defterm{Feynman measure}, depends on $\tau$ 
and end points $x_\fix$ and $x_\iix$; 
maybe only in a \vague{trivial} way. 

In other words, an amplitude density of 
an elementary history on the space 
$[\tau,\trajset(x_\fix|x_\iix)]$ is
\begin{equation}\label{freepartelhistampl}
  \ampl{A}(\tau,\traj{x}) = \fmsr(\tau,x_\fix|x_\iix)[\traj{x}]
  \,\exp\bigl(-I(\tau,\traj{x})\bigr) \period
\end{equation}

It is well known 
\cite{FeynmanHibbs:book,Parker:1979,DeWitt:1975,Hartle:LH1992}
that with the right choice of the measure 
$\fmsr(\tau, x_\fix|x_\iix)$ the propagator 
satisfies the equation\note{nt:DotsConv}
\begin{equation}\label{hkrnleq}
  - \tder\hkrnl(\tau) = 
  \frac \isign2 \scdst{F}\stint\hkrnl(\tau)\comma
  \hkrnl(0) = \stbivol^{\dash1}\commae
\end{equation}
where $\scdst{F}$ is a wave operator fixed by the action 
and the measure (see later), $\stbivol = \stvol\deltadst$ is 
a delta distribution\note{nt:DistrNot} on $\stmfld$ 
normalized to the metric volume element $\stvol$. 
In other words, $\hkrnl$ is the exponential of $\scdst{F}$
\begin{equation}\label{hkrnlisexp}
  \hkrnl(\tau) = 
  \exp\left( - \frac{\isign\tau}2 \scdst{F} \right)
  \stint \stbivol^{\dash1}\period
\end{equation}

We have not specified the \vague{right choice} 
of the measure yet. It can be a very problematic task 
from the pure mathematical point of view. 
Instead of trying to develop a measure theory 
on infinite dimensional spaces for oscillatory integrals 
(where the main problem lies), we take the usual approach 
of formal manipulations, and we define the measure 
by its decomposition properties and approximation 
for small time intervals. The former is given in 
equation \eqref{fmsrdecomp}, and the latter is given 
in equation \eqref{hkrnlstampl}.

The idea of the proof of the relations \eqref{hkrnleq} 
is in proving key properties of the exponential,
\begin{gather}
  \hkrnl(\tau_\fix) \stint \stbivol \stint \hkrnl(\tau_\iix) \;=\;
  \hkrnl(\tau_\fix + \tau_\iix) \commae \label{hkrnldecomp}\\
  \stbivol \stint \hkrnl(\tau) \stint \stbivol \;=\;
  \stbivol - \frac{\isign\tau}2 \biF + \order{\tau^2} \commae 
  \label{hkrnlapprox}
\end{gather}
where $\biF = \stbivol \stint \scdst{F}$ 
is the quadratic form of the differential operator $\scdst{F}$.

\subsection{Composition law}

The first condition \eqref{hkrnldecomp} is a composition law 
for the amplitude $\hkrnl$. This law reflects the possibility 
to decompose a history $[\tau,\traj{x}]$ into histories 
$[\tau_\iix,\traj{x}_\iix]$ during an initial amount of inner 
time $\tau_\iix$ and $[\tau_\fix,\traj{x}_\fix]$ during 
a final amount of inner time $\tau_\fix$. We say that a history 
$[\tau,\traj{x}] = 
[\tau_\fix,\traj{x}_\fix] \join [\tau_\iix,\traj{x}_\iix]$ 
is given by \defterm{joining} of histories
$[\tau_\fix,\traj{x}_\fix]$ and $[\tau_\iix,\traj{x}_\iix]$ if
\begin{equation}
    \tau = \tau_\fix + \tau_\iix \comma 
	\traj{x}_\fix(0) = \traj{x}_\iix(1)\comma
    \traj{x}_\fix = 
	\traj{x}(\frac{\tau_\iix + \eta \tau_\fix}{\tau}) \comma
    \traj{x}_\iix = \traj{x}(\frac{\eta \tau_\iix}{\tau}) \period
\end{equation}
The actions is additive with respect of joining histories.

We have a natural decomposition of the set of histories 
$[\tau,\trajset(x_\fix|x_\iix)]$ which defines the propagator 
$\hkrnl(\tau, x_\fix|x_\iix)$ to disjoint sets 
$[\tau_\fix,\trajset(x_\fix|x_\oix)] 
\times [\tau_\iix,\trajset(x_\oix|x_\iix)]$
\begin{equation}
  [\tau,\trajset(x_\fix|x_\iix)] = \bigcup_{x_\oix \in \stmfld}
  [\tau_\fix,\trajset(x_\fix|x_\oix)] 
  \times [\tau_\iix,\trajset(x_\oix|x_\iix)]\period
\end{equation}
If the measures on these sets are related by
\begin{equation}\label{fmsrdecomp}
  \fmsr(\tau, x_\fix|x_\iix)[\traj{x}] =
  \fmsr(\tau_\fix, x_\fix|x_\oix)[\traj{x}_\fix] \;\stsign\stvol(x_\oix)\;
  \fmsr(\tau_\iix, x_\oix|x_\iix)[\traj{x}_\iix] \commae
\end{equation}
we get \eqref{hkrnldecomp} by a straigthforward calculation.

The condition \eqref{fmsrdecomp} represents a reasonable 
assumption of the locality of the measure $\fmsr$. 
Together with the additivity of the action 
it reflects the rule of the sum-over-histories approach 
to quantum mechanics --- that the amplitude of independent 
(here consequent) events is given by multiplication of 
individual amplitudes. This condition is the first part 
of our definition of the measure. Now we know how to construct 
the measure $\fmsr(\tau)$ for some time $\tau$ from measures 
for shorter time intervals. To conclude the definition of 
the measure, we need to specify it for an infinitesimally 
short inner time interval. This moves us to an investigation 
of the short time behavior of the heat kernel.

\subsection{Short time amplitude}

Now we turn to prove equation \eqref{hkrnlapprox}. 
It can be found in the literature (e.g. \cite{Parker:1979,DeWitt:1975}), 
but we present it here to show how the measure is actually 
determined and how the operator $\biF$ depends on this choice.

We ignore technical difficulties in the definition of 
the path integral, and we assume that this integral has 
most of the properties of a usual integral in 
a finite-dimensional manifold. This allows us to find 
the short time behavior for the propagator.

First we write an expansion of the action for small $\tau$
\begin{equation}
  I(\tau,\traj{x}) = \frac1\tau\, I_{-1}(\traj{x}) + I_{0}(\traj{x})
  + \tau\, I_{1}(\traj{x}) + \dots \period
\end{equation}
For the action we are using it means
\begin{align}
  I_{-1}(\traj{x}) &= \frac1{2\isign} \int_{\eta\in\langle0,1\rangle}
     \tder{\traj{x}}^{\absidx{\alpha}} \tder{\traj{x}}^{\absidx{\beta}}
     \stmtrc_{\absidx{\alpha\beta}}(\traj{x}) \dvol\eta \commae\label{actkin}\\
  I_{0}(\traj{x}) &= 0 \commae\\
  I_{1}(\traj{x}) &= \frac\isign2 \int_{\eta\in\langle0,1\rangle}
     \stpot(\traj{x}) \dvol\eta \period
\end{align}
We assume that the measure is slowly changing in 
$\tau$ compared to the leading term in the action.

The dominant contribution to the integral \eqref{hkrnlampl} 
comes from an extremum $\extr{\traj{x}}(x_\fix|x_\iix)$ of 
the leading term $I_{-1}$ in the exponent. 
But the extremum of the functional \eqref{actkin} 
is clearly a geodesic of the metric $\stmtrc$. 
We expand all expressions around this extremum
\begin{equation}\label{changextoX}
  \traj{x} = \extr{\traj{x}}(x_\fix|x_\iix) + \sqrt{\tau} \vec{\traj{x}} \comma
\end{equation}
where $\vec{\traj{x}}$ is a tangent vector to the space of trajectories 
$\trajset(x_\fix|x_\iix)$ at the extremum $\extr{\traj{x}}(x_\fix|x_\iix)$. 
We actually need to specify what the addition in the last equation means. 
It will be done more carefully in a similar situation in appendix 
\ref{apx:AEHK} (see eq. \eqref{ChangezToZ}). But now we are interested 
more in a qualitative answer, so we skip these details here. 
The expanded integral \eqref{hkrnlampl} has the structure
\begin{equation}\label{hkrnlamplexpand}
\begin{split}
  \frac1\stsign\,\hkrnl(\tau,x_\fix|x_\iix) =
     \exp\Bigl(- I\bigl(\extr{\traj{x}}(x_\fix|x_\iix)\bigr)\Bigr)
     &\int_{\vec{\traj{x}} \in \tens_{\extr{\traj{x}}}\,\trajset}
     \fmsr_{*}(\tau,x_\fix|x_\iix)
     \exp\left( - \frac12 \vec{\traj{x}}\cdot
     \variat^2 I_{-1}(\extr{\traj{x}}(x_\fix|x_\iix))
     \cdot\vec{\traj{x}} \right) \stimes\\
  &\qquad\stimes\Bigl( 1
     + \sqrt{\tau}\, ( \vec{\traj{x}}^{\text{odd}}\text{- terms} )
     + \tau\, ( \vec{\traj{x}}^{\text{even}}\text{- terms} )
     + \dots \Bigr) \period
\end{split}
\end{equation}

Here $\fmsr_{*}(\tau,x_\fix|x_\iix)$ is a leading term in the 
$\tau$ and $\vec{\traj{x}}$-expansion of the measure 
$\fmsr(\tau,x_\fix|x_\iix)$ after change of variables 
$\traj{x} \rightarrow \vec{\traj{x}}$. 
$\fmsr_{*}$ is a constant measure on the tangent vector space 
to the space of trajectories $\trajset(x_\fix|x_\iix)$. 
The actual dependence on $\vec{\traj{x}}$ is hidden in higher 
terms of the $\vec{\traj{x}}$-expansion. As a leading term in the 
$\tau$-expansion, $\fmsr_{*}$ depends on $\tau$ in a trivial way 
--- it is proportional to a power of $\tau$. Of course, this statement 
is formal --- the exponent of $\tau$ in $\fmsr_{*}$ is of the order 
of the dimension of the tangent space, which is infinite. 

\vague{$\vec{\traj{x}}$-terms} in the last equation represents 
terms resulting from the expansion of the action and the measure; 
$\vec{\traj{x}}^{\text{odd}}$ or $\vec{\traj{x}}^{\text{even}}$ 
suggest that $\vec{\traj{x}}$ occurs in these terms in odd or 
even power. For convenience we combined the term $\tau I_{1}$ 
into the prefactor despite the fact that it could be 
included among terms proportional to $\tau$.

The value $\isign I_{-1}(\extr{\traj{x}}(x_\fix|x_\iix))$ 
is a well-known quantity called the world function, 
or half the squared geodesic distance\note{nt:DotsConv},
\begin{equation}\label{stwfDef}
  \stwf(x_\fix|x_\iix) = \isign I_{-1}(\extr{\traj{x}}(x_\fix|x_\iix)) =
  \frac12 \int_{\langle0,1\rangle}
  \tder{\extr{\traj{x}}}(x_\fix|x_\iix) \stctr \stmtrc(\extr{\traj{x}}(x_\fix|x_\iix))
  \stctr \tder{\extr{\traj{x}}}(x_\fix|x_\iix)\,\dvol\eta\period
\end{equation}
We also use the notation
\begin{equation}
  \extr{\stpot}(x_\fix|x_\iix) = 
    \frac2\isign I_{1}(\extr{\traj{x}}(x_\fix|x_\iix)) =
    \int_{\langle0,1\rangle} \stpot(\extr{\traj{x}}(x_\fix|x_\iix))\, \dvol\eta \period
\end{equation}

The integral \eqref{hkrnlamplexpand} is a simple Gaussian integration. 
(In fact, one approach to defining infinite-dimensional 
integrals is through the definition of a \vague{Gaussian} 
measure which in our case would be 
$\fmsr_{*}\exp(-\frac12\,\vec{\traj{x}}\cdot\variat^2 I_{-1} 
\cdot \vec{\traj{x}})$). 
The integration can be performed, at least formally. If the measure
$\fmsr_{*}(\tau)$ has the already mentioned $\tau$-dependence, 
the result can be written
\begin{equation}\label{hkrnlstampl}
\begin{split}
  \hkrnl(\tau,x_\fix|x_\iix) \;=\;
    \frac\stsign{(2\pi\isign\tau)^{\frac\stdim2}}  \stvvmd(x_\fix|x_\iix) \; &
    \Bigl( \alpha_0(x_\fix|x_\iix) - \tau \frac\isign2\, \alpha_1(x_\fix|x_\iix)
    + \order{\tau^2} \Bigr)\stimes\\
    & \quad\stimes\exp\left( - \frac1{\isign\tau} \stwf(x_\fix|x_\iix)
    - \frac{\isign\tau}2 \extr{\stpot}(x_\fix|x_\iix) \right)\commae
\end{split}
\end{equation}
where $\alpha_0(x_\fix|x_\iix)$ should satisfy
\begin{equation}\label{alphazeroCL}
  \alpha_0(x|x) = 1 \period
\end{equation}
Here $\stvvmd(x_\fix|x_\iix)$ is Van Vleck-Morette determinant 
(see \eqref{stvvmdDef}). 

The terms proportional to $\sqrt{\tau}$ disappeared during 
integration, thanks to the odd power of $\vec{\traj{x}}$. 
The particular behaviour a coincidence limit of the coefficient 
$\alpha_0$ will be needed for a proper normalization in \eqref{hkrnleq}.
To obtain this behaviour we need the mentioned $\tau$-dependence 
of the measure, which can be expressed by condition
\begin{equation}\label{fmsrCL}
  \fmsr_{*}(\tau,x|x)
  \Det\left(\frac{\variat^2 I_{-1}(\extr{\traj{x}}(x|x))}
  {2\pi\isign\tau}\right)^{\!-\frac12} =
  \frac1{(2\pi\isign\tau)^{\frac\stdim2}} \period
\end{equation}
I.e., the meassure must satisfy this condition to conlude proof of 
equation \eqref{hkrnlapprox}.

Coefficients in front of 
powers of $\tau$ could be expressed in terms of variations of 
the action and the measure. But because we did not specify 
the measure precisely yet, we can do it now by fixing 
these coefficients. I.e., we can define the measure 
$\fmsr$ by choosing functions $\alpha_0(x_\fix|x_\iix)$ 
and $\alpha_1(x_\fix|x_\iix)$.

In the following we will prove that a form of the operator 
$\biF$ in \eqref{hkrnlapprox} depends only on the 
coincidence limits of $\alpha_1$ and the first two derivatives 
of $\alpha_0$. So we can ignore terms with higher power 
of $\tau$ in eq. \eqref{hkrnlstampl}. As discussed before, 
equations \eqref{hkrnlapprox} together with composition 
law \eqref{hkrnldecomp} determines the propagator $\hkrnl$, 
i.e. also all important information hidden in the measure $\fmsr$. 
This means that a knowledge of the mentioned coincidence 
limits concludes our definition of the measure and 
path integral itself.

Let us note that this argument is some kind of justification 
of the usual time-discretization of the path integral 
and of a priori choice of the short time amplitude 
in the form \eqref{hkrnlstampl}. But in principle it 
would be possible to define the measure $\fmsr$ in some 
more compact way and compute exactly the form of 
the functions $\alpha_0$ and $\alpha_1$ in terms 
of variation of the action and the measure.

\subsection{Short time behavior of the heat kernel}

Now we continue with the proof of equation \eqref{hkrnlapprox}. 
We show that for small $\tau$ the amplitude \eqref{hkrnlstampl} 
has the desired behavior in a distributional sense. 
Most of the technical work is done in appendix \ref{apx:AEHK} 
where it is shown that for small $\tau$ the following expansion 
holds (equation \eqref{hkrnlstamplexpres})
\begin{equation}\label{hkrnlstamplexpsm}
\begin{split}
  \frac\stsign{(2\pi\isign\tau)^{\frac\stdim2}}\, 
  \int_{x,z\in\stmfld} \stvol(x) \stvol(z)\;  \stvvmd(x|z)\; &
  \exp\Bigl( - \frac1{\isign\tau} \stwf(x|z)\Bigr) \varphi(x) \psi(z) = \\
  &= \varphi \stint \stbivol \stint \psi \;-\; 
  \tau \frac\isign2\; \varphi\stint\bidst{L}\stint\psi \;+\; \order{\tau^2} \commae
\end{split}
\end{equation}
where $\varphi$ and $\psi$ are smooth test functions 
and $\bidst{L}$ is the Laplace operator quadratic 
form\note{nt:DotsConv}
\begin{equation}
\begin{split}
  \varphi\stint\bidst{L}\stint\psi &=
  \int_{\stmfld} \stvol\;
  (\grad\varphi)\stctr\stmtrc^{\dash1}\stctr (\grad\psi) =\\
  &= - \int_{\stmfld} \stvol \varphi \left( \stlaplace\psi \right) 
  \;=\; - \int_{\stmfld} \stvol \psi \left( \stlaplace\varphi \right) \period 
\end{split}
\end{equation}
Let us remember that now we are discussing the case of 
a manifold without boundary, and therefore we 
do not have to worry about boundary conditions 
for the Laplace operator and integration by parts.

Using this result it is easy to show that
\begin{align}
\begin{split}
  {}&\varphi\stint\stbivol\stint\hkrnl(\tau)\stint\stbivol\stint\psi \;=
  \nonumber\\
  &\qquad\qquad= \frac\stsign{(2\pi\isign\tau)^{\frac\stdim2}}\, 
  \int_{x,z\in\stmfld} \stvol(x) \stvol(z) \stvvmd(x|z)
  \left(\alpha_0(x|z) - \tau\frac\isign2\,\alpha_1(x|z)
  + \order{\tau^2}\right) \stimes\nonumber\\
  &\mspace{300mu}\stimes\exp\left( - \frac1{\isign\tau} \stwf(x|z) -
  \frac{\isign\tau}2\extr{\stpot}(x|z)\right)\varphi(x) \psi(z) = 
  \end{split}\displaybreak[0]\\
\begin{split}
  {}&\qquad\qquad=\;\varphi \stint \stbivol \stint \psi \,-\\
  &\qquad\qquad\qquad -\, \tau \frac\isign2\; \varphi\stint\Bigl(
  \bidst{L} + \bigl(\stpot\stbivol\bigr) +
  \bigl(\stmtrc^{\dash1\absidx{\mu\nu}} 
  \stCL{\grad_{\argl\absidx{\mu}}\grad_{\argr\absidx{\nu}}\alpha_0} \stbivol
  \bigr) + \bigl(\stCL{\alpha_1}\stbivol\bigr)\Bigr)\stint\psi \,-\\
  &\qquad\qquad\qquad -\, \tau \frac\isign2\; \varphi\stint\Bigl(
  \bigl(\bigradl_{\absidx{\mu}}\stmtrc^{\dash1\absidx{\mu\nu}}
  \stCL{\grad_{\argl\absidx{\nu}}\alpha_0}\bigr) \stint \stbivol +
  \stbivol\stint \bigl( \stCL{\grad_{\argr\absidx{\mu}}\alpha_0}
  \stmtrc^{\dash1\absidx{\mu\nu}} \bigradr_{\absidx{\nu}}\bigr) \Bigr)\stint\psi \,+\\
  &\qquad\qquad\qquad +\, \order{\tau^2} \commae
  \end{split}\nonumber
\end{align}
where we used $\stCL{\extr{\stpot}} = \stpot$ and $\stCL{\alpha_0} = 1$. 
Here the $\stCL{A}$ denotes a coincidence limit of a bitensor $A$, 
$\grad_\argr A$ and $\grad_\argl A$ are derivatives with respect 
of the right and left arguments and the bi-distributions 
$\bigradl$ and $\bigradr$ are derivatives acting to the 
left and to the right\note{nt:DistrNot}.

If the condition
\begin{equation}\label{minimalfmsr}
  \stCL{\grad_\argr\alpha_0} = \stCL{\grad_\argl\alpha_0} = 0
\end{equation}
is satisfied, we see that the propagator $\hkrnl(\tau)$ 
has really the form \eqref{hkrnlapprox} with
\begin{gather}
  \biF = \bidst{L} + \bidst{V}\commae\\
  \bidst{V} = \bigl( \stpot + \stCL{\alpha_1} +
  \stmtrc^{\dash1\,\absidx{\mu\nu}}
  \stCL{\grad_{\argl\absidx{\mu}}\grad_{\argr\absidx{\nu}}\alpha_0}
  \bigr)\stbivol\period \label{bistpot}
\end{gather}
I.e., $\biF$ is a Laplace operator with a potential 
term which include the original potential $\stpot$ 
from the action and additional parts 
depending on the choice of the measure.

A common choice for $\alpha_0$  is 
a power of the Van Vleck-Morette determinant
\begin{equation}\label{PowerOfstvvmd}
  \alpha_0 = \stvvmd^{\!- p}\commae
\end{equation}
which satisfies the condition \eqref{minimalfmsr}. 
It leads to an additional part in the potential,
\begin{equation}
  \stmtrc^{\dash1\,\absidx{\mu\nu}}
  \stCL{\grad_{\argl\absidx{\mu}}\grad_{\argr\absidx{\nu}}\stvvmd^{\!- p}}
   = \frac p3\stscur \commae
\end{equation}
where $\stscur$ is a scalar curvature of the metric $\stmtrc$.

The condition \eqref{minimalfmsr} is actually 
a consequence of \eqref{alphazeroCL} 
and an assumption of the symmetry of $\alpha_0$
\begin{equation}\label{alphazerosym}
  \alpha_0(x|z) = \alpha_0(z|x)\period
\end{equation}
It is a natural assumption in the case when 
the theory is symmetric under trajectory reversal. 
However this condition does not have to be satisfied 
if there is a preferred path direction as for example 
in the case of interaction with an electromagnetic field. 
But we will not discuss such a situation, 
and in the following we will assume that the conditions 
\eqref{alphazerosym} and \eqref{minimalfmsr} are satisfied.   

In summary, we have seen that for small $\tau$ the propagator 
$\hkrnl(\tau)$ has the behavior given by \eqref{hkrnlapprox}.
If the measure is defined using the decomposition property 
\eqref{hkrnldecomp} and the short time amplitude \eqref{hkrnlstampl}, 
the operator $\biF$ is fixed by knowledge of the coincidence 
limits of $\alpha_1$ and the first two derivatives of $\alpha_0$.

\subsection{Feynman Green Function}

For the relativistic particle the inner time is physically 
undetectable and therefore any physical set of histories 
will include elementary histories with all possible inner times. 
Therefore we are interested in the amplitude called 
the \defterm{Feynman Green function}
$\frac1\stsign\GFF(x_\fix|x_\iix)$ associated with 
the set of histories restricted only by the initial 
and final points $x_\fix, x_\iix$. 
We can obtain it from the propagator 
$\frac1\stsign\hkrnl(\tau,x_\fix|x_\iix)$ 
by summing over all possible inner times $\tau$ 
using the measure $\frac\isign2\dvol\tau$
\begin{equation}
  \frac1\stsign\GFF(x_\fix|x_\iix) =
  \int_{\realn^+} \frac1\stsign\hkrnl(\tau,x_\fix|x_\iix)
  \,\frac\isign2\dvol\tau\period
\end{equation}
Using eq. \eqref{hkrnlisexp} we immediately get that 
the Feynman Green function is 
the inverse of the wave operator $\biF$.

\subsection{Boundary conditions}

In this section we completely ignored the question 
of boundary conditions using as excuse that we are 
working in a manifold without boundary. Certainly 
this is correct, if the manifold $\stmfld$ is compact. 
But it is also correct in the case of a non-compact 
manifold with a sufficiently \vague{nice} metric 
$\stmtrc$ at infinities. In such cases there exists 
a canonical choice of boundary conditions 
for differential operators used above, and these boundary 
conditions usually allow us to integrate by parts. 
But problem arise for the relativistic particle 
when the manifold $\stmfld$ is Lorentzian and 
operators as $\bidst{L}$, $\biF$ are hyperbolic. 
In this case the choice of boundary conditions 
at the temporal infinities plays an important 
physical role, and it is worth further investigation. 
First let us note that in a special situations 
(e.g. existence of a time-like Killing vector 
in the distant past and future) a canonical choice 
of boundary conditions still exists. 
But \emph{canonical} here essentially means 
the most natural physical choice. 
In a general spacetime we do not have this special choice, 
and we have to address the question of 
the boundary conditions. To deal with this problem, 
in the next section we will investigate our theory 
in a bounded domain $\Omega$ of the manifold $\stmfld$.


\section{Particle in a curved space with boundary}
\label{sc:SOH-partwbnd}

\subsection{General consideration}

In this section our goal is a better understanding 
of the physical meaning of boundary conditions for 
the differential operators in the equation for 
the propagator and Green function. 
Therefore we restrict ourselves to a domain $\Omega$ 
in the manifold $\stmfld$ which is bounded 
\vague{in all physically interesting directions}. 
This means that we  will investigate the boundary 
conditions on boundaries which are not at infinity. 

If the target space metric is positive definite
(i.e. in Euclidian and non-relativistic version of the theory)
all problems with infinities could be solved by 
a restriction to a compact domain.
But in the case of relativistic particle, 
which we are mostly intersted in, 
we also allow domains which do not have to be compact,
we allow the domain $\Omega$ to be unbounded 
if we know that its infinity is \vague{safe}.

The situation we have in the mind is the 
Lorentzian globally hyperbolic manifold with 
asymptotically flat spatial infinity. In this case 
we can ignore spatial infinity because it makes a 
sense to restrict ourselves to situations in 
which spacetime is \vague{empty} sufficiently 
far in space directions. But because of the hyperbolic 
nature of the evolution equation we cannot ignore 
boundary conditions in the time directions. 
They represent \vague{initial} and \vague{final} 
conditions of the system. And we want 
to understand exactly this relationship.

Therefore in the case of a Lorentzian globally 
hyperbolic manifold, the typical choice of 
the  domain will be a sandwich domain between 
two Cauchy surfaces. But whole discussion is also
valid for Euclidian version of the theory restricted
to a compact domain.

\subsection{Restriction to a domain --- naive approach}

Let us start with a straighforward restriction to a 
domain $\Omega$. We want to compute an amplitude 
$\hkrnl_\bdo(\tau,x_\fix|x_\iix)$ which corresponds 
to a set of histories with inner time $\tau$, 
endpoints $x_\iix$, $x_\fix$ and which wholly belong 
to the domain $\Omega$. We can repeat the derivation 
of the short time amplitude \eqref{hkrnlstampl}, 
at least for $x_\fix$, $x_\iix$ sufficiently 
far from the boundary, because for small $\tau$ 
only trajectories near to the geodesic 
between $x_\fix$ and $x_\iix$ contribute to the amplitude. 

If we do this calculation, we find that there is 
a new term in the expansion. As can be seen in 
eq. \eqref{hkrnlstamplexpcomp}, the smoothed 
short time amplitude leads to a Gauss integration 
in a variable $Z$ from a tangent space at a point $x$, 
and in the case of a space without boundary 
the integration of odd powers of $Z$ disappears. 
But in the case of a domain with a boundary for 
a point $x$ near the boundary the Gauss integration 
is not always over the whole tangent space and 
therefore the integral of odd powers of $Z$ 
does not disappear. As shown in appendix \ref{apx:AEHK} 
(equation \eqref{hkrnlwbstamplexp}), the correct 
asymptotic expansion of the leading term of 
the short time amplitude 
(equivalent of \eqref{hkrnlstamplexpsm}) is given by
\begin{equation}\label{hkrnlwbstamplexpsm}
\begin{split}
  \frac\stsign{(2\pi\isign\tau)^{\frac\stdim2}}\, 
  \int_{x,z\in\stmfld} \stvol(x) &\stvol(z)   \stvvmd(x|z)
  \exp\left( - \frac1{\isign\tau} \stwf(x|z)\right) \varphi(x) \psi(z) = \\
  &= \varphi \stint \left( \stbivol \,+\,
  \sqrt{\tau} \bigl(-\frac1\stsign\sqrt{\frac\isign{2\pi}}\bigr)\bbivol\,-\,
  \tau \frac\isign2\, \acttolr{\bidst{L}} \,+\,
  \order{\tau^{\frac32}} \right) \stint\psi\commae
\end{split}
\end{equation}
where $\bbivol[\bound\Omega]$ is a delta bi-distribution 
localized on the boundary normalized to the boundary 
volume element $\svol$ understood as a distribution on spacetime,  
\begin{equation}
  \varphi\stint\bbivol\stint\psi =
  \int_{\bound\Omega} \varphi\psi\,\svol\commae
\end{equation}
and $\acttolr{\bidst{L}}$ is a particular 
ordering of the Laplace operator given by
\begin{gather}
  \acttolr{\bidst{L}} =
  \frac12\,\bigl(\,\acttol{\bidst{L}} + \acttor{\bidst{L}}\,\bigr)\commae\\
  \varphi\stint\acttor{\bidst{L}}\stint\psi =
  - \int_{\Omega} \varphi\bigl(\stlaplace\psi\bigr) \stvol\period
\end{gather}

Using this result it is easy to show that 
the expansion of the propagator $\hkrnl_\bdo$ is
\begin{equation}
  \stbivol\stint\hkrnl_\bdo(\tau)\stint\stbivol = \stbivol +
  \sqrt{\tau} \bigl(-\frac1\stsign\sqrt{\frac\isign{2\pi}}\bigr)\bbivol
  \,-\,\tau \frac\isign2\, \biFlr\,+\,
  \order{\tau^{\frac32}} 
\end{equation}
and $\biFlr$ is a Laplace-like quadratic form with potential,
\begin{gather}
  \biFlr=\frac12\bigl(\biFl + \biFr\bigr) =
  \acttolr{\bidst{L}} + \bidst{V}\commae\\
  \biFl{}^\qformT = \biFr = \acttor{\bidst{L}} + \bidst{V}\period
\end{gather}
The corrected potential $\bidst{V}$ is given again 
by the expression \eqref{bistpot} and 
we have assumed that the condition \eqref{minimalfmsr} is satisfied.

We see that the expansion of the propagator has 
an additional term localized on the boundary 
$\bound\Omega$ proportional to $\sqrt{\tau}$. 
This $\tau$-dependence causes a problem because 
$\tder\hkrnl_\bdo(0)$ is singular on the boundary. 
An origin of the singular term on the boundary 
lies in our careless approximation of the propagator 
by the short time amplitude \eqref{hkrnlstampl}. 
This approximation is correct only for endpoints 
sufficiently far from the boundary. For points 
near the boundary we have to investigate 
the structure of the propagator more thoroughly.

\subsection{Boundary correction term}

The short time amplitude \eqref{hkrnlstampl} 
represents the dominant contribution to 
the heat kernel from trajectories near 
the geodesic joining endpoints $x_\fix$ and $x_\iix$. 
But in the case of a sum over trajectories restricted 
to the domain $\Omega$ there are other dominant 
terms given by contributions of trajectories 
near extremal paths which reflect on the boundary.

\begin{figure}
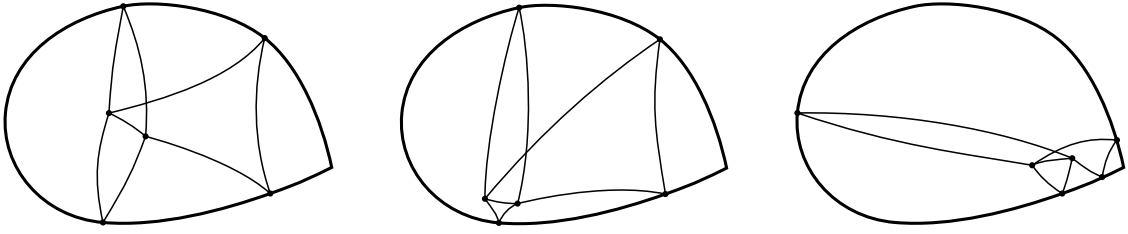

  \begin{center}
  \epsfig{file=ENDPTIN.AI,width=0.3\textwidth}
  \hspace{0.03\textwidth}
  \epsfig{file=ENDPTBD.AI,width=0.3\textwidth}
  \hspace{0.03\textwidth}
  \epsfig{file=ENDPTCR.AI,width=0.3\textwidth}
  \end{center}
  \caption{Example of extremal trajectories.}
  \figdescription{
    Dominant terms to sum over trajectories 
	are given by trajectories near to extreme 
	trajectories, possibly reflected from the boundary. 
	If close endpoints are sufficiently far from 
	the boundary, the reflected geodesics are 
	longer then the straight geodesic. 
	If the endpoints are near the boundary, 
	there is a reflected geodesic with 
	the length comparable to the length of 
	straight one. Near the corner there are 
	more reflected geodesics with comparable length.
	}
  \label{fig:EndPointFar}
\end{figure}

In general we should take into account trajectories 
with an arbitrary number of reflections on 
the boundary and compute the dominant contributions 
from all of them. However, for endpoints sufficiently 
far from the boundary the contributions from the 
reflected paths are negligible compared 
to the straight geodesic --- for small $\tau$ 
only short paths contribute to the sum, 
and any trajectory with a reflection on the boundary 
is too long (see figure \ref{fig:EndPointFar}).

But for endpoints near to the boundary 
the contributions from the reflected trajectories 
can be comparable with the leading term. 
For the endpoints near a smooth boundary 
there exists exactly one extreme trajectory 
$\extr{\traj{x}}_\bnd(x|z))$ with one reflection 
which gives a contribution comparable to 
the contribution from the straight geodesic 
$\extr{\traj{x}}(x|z))$ (see figure \ref{fig:EndPointFar}). 

The \defterm{reflected extreme trajectory 
$\extr{\traj{x}}_\bnd(x|z))$} is an extremum of 
the leading term of the action \eqref{actkin} 
with the additional condition that 
the trajectory reflects on the boundary. 
Let's denote the \defterm{point of the reflection 
$\rflpt(x|z)$} and the \defterm{parameter for which 
the reflection occurs} $\rfltmr(x|z)$ and 
its complement $\rfltml(x|z)$.
Clearly the trajectory is a joining of two geodesics
\begin{equation}
  [\tau,\extr{\traj{x}}_\bnd(x|z)] =
  [\rfltml(x|z)\tau,\extr{\traj{x}}(x|\rflpt(x|z))] \join
  [\rfltmr(x|z)\tau,\extr{\traj{x}}(\rflpt(x|z)|z)]\period 
\end{equation}
Using additivity of the action we get the value of its leading term 
\begin{equation}
  \stwf_\bnd(x|z) \defeq \isign I_{-1}(\extr{\traj{x}}_\bnd(x|z)) =
  \frac{\stwf_\bdl(x|z)}{\rfltml(x|z)}  +
  \frac{\stwf_\bdr(x|z)}{\rfltmr(x|z)} \commae
\end{equation}
where, following the convention \eqref{lrIdxConvention},
\begin{equation}
  \stwf_\bdl(x|z) = \stwf(x|\rflpt(x|z)) \comma 
  \stwf_\bdr(x|z) = \stwf(\rflpt(x|z)|z) \period
\end{equation}
The extremum requirement gives us conditions 
on $\rflpt$ and $\rfltml$, $\rfltml$,
\begin{gather}
  \frac{\sgrad\stwf(x|\rflpt(x|z))}{\rfltml(x|z)} +
  \frac{\sgrad\stwf(\rflpt(x|z)|z)}{\rfltmr(x|z)}
  = 0 \commae\\
  \frac{\stwf_\bdl}{\rfltml^2} = \frac{\stwf_\bdr}{\rfltmr^2}\commae
\end{gather}
where $\sgrad$ denotes the orthogonal projection 
of the gradient on the boundary (the gradient 
with respect of the argument on the boundary).
See appendix \ref{apx:GdTh} for more details 
and other quantities defined on the boundary.

Now we can estimate the contribution from 
the trajectories near to the reflected geodesic 
$\extr{\traj{x}}_\bnd(x|z))$. 
Using reasoning similar to that used for deriving 
\eqref{hkrnlstampl}, we can write an approximation 
of the short time amplitude associated 
with the reflected geodesic as
\begin{equation}\label{boundhkrnlstampl}
  \hkrnl_\bnd(\tau,x_\fix|x_\iix) \;=\;
    \frac\stsign{(2\pi\isign\tau)^{\frac\stdim2}}
    \stvvmd_\bnd^{1-p}(x_\fix|x_\iix) \;\beta(\tau,x_\fix|x_\iix) 
    \exp\Bigl( - \frac1{\isign\tau} \stwf_\bnd(x_\fix|x_\iix)
    \Bigr)\commae
\end{equation}
where $\stvvmd_\bnd$ is Van Vleck-Morette determinant 
associated with the reflected geodesic (see \eqref{stvvmdBndDef}). 
The coefficient $\beta$ is an analog of 
the coefficients $\alpha_0$, $\alpha_1$, 
only in this case we have to expect 
an expansion in powers of $\sqrt{\tau}$:
\begin{equation}
  \beta(\tau,x|z) = \beta_0(x|z) +
  \sqrt{\tau} \beta_{\frac12}(x|z) + 
  \order{\tau}\period
\end{equation}
As we will see, the right normalization 
relative to the leading term $\hkrnl_\bdo$ requires
\begin{equation}\label{betazeroCL}
  \beta_0(x|x) = 1 \period
\end{equation}
We did not bother to write down a potential term, 
because terms of order $\order{\tau}$ 
are negligible in the approximation we need, 
as can be seen in the calculation in appendix \ref{apx:AEHK}. 
We also already anticipated an arbitrary power of 
the Van Vleck-Morette determinant, 
similarly to the choice \eqref{PowerOfstvvmd}.

Fixing this short time amplitude 
(i.e. specification of coefficients $p$ and $\beta$, 
or more precisely its coincidence limits as we will see below) 
together with amplitude \eqref{hkrnlstampl}  
concludes the definition of the path integral 
in the domain with a smooth boundary.

\subsection{Short time behavior of the heat kernel}

Next we proceed to derive the short 
time behavior of the propagator. 
Again, the technical work is done in appendix \ref{apx:AEHK}, 
where it is shown that for small $\tau$ we have the expansion 
(see eq. \eqref{hkrnlwbstamplbndexp})
\begin{equation}
  \begin{split}
  &\varphi\stint\stbivol\stint\hkrnl_\bnd(\tau)
  \stint\stbivol\stint\psi =\\
  &\qquad\qquad= 
  \sqrt{\tau} \frac1\stsign\sqrt{\frac\isign{2\pi}} \,
  \varphi\stint\bbivol\stint\psi -\\
  &\qquad\qquad\qquad - \tau \frac\isign2\, \frac12
  \varphi\stint(\biFdl_{\revidx\bbcnd{n}} + \biFdr_{\revidx\bbcnd{n}}) \stint\psi -\\
  &\qquad\qquad\qquad - \tau \frac\isign2\, \varphi\stint
  \Bigl(\frac{1+p}3\extscur + \beta\text{-terms}\Bigr)\bbivol
  \stint\psi + \order{\tau^\frac32}\period
\end{split}
\end{equation}
The $\beta$-terms contains coincidence limits of 
the first two derivatives of the coefficient $\beta$ 
on the boundary, and the exact form can be found in 
\eqref{BetaTerms}. $\extscur$ is the trace of 
the external curvature \eqref{extcurDef} and 
$\biFdl_{\revidx\bbcnd{n}}$, $\biFdr_{\revidx\bbcnd{n}}$
are defined in \eqref{biFdrldefexplicit}.
The normalization \eqref{betazeroCL} ensures 
that the $\sqrt{\tau}$-term in $\hkrnl_\bnd$ 
cancels exactly with such a term in $\hkrnl_\bdo$.

So, if we add both dominant terms we get
\begin{gather}
  \hkrnl_{\bbcnd{k}}(\tau) = 
  \hkrnl_\bdo(\tau) + \hkrnl_\bnd(\tau) \commae \\
  \stbivol\stint\hkrnl_{\bbcnd{k}}(\tau)
  \stint\stbivol = 
  \stbivol - \tau \frac\isign2 \biF_{\revidx\bbcnd{k}} +
  \order{\tau^\frac32}\commae \label{ShTmHKBehWBound}
\end{gather}
where $\biF_{\revidx\bbcnd{k}}$ is the quadratic form 
of Laplace-like operator with the boundary conditions 
given by the choice of $\beta$ coefficients\note{nt:RevIndx}
\begin{gather}
  \biF_{\revidx\bbcnd{k}} = \biF_{\revidx\bbcnd{n}} - \biTh_{\bbcnd{k}}\comma
  \label{biFkUsingbiFnTh}\\
  \biF_{\revidx\bbcnd{n}} = \bigradl_{\absidx{\alpha}} \stint
  (\charfc \stmtrc^{\dash 1\,\absidx{\alpha\beta}} \,\stbivol)
  \stint \bigradr_{\absidx{\beta}} +
  (\charfc \bidst{V})\commae\label{biFdomega}\\
  \biTh_{\bbcnd{k}} \defeq
  - \Bigl(\frac{1+p}{3}\extscur +
  \beta\text{-terms}\Bigr) \bbivol \comma
\end{gather}
where $\charfc$ is the characteristic 
function of the domain $\Omega$. Here $\biF_{\revidx\bbcnd{n}}$ is
the standard quadratic form which appears in the action for
non-interacting scalar field with a potential evaluated
on the domain $\Omega$ and $\biTh_{\bbcnd{k}}$ is a bi-distribution
localized on the boundary.

The short time behavior \eqref{ShTmHKBehWBound} 
together with the composition law again prove 
the heat equation for the propagator:
\begin{equation}
  -\stbivol\stint\tder\hkrnl_{\bbcnd{k}}(\tau) = \frac\isign2
  \biF_{\revidx\bbcnd{k}}\stint\hkrnl_{\bbcnd{k}}(\tau)
\end{equation}

The quadratic form $\biF_{\revidx\bbcnd{k}}$ identifies 
what kind of boundary conditions the propagator 
$\hkrnl_{\bbcnd{k}}$ satisfies. If we compare 
$\biF_{\revidx\bbcnd{k}}$ with the operator $\biFr$
(all derivatives act to the right) we find
\begin{gather}
  \biF_{\revidx\bbcnd{k}} = \biFr + \biFdr_{\revidx\bbcnd{k}}
  \comma\label{biFRels} \\
  \biFdr_{\revidx\bbcnd{k}} =
  \biFdr_{\revidx\bbcnd{n}} - \biTh_{\bbcnd{k}}\comma
\end{gather}
where the bi-distribution $\biFdr_{\revidx\bbcnd{n}}$ is
essentially integration of a value of the left argument
and a momenta of the right argument over boundary of the domain
\begin{equation}
  \varphi\stint\biFdr_{\revidx\bbcnd{n}}\stint\psi =
  \int_{\bound\Omega} \varphi\,\normv^{\absidx{\alpha}}
  \grad_{\absidx{\alpha}}\psi\;\svol\period
  \label{biFdrldefexplicit}
\end{equation}
To see clearly what kind of boundary conditions 
the propagator satisfies, we write 
the heat equation in the following way
\begin{equation}
  - \stbivol\stint\tder{\hkrnl}_{\bbcnd{k}}(\tau) =
  \frac\isign2\:\biFr\stint\hkrnl_{\bbcnd{k}}(\tau) +
  \frac\isign2\:\biFdr_{\revidx\bbcnd{k}}
  \stint\hkrnl_{\bbcnd{k}}(\tau)\period
\end{equation}
The solution of the heat equation 
is smooth for non-zero time $\tau$. 
Therefore the left-hand side is smooth, 
as well as the first term on the right-hand side. 
The second term is localized on the boundary 
and therefore it has to vanish. So we find
\begin{equation}
  \biFdr_{\revidx\bbcnd{k}}\stint\hkrnl_{\bbcnd{k}}(\tau) = 0 \period
\end{equation}
Or, if we define maps $\val$ and $\mom$ which assign a value 
$\varphi = \val\stint\phi$ and a momenta $\pi = \mom\stint\phi$
on the boundary to a spacetime function $\phi$, 
we can write\note{nt:DotsConv}
\begin{gather}
  \biFdr_{\revidx\bbcnd{n}} = \val\sint\mom\comma\\
  \biTh_{\bbcnd{k}} = - \val\sint
  \Bigl( ({\textstyle\frac{1+p}{3}}\extscur +
  \beta\text{-terms})\, \svol\deltadst\Bigr)
  \sint\val\comma
\end{gather}
and the boundary conditions get the form
\begin{equation}
  \Bigl(\mom + ({\textstyle\frac{1+p}{3}}\extscur +
  \beta\text{-terms}) \val\Bigr)\stint
  \hkrnl_{\bbcnd{k}}(\tau) = 0\period
\end{equation}
We see that $\hkrnl_{\bbcnd{k}}$ satisfies Robin-like 
boundary conditions.

Let us summarize. We have found that the propagator 
given by the sum of amplitudes over histories in 
the domain $\Omega$ with fixed endpoints and inner 
time $\tau$ is a solution of the heat equation 
with specific boundary conditions. The boundary 
conditions depend on the definition 
of the path integral through the coincidence limits 
of derivatives of coefficients $\beta$ in 
the short time amplitude \eqref{boundhkrnlstampl}. 
In general, they are Robin-like conditions with 
a non-degenerate coefficient in front of momentum.

\subsection{Green function}

In the case of a relativistic particle the inner 
time is an unphysical quantity, and all physically 
distinguishable sets of histories should contain 
histories with all possible inner times. 
Therefore we will compute the amplitude associated 
with the set of histories with fixed endpoints 
but without a restriction on the inner time. 
As before, we will call this amplitude 
the Feynman Green function\note{nt:stsignAndProp}
\begin{equation}
  \frac1\stsign \GFF_{\bbcnd{k}}(x|z) =
  \int_{\tau\in\realn^+} \frac1\stsign 
  \hkrnl_{\bbcnd{k}}(\tau,x|z)\frac\isign2\dvol\tau\period
\end{equation}
Using the heat equation and initial conditions 
for the heat kernel, the integration gives
\begin{equation}
  \biF_{\revidx\bbcnd{k}} \stint \GFF_{\bbcnd{k}} = \deltadst\period
\end{equation}
So, $\GFF_{\bbcnd{k}}$ is the inverse of 
$\biF_{\revidx\bbcnd{k}}$ and restricted to smooth sources 
it satisfies the same boundary conditions as 
the propagator $\hkrnl_{\bbcnd{k}}$.

\subsection{Amplitude of particles emitted by a source}

Let us compute the  amplitude $\trac{1}{\bbcnd{k}}(J)$ 
of a set of histories which end at a given point $x$ 
and are emitted by a source described by a spacetime 
dependent amplitude\note{nt:stsignAndSrc} $\stsign J$. 
We will call it the one-particle amplitude. 
It is clearly given by 
\begin{equation}
  \trac{1}{\bbcnd{k}}(J) = \GFF_{\bbcnd{k}}\stint J = 
  \sol_{\bbcnd{k}}(J)\period
\end{equation}
It satisfies the same boundary conditions as 
the Feynman Green function.

We will interpret the boundary conditions as 
a consequence of the fact that we have not 
allowed particles to start on the boundary. 
More precisely, we have allowed the smooth source 
to be non-zero up to the boundary, but we have not 
allowed an emission of particles from the boundary 
comparable to an emission from a finite volume. 

We can ask why some particular boundary conditions 
means that no particles are emitted from the boundary. 
What about different boundary conditions? 
Why is the choice of the conditions above special? 
We are touching a question of what kind of particles 
we are dealing with. What does it mean that 
no particles are emitted (or absorbed --- for scalar 
particle the meanings are interchangeable
if we do not distinguish initial and 
final parts of the boundary).

First we have to realize that the statement 
``no particles on the boundary'' has to be interpreted 
as a result of a measurement on the boundary. 
We have to arrange apparatuses on the whole boundary 
which are sensitive to particles, and when all these 
devices measure no particle we can speak about 
no emission or absorption. Clearly this is very 
complicated global measurement. It depends on 
an exact arrangement of experimental devices 
on the whole boundary and on an interaction of 
particles with devices. We have hidden this dependence 
in the definition of the path integral through 
the non-specified $\beta$-terms. Therefore we see 
that we cannot expect a unique canonical meaning for 
the statement ``no particles on the boundary''. 
Only if we specify the kind of measurement 
we are performing we do have a meaning for this statement. 
And necessary information about experimental devices 
can be phenomenologically characterized by 
the choice of boundary conditions of the 
type we encountered above.

In the following paper \cite{Krtous:QRPBC} we will discuss what
interaction of particles with boundary leads to
different boundary conditions. 

So, the different boundary conditions correspond to
different type of detections of particles on
the boundary and in the following paper 
we will see that it can correspond to
different definitions of vacua in the standard
particle theory - in the scalar field theory in 
curved spacetime.

\subsection{Emission from the boundary}

Of course, we can ask what is the amplitude 
to find a particle at a point $x$ if 
we allow an emission from the boundary. 
Let us assume that the amplitude of 
the emission from the boundary is given 
by a density $\stsign j$ on the boundary manifold, 
which we call the \defterm{boundary source}. 
The amplitude $\trac{1}{\bbcnd{k}}(\tau;j)$ 
associated with the set of one-particle 
histories which are emitted by this boundary 
source and end in time $\tau$ at a point $x$, 
can be written using the 
\defterm{boundary propagator $\hkrnl^\dashv_{\bbcnd{k}}$}
\begin{equation}
  \trac{1}{\bbcnd{k}}(\tau;j) =
  \hkrnl^\dashv_{\bbcnd{k}}(\tau)\sint j\period
\end{equation}
The boundary propagator propagates between 
points inside of the domain and boundary sources. 
It has the character of a function on the domain 
$\Omega$ in the left argument and the function 
on the boundary manifold $\bound\Omega$ 
in the right argument. 

Clearly, the boundary propagator satisfies 
a composition law similar to \eqref{hkrnldecomp}
\begin{equation}
  \hkrnl^\dashv_{\bbcnd{k}}(\tau) =
  \hkrnl_{\bbcnd{k}}(\tau-\epsilon)\stint \stbivol\stint
  \hkrnl^\dashv_{\bbcnd{k}}(\epsilon)\period
\end{equation}
We can take a limit $\epsilon\rightarrow0$ and get
\begin{gather}
  \hkrnl^\dashv_{\bbcnd{k}}(\tau) =
  \hkrnl_{\bbcnd{k}}(\tau) \stint
  \acttol{\hkrnl}_{\bbcnd{k}}\commae\\
  \acttol{\hkrnl}_{\bbcnd{k}} \defeq \stbivol\stint
  \hkrnl^\dashv_{\bbcnd{k}}(0)\period
\end{gather}
We see that the amplitude is given by the 
propagator $\hkrnl_{\bbcnd{k}}(\tau)$ with 
no emission from the boundary, and by the boundary 
term $\acttol{\hkrnl}_{\bbcnd{k}}$ which 
\vague{translates} between the space of sources 
on the boundary and amplitudes in the domain. 
Similarly, if we sum over all possible inner times we get
\begin{equation}
  \trac{1}{\bbcnd{k}}(j) =
  \GFF_{\bbcnd{k}} \stint \acttol{\hkrnl}_{\bbcnd{k}}
  \sint j \period
\end{equation}

The boundary term $\acttol{\hkrnl}_{\bbcnd{k}}$ is 
a zero-time amplitude, so it is straightforward 
to estimate it. The short time amplitude approximation 
similar to \eqref{hkrnlstampl} for the boundary propagator is
\begin{equation}
  \begin{split}
  &\phi\stint\stbivol\stint 
  \hkrnl^\dashv_{\bbcnd{k}}(\tau)\sint j = \\
  &\qquad\qquad = \frac\stsign{(2\pi\isign\tau)^{\frac\stdim2}}\, 
  \int_{\substack{x\in\Omega\\ \yonb\in \bound\Omega}}
  \stvol(x) \phi(x) j(\yonb)\, \stvvmd(x|\yonb)\;
  \exp\Bigl( - \frac1{\isign\tau} \stwf(x|\yonb)\Bigr)
  \bigl(1 + \order{\sqrt{\tau}}\bigr) = \\
  &\qquad\qquad = \int_{\yonb\in \bound\Omega}
  \phi(\yonb) j(\yonb) \;\bigl(1 + \order{\sqrt{\tau}}\bigr) =
  \phi\stint\val\sint j\;\,
  \bigl(1 + \order{\sqrt{\tau}}\bigr) \period
\end{split}
\end{equation}
Therefore, for zero inner time we get
\begin{equation}
  \acttol{\hkrnl}_{\bbcnd{k}} = \val \period
\end{equation}
It means that the emission from the boundary 
is equivalent to the emission of particles inside 
of the domain but with a distributional source 
$\bound J = j\sint\val$ with support on the boundary

Allowing both boundary sources and sources inside 
of the domain, we find that the one-particle amplitude is
\begin{equation}
  \trac{1}{\bbcnd{k}}(J,\bound J) =
  \GFF_{\bbcnd{k}}\stint (J + \bound J) =
  \sol_{\bbcnd{k}}(J+\bound J)\period
\end{equation}
A careful discussion of the distributional character
of introduced boundary sources, differential operators
and Green functions shows (see \cite{Krtous:thesis}) 
that it satisfies the equation of motion in 
the expected form
\begin{equation}
  \biF_{\revidx\bbcnd{k}}\stint \sol_{\bbcnd{k}}(J+\bound J) =
  J + \bound J
\end{equation}
with boundary conditions fixed by the boundary source
\begin{equation}
  \bound J = \biFdr_{\revidx\bbcnd{k}}\stint
  \sol_{\bbcnd{k}}(J+\bound J) \period
\end{equation}


\section*{Conclusion}

In this paper we studied sum-over-histories quantization of
relativistic particle on a bounded domain of the spacetime. We showed
that we have to modify a definition of the path integral
by adding terms corresponding to paths reflected on the boundary
of the domain. Such contributions can be dominant in the short time
approximation near the boundary in addition to the usual dominant
contributions from the straight geodesic. They compensate other terms
localized on the boundary which arise from restriction
of non-reflected paths to the domain. They also specify
the exact form of the boundary conditions for the propagator
and Green functions. We found that boundary conditions
have Robin-like form and their exact form depends on the details
of the definition of the path integral --- a non-uniqueness is hidden
in the specification of $\beta$-coefficients in 
the short time amplitude \eqref{boundhkrnlstampl}.

We interpreted the specific boundary conditions as
a consequence of an interaction with apparatuses localized on 
the boundary which allow us to define a notion of particles.
Because of the non-uniqueness of the definition of the path integral
we do not have a uniqueness in the definition of particles.
In this correspondence the boundary condition 
can be also viewed as a phenomenological
description of the specific kind of particles.

In the next paper \cite{Krtous:QRPBC} we will discuss
a freedom in boundary conditions in more detail.
We will show that inclusion of some additional interaction
with measurement apparatuses on the boundary can
lead to general boundary conditions which include
the conditions corresponding to a definition
of particles in the quantum scalar field theory.


\appendices


\section{Asymptotic expansion of the leading term in the heat kernel}
\label{apx:AEHK}


\subsection{Vector space}

In a vector space $V$ equipped with a positive 
nondegenerate quadratic form $\stmtrc$ 
a simple Gaussian integration gives
\begin{equation}
\begin{aligned}
  \frac1{(2\pi\isign\tau)^{\frac\stdim2}}&
  \int_{X,Z\in V}\stvol(X)\stvol(Z)
  \varphi(X)\psi(Z)
  \exp\Bigl(-\frac1{2\isign\tau}\, (X-Z)\stctr\stmtrc\stctr(X-Z)\Bigr) = \\
  &=  \frac1{(2\pi\isign)^{\frac\stdim2}}
  \int_{X\in V}\stvol(X) \varphi(X)
  \int_{Y\in V}\stvol(Y) \psi(X+\sqrt{\tau}Y)
  \exp\Bigl(-\frac1{2\isign}\, Y\stctr\stmtrc\stctr Y\Bigr) = \\
  &=  \frac1{(2\pi\isign)^{\frac\stdim2}}
  \int_{X\in V}\stvol(X) \varphi(X)
  \sum_{k\in\naturaln_0} \tau^{\frac k2} 
  \bigl[\vectgrad_{\absidx{\alpha_1}}\dots
  \vectgrad_{\absidx{\alpha_k}}\psi\bigr](X)\stimes\\
  &\mspace{200mu}\stimes\frac1{k!}\int_{Y\in V}\stvol(Y)\,
  Y^{\absidx{\alpha_1}}\dots Y^{\absidx{\alpha_k}}\,
  \exp\Bigl(-\frac1{2\isign} Y\stctr\stmtrc\stctr Y\Bigr) = \\
  &= \int_{X\in V}\stvol(X) \varphi(X)
  \sum_{m\in\naturaln_0} \frac{(2m-1)!}{(2m)!} (\tau\isign)^m 
  \;\bigl[(\stmtrc^{\dash1\,\absidx{\alpha\beta}} \vectgrad_{\absidx{\alpha}}
  \vectgrad_{\absidx{\beta}})^m\psi\bigr](X) = \\
  &= \sum_{m\in\naturaln_0} \frac1{m!}
  \left(-\frac{\isign\tau}2\right)^m
  \varphi \stint \bidst{L}^m \stint \psi\period
\end{aligned}
\end{equation}
Here $\varphi$, $\psi$ are test functions, 
$\stvol$ is the constant volume element of 
the metric $\stmtrc$, $\isign$ is a number 
($\re\frac1\isign \geq 0$) and $\bidst{L}^m$ 
represents a bi-distribution
\begin{equation}
  \varphi \stint \bidst{L}^m \stint \psi =
  \int_{V}\stvol \varphi
  \bigl[(\stmtrc^{\dash1\,\absidx{\alpha\beta}} \vectgrad_{\absidx{\alpha}}
  \vectgrad_{\absidx{\beta}})^m\psi\bigr] =
  \int_{V}\stvol\psi
  \bigl[(\stmtrc^{\dash1\,\absidx{\alpha\beta}} \vectgrad_{\absidx{\alpha}}
  \vectgrad_{\absidx{\beta}})^m\varphi\bigr] \period
\end{equation}
It can be viewed as an \vague{$m$-th} power of the Laplace quadratic form $\bidst{L}$ associated with the metric $\stmtrc$ on the vector space $V$.

So we can write the asymptotic expansion for small $\tau$ as
\begin{equation}\label{hkrnlexpvctsp}
  \frac \stsign{(2\pi\isign\tau)^{\frac\stdim2}}
  \stvol(X)\stvol(Z)
  \exp\Bigl(-\frac1{2\isign\tau} (X-Z)\stctr\stmtrc\stctr(X-Z)\Bigr) = 
  \sum_{m\in\naturaln_0} \frac1{m!}
  \left(-\frac{\tau\isign}2\right)^m \bidst{L}^m\period
\end{equation}
Here we allowed the metric $\stmtrc$ to be Lorentzian 
--- this case can be obtained by analytical continuation in 
the phase factor $\stsign$ which characterizes the signature of
the metric.

In the case of the vector space the expression on the left side of 
eq. \eqref{hkrnlexpvctsp} is the heat kernel of 
the operator $\bidst{L}$. Because the right side is 
formally the exponential we see that the expansion is exact. 
To say more about convergence it is necessary to specify 
functional spaces on which all the operators act and we will 
not do this here. But see e.g. \cite{ReedSimon:book} for some details.

\subsection{Manifold without a boundary}

Now we would like to find the expansion of the similar 
expression in a general manifold $\stmfld$ without boundary. 
More precisely, we want to expand
\begin{equation}
  \frac\stsign{(2\pi\isign\tau)^{\frac\stdim2}}
  \stvvmd(x|z) \stvol(x)\stvol(z)
  \exp\Bigl(-\frac1{\isign\tau} \stwf(x|z)\Bigr) 
\end{equation}
for small $\tau$.

We smooth both arguments with test functions $\varphi$, 
$\psi$ and note that for a small $\tau$ the integration 
over $x$ and $z$ is dominated by a diagonal $x \approx z$ 
thanks to $\stwf(x|z) \approx 0$ for $x \approx z$. 
Therefore for a fixed $x$ we can restrict integration 
over $z$ to a normal neighborhood of $x$. 
In this neighborhood we can change variables 
$z \rightarrow Z$ with
\begin{equation}
  z = \stgeod_x(\sqrt{\tau}Z) \approx x + \sqrt{\tau}Z
  \commae\label{ChangezToZ}
\end{equation}
where $\stgeod_x(\epsilon Z)$ is a geodesic 
with an origin $x$ and initial tangent vector $Z$ 
(see \eqref{umapDef}). This is the exact meaning 
of \vague{adding} of a vector to a point as mentioned 
after equation \eqref{changextoX}. 

The Jacobian associated with this change of variables 
is given by Van-Vleck Morette determinant 
(see \eqref{vvmdetisjac})
\begin{equation}\label{jacisvvmdetis}
  (\stgeod_x^{\dash1\star}\stvol)(\sqrt{\tau}Z) =
  \tau^{\frac\stdim2}\stvol(x)[Z]\,\stvvmd^{\!-1}(x|z) \commae
\end{equation}
where $\stvol(x)[Z]$ is understood as a constant 
measure on the target vector space $\tens_x\stmfld$. 
After a change of variables, using \eqref{wfofgeod}, 
expanding $\psi$ and performing a Gaussian integration, we get
\begin{equation}\label{hkrnlstamplexpcomp}
\begin{split}
  &\frac\stsign{(2\pi\isign\tau)^{\frac\stdim2}}
  \int_{x,z\in\stmfld} \stvvmd(x|z) \stvol(x)\stvol(z) \varphi(x)\psi(z)
  \exp\Bigl(-\frac1{\isign\tau} \stwf(x|z)\Bigr) = \\
  &\qquad= \frac\stsign{(2\pi\isign\tau)^{\frac\stdim2}}
  \int_{x\in\stmfld} \stvol(x) \varphi(x)
  \int_{Z\in\tens_x\stmfld} \stvol(x)[Z]\stimes\\
  &\mspace{200mu}\stimes\Bigl(\sum_{k\in\naturaln_0}
  \frac1{k!} \tau^{\frac k2}
  \psi_{k\,{\absidx{\alpha_1}\dots \absidx{\alpha_k}}}(x)\;
  Z^{\absidx{\alpha_1}}\dots Z^{\absidx{\alpha_k}}\Bigr)\;
  \exp\Bigl(-\frac1{\isign\tau} Z\stctr\stmtrc(x)\stctr Z\Bigr) = \\
  &\qquad= \int_{\stmfld}\stvol\varphi\;\Bigl(\psi +
  \frac{\isign\tau}2\,\stmtrc^{\dash1\,\absidx{\alpha\beta}}
  \psi_{2\,{\absidx{\alpha\beta}}} +
  \order{\tau^2}\Bigr)\period
\end{split}
\end{equation}
Using \eqref{covexpandsc} we have $\psi_2 = \stcnx\grad\psi$, so we get
\begin{equation}\label{hkrnlstamplexpres}
  \frac\stsign{(2\pi\isign\tau)^{\frac\stdim2}}
  \stvvmd(x|z) \stvol(x)\stvol(z)
  \exp\Bigl(-\frac1{\isign\tau} \stwf(x|z)\Bigr) =
  \stbivol - \frac{\isign\tau}2 \bidst{L} + \order{\tau^2}
  \period
\end{equation}

\subsection{Half line}

Next we will investigate the simplest case of 
the manifold with the boundary --- half line $\realn^+$. 
We assume it is equipped with a special coordinate 
$\eta$ which selects a measure and derivative
\begin{equation}
  \mu = \dvol\eta \comma
  \bidst{M} = \mu\deltadst\comma 
  \etader = \frac\partial{\partial\eta}\period 
\end{equation}
We define bi-distributions of the $m$-th derivative
\begin{equation}
  \omega\stint\acttor{\etader}{}^{\langle m \rangle}\stint\varphi 
  = \omega\stint(\etader^m\varphi) 
  = \int_{\realn^+} \omega (\etader^m\varphi) \comma
  \acttol{\etader}{}^{\langle m \rangle} =
  \acttor{\etader}{}^{\langle m \rangle \qformT}\commae
\end{equation}
and
\begin{equation}
  \acttor{\etader} = \acttor{\etader}{}^{\langle 1 \rangle}\comma
  \acttol{\etader} = \acttol{\etader}{}^{\langle 1 \rangle}\period
\end{equation}

We can define also a boundary delta bi-distribution $\bnddelta$ as
(the boundary is one point now)
\begin{equation}
  \varphi\stint\bnddelta\stint\psi =
  (\varphi \psi)|_{\text{boundary}}\period
\end{equation}
Integration by parts can be expressed by the relation
\begin{equation}\label{HLIntByParts}
\begin{gathered}
  \acttol{\etader}\stint\bidst{M} + \bidst{M}\stint\acttor{\etader} =
  - \bnddelta \commae\\
  \acttol{\etader}{}^{\langle m+1 \rangle} \stint\bidst{M} +
  (-1)^m \bidst{M}\stint\acttor{\etader}{}^{\langle m+1 \rangle} =
  - \sum_{k = 0,\dots,m}  \acttol{\etader}{}^{\langle m-k \rangle}
  \stint\bnddelta\stint \acttor{\etader}{}^{\langle k \rangle} \period
\end{gathered}
\end{equation}

Next we define quadratic forms of powers of the Laplace operator
\begin{equation}
\begin{gathered}
  \acttor{\bidst{L}} = - \bidst{M}\stint
  \acttor{\etader}{}^{\langle 2 \rangle}\comma
  \acttor{\bidst{L}}{}^{\langle m \rangle} =
  (-1)^m \bidst{M}\stint\acttor{\etader}{}^{\langle 2m \rangle}\commae\\
  \acttol{\bidst{L}} = - \acttol{\etader}{}^{\langle 2 \rangle}
  \stint\bidst{M}\comma
  \acttol{\bidst{L}}{}^{\langle m \rangle} =
  (-1)^m \acttol{\etader}{}^{\langle 2m \rangle}\stint\bidst{M}\commae\\
  \acttolr{\bidst{L}} =
  \frac12\bigl(\acttol{\bidst{L}}+\acttor{\bidst{L}}\bigr)\comma
  \acttolr{\bidst{L}}{}^{\langle m \rangle} =
  \frac12\bigl(\acttol{\bidst{L}}{}^{\langle m \rangle}
  +\acttor{\bidst{L}}{}^{\langle m \rangle}\bigr)\commae\\
  \bidst{L} = \acttol{\etader}\stint\bidst{M}\stint\acttor{\etader}\period
\end{gathered}
\end{equation}
The symplectic form on the boundary is
\begin{equation}
  \bound\bidst{L} = \acttor{\bidst{L}} - \acttol{\bidst{L}} =
  - \acttol{\etader}\stint\bnddelta +
  \bnddelta\stint\acttor{\etader}\period
\end{equation}
It is straightforward to check that
\begin{equation}
  \acttor{\bidst{L}}{}^{\langle m \rangle} -
  \acttol{\bidst{L}}{}^{\langle m \rangle} =
  (-1)^m \sum_{\substack{k,l\in\naturaln_0\\k+l+1 = 2m}}
  \acttol{\etader}{}^{\langle k \rangle}\stint
  \bnddelta\stint\acttor{\etader}{}^{\langle l \rangle}\period
\end{equation}

Now we prove the following expansion for small $\tau$
\begin{equation}\label{SmHLExpLemma}
\begin{split}
  &\frac1{\sqrt{2\pi\tau\isign}}\; \exp\Bigl(
  -\frac1{2\tau\isign}(\xi-\zeta)^2\Bigr)\; \mu(\xi)\mu(\zeta) = \\
  &\qquad =\sum_{m\in\naturaln_0} \frac1{\Gammafc\bigl(\frac{m}{2}+1\bigr)}
  \Bigl(\frac{\tau\isign}{2}\Bigr)^{\frac{m}{2}} \frac12
  \Bigl(\acttol{\etader}{}^{\langle m \rangle} \stint\bidst{M}+
  \bidst{M}\stint \acttor{\etader}{}^{\langle m \rangle}\Bigr) = \\
  &\qquad = \sum_{m\in\naturaln_0} \frac{(-1)^m}{m!}
  \Bigl(\frac{\tau\isign}{2}\Bigr)^{m}\:
  \acttolr{\bidst{L}}{}^{\langle m \rangle} + \\
  &\qquad\qquad + \sum_{m\in\naturaln_0}
  \frac{(-1)^{m+1}}{\Gammafc\bigl(m + \frac32\bigr)}
  \Bigl(\frac{\tau\isign}{2}\Bigr)^{m+\frac12} \;\frac12
  \sum_{\substack{k,l\in\naturaln_0\\k+l=2m}}
  (-1)^{\frac{k-l}2} \acttol{\etader}{}^{\langle k \rangle}\stint
  \bnddelta\stint\acttor{\etader}{}^{\langle l \rangle}\period
\end{split}
\end{equation}
Here $\isign$ is a complex number such that $\frac1\isign$ 
has non-negative real part. Strictly speaking, the following 
derivation needs a positive real part; but for an imaginary 
value of $\isign$ the relation can be obtained 
by limiting procedure. Because only a combination 
the $\tau\isign$ appears in the equation, we drop 
$\isign$ in the following derivation --- it can be easily 
restored by inspection of the $\tau$-dependence.

Clearly, the second equality follows from integration 
by parts \eqref{HLIntByParts}. To prove the first one 
we smooth it with test functions $\varphi$ and $\psi$ and get
\begin{equation}\label{SmHLExpProof}
\begin{split}
  \frac1{\sqrt{2\pi\tau}}\; & \int_{\xi,\zeta\in\realn^+}
  \exp\Bigl( -\frac1{2\tau}(\xi-\zeta)^2\Bigr)\;
  \varphi(\xi) \psi(\zeta)\: \mu(\xi)\mu(\zeta) = \\
  = &\frac1{\sqrt{2\pi\tau}}\;
  \int_{\xi\in\langle0,\epsilon\rangle}\dvol\xi
  \int_{\zeta\in\realn^+}\dvol\zeta
  \varphi(\xi) \psi(\zeta)
  \exp\Bigl( -\frac1{2\tau}(\xi-\zeta)^2\Bigr) + \\
  + &\frac1{\sqrt{2\pi\tau}}\;
  \int_{\xi\in\langle\epsilon,\infty\rangle}\dvol\xi
  \int_{\zeta\in\realn^+}\dvol\zeta
  \varphi(\xi) \psi(\zeta)
  \exp\Bigl( -\frac1{2\tau}(\xi-\zeta)^2\Bigr) 
\end{split}
\end{equation}
for some $\epsilon\in\realn^+$.

For a small $\tau$ the exponential suppresses 
any contribution except from $\xi\approx\zeta$. 
Therefore for small $\epsilon$ only small values of 
$\xi$ and $\zeta$ contribute to the first 
term of the last equation. We can rescale variables 
by factor $\sqrt{\tau}$ and expand $\varphi$ 
and $\psi$ at zero and  obtain
\begin{align}
  &\frac1{\sqrt{2\pi\tau}}\;
  \int_{\xi\in\langle0,\epsilon\rangle}\dvol\xi
  \int_{\zeta\in\realn^+}\dvol\zeta
  \varphi(\xi) \psi(\zeta)
  \exp\Bigl( -\frac1{2\tau}(\xi-\zeta)^2\Bigr) =
  \displaybreak[0]\nonumber\\
  &\qquad = \frac{\tau}{\sqrt{2\pi\tau}}
  \int_{\xi\in\langle0,\frac\epsilon{\sqrt\tau}\rangle}\dvol\xi
  \varphi(\sqrt\tau\xi)
  \sum_{l\in\naturaln_0} \tau^{\frac{l}{2}} \psi{}^{\langle l\rangle}\!(0)\;
  \frac1{l!} \int_{l\in\realn^+}\dvol\zeta \zeta^l 
  \exp\Bigl(-\frac1{2}(\xi-\zeta)^2\Bigr) =
  \displaybreak[1]\nonumber\\
  &\qquad = \frac1{\sqrt{2\pi}}
  \sum_{k,l\in\naturaln_0} \tau^{\frac{k+l+1}2}\:
  \varphi{}^{\langle k\rangle}\!(0)\:\psi{}^{\langle l\rangle}\!(0)\;
  \frac1{k!} \int_{\xi\in\langle0,\frac\epsilon{\sqrt\tau}\rangle}
  \dvol\xi\xi^k \rfc_{l+1}(\xi)\commae
\end{align}
where we have used the definition of special functions 
$\rfc_{l}$ \eqref{rfcDef}. Properties of these special 
functions are summarized in appendix \ref{apx:SFcR}. 
Using equation \eqref{intZeroXofrfc} the last expression gives
\begin{align}
\begin{split}
  &\frac1{\sqrt{2\pi}}
  \sum_{k,l\in\naturaln_0} \tau^{\frac{k+l+1}2}\:
  \varphi{}^{\langle k\rangle}\!(0)\:\psi{}^{\langle l\rangle}\!(0) \stimes\\
  &\qquad\stimes\biggl(\;\; (-1)^{l+1} \rfc_{k+l+2}(0) + \\
  &\qquad\qquad + \frac{\sqrt{2\pi}}{l! k!}
  \sum_{\substack{m\in\naturaln_0\\ 2m\leq l}}
  \frac{(2m-1)!!}{k+l-2m+1} \binom{l}{2m}
  \Bigl(\frac\epsilon{\sqrt\tau}\Bigr)^{k+l-2m+1} + \\
  &\qquad\qquad + (-1)^l \sum_{m=0,\dots,k} \frac1{m!} 
  \Bigl(\frac\epsilon{\sqrt\tau}\Bigr)^{m}\;
  \rfc_{\scriptscriptstyle k+l-m+2}
  \bigl(-\tfrac\epsilon{\sqrt\tau}\bigr)\biggr) =
  \end{split} \displaybreak[0]\nonumber\\
\begin{split}
  &= \sum_{m\in\naturaln_0} \tau^{\frac{m+1}2}
  \Bigl(2^{\frac{m+1}2}\Gammafc\bigl(\frac{m+1}2+1\bigr)\Bigr)^{-1}\;
  \frac12 \sum_{l=0,\dots,m} (-1)^{l+1} \:
  \varphi{}^{\langle m-l\rangle}\!(0)\:\psi{}^{\langle l\rangle}\!(0) +\\
  &\qquad + \sum_{m\in\naturaln_0} \frac{\tau^m}{(2m)!!}
  \sum_{k,l\in\naturaln_0}
  \frac{\epsilon^{k+l+1}}{k!\,l!\,(k+l+1)}\:
  \varphi{}^{\langle k\rangle}\!(0)\:\psi{}^{\langle l+2m\rangle}\!(0) +\\
  &\qquad + \exp\Bigl(-\frac12\frac{\epsilon^2}{\tau}\Bigr)\:
  \order{\tau} =
  \end{split} \displaybreak[1]\nonumber\\
\begin{split} 
  &= - \sum_{m\in\naturaln} \frac{\tau^m}{(2m)!!} 
  \frac12 \sum_{l=0,\dots,2m-1} (-1)^{l} \:
  \varphi{}^{\langle 2m-l-1\rangle}\!(0)\:\psi{}^{\langle l\rangle}\!(0) + \\
  &\qquad + \sum_{m\in\naturaln_0} \frac{\tau^{m+\frac12}}{(2m+1)!!} 
  \frac{-1}{\sqrt{2\pi}} \sum_{l=0,\dots,2m} (-1)^{l} \:
  \varphi{}^{\langle 2m-l\rangle}\!(0)\:\psi{}^{\langle l\rangle}\!(0) + \\
  &\qquad + \sum_{m\in\naturaln_0} \frac{\tau^m}{(2m)!!}
  \sum_{k,l\in\naturaln_0}
  \frac{\epsilon^{k+l+1}}{k!\,l!\,(k+l+1)}\:
  \varphi{}^{\langle k\rangle}\!(0)\:\psi{}^{\langle l+2m\rangle}\!(0) + \\
  &\qquad + \exp\Bigl(-\frac12\frac{\epsilon^2}{\tau}\Bigr)\:
  \order{\tau} \period
  \end{split}\label{HLFirstTerm}
\end{align}
Here we have used relations \eqref{ZeroOfrfc} and \eqref{rfcAsymptB}.

In the second term of the expression \eqref{SmHLExpProof} 
we change variables 
$\zeta \rightarrow \eta = \frac1{\sqrt\tau}(\xi-\zeta)$ 
using again the fact that only the contribution from 
$\zeta\approx\xi$ is not suppressed by the exponential. 
Next we expand $\psi$ around $\eta=0$ and obtain
\begin{align}
  &\frac1{\sqrt{2\pi\tau}}\;
  \int_{\xi\in\langle\epsilon,\infty\rangle}\dvol\xi
  \int_{\zeta\in\realn^+}\dvol\zeta
  \varphi(\xi) \psi(\zeta)
  \exp\Bigl( -\frac1{2\tau}(\xi-\zeta)^2\Bigr) =
  \nonumber\\
\begin{split}
  &\qquad = \frac1{\sqrt{2\pi\tau}}\;
  \int_{\xi\in\langle\epsilon,\infty\rangle}\dvol\xi
  \varphi(\xi) \sum_{k\in\naturaln_0} 
  \psi{}^{\langle k\rangle}\!(\xi)\: \tau^{\frac{k}2} \stimes\\
  &\qquad\qquad\stimes\Biggl(
  \frac1{k!} \int_{\eta\in\realn}\dvol\eta \eta^k
  \exp\Bigl(-\frac12\eta^2\Bigr) -
  \frac1{k!} \int_{\eta\in\langle-\infty,\frac\xi{\sqrt\tau}\rangle}
  \dvol\eta \eta^k \exp\Bigl(-\frac12\eta^2\Bigr)\Biggr) =
  \end{split}\displaybreak[0]\nonumber\\
\begin{split}
  &\qquad = \sum_{m=\naturaln_0} \frac{\tau^m}{(2m)!!}
  \int_{\xi\in\langle\epsilon,\infty\rangle} \dvol\xi
  \varphi(\xi)\psi{}^{\langle 2m\rangle}\!(\xi) - \\
  &\qquad\qquad - \frac1{\sqrt{2\pi\tau}}\; 
  \sum_{k\in\naturaln_0} \tau^{\frac{k}2}
  \sum_{l=0,\dots,k} \frac{(-1)^k}{l!} 
  \int_{\langle\epsilon,\infty\rangle}\dvol\xi
  \Bigl(\frac\xi{\sqrt\tau}\Bigr)^l\;
  \rfc_{k-l+1}\bigl(-\tfrac\xi{\sqrt\tau}\bigr)\:
  \varphi(\xi)\:\psi{}^{\langle k\rangle}\!(\xi) =
  \end{split}\displaybreak[0]\nonumber\\
\begin{split}
  &\qquad = \sum_{m=\naturaln_0} \frac{\tau^m}{(2m)!!}
  \int_{\xi\in\realn^+} \dvol\xi
  \varphi(\xi)\psi{}^{\langle 2m\rangle}\!(\xi) -
  \sum_{m=\naturaln_0} \frac{\tau^m}{(2m)!!}
  \int_{\xi\in\langle 0,\epsilon\rangle} \dvol\xi
  \varphi(\xi)\psi{}^{\langle 2m\rangle}\!(\xi) + \\
  &\qquad\qquad + \exp\Bigl(-\frac12\frac{\epsilon^2}{\tau}\Bigr)\:
  \order{\tau}\period
  \end{split}
\end{align}
Here we used \eqref{pfcIntDef}, \eqref{ZeroOfpfc}, 
\eqref{intInftyXofrfc} and \eqref{rfcAsymptB}. 
For small $\epsilon$ we can expand $\varphi$ 
and $\psi{}^{\langle 2m\rangle}$ about zero 
in the second term of the last expression. 
Performing also an integration by parts in 
the first term transforms the last expression to
\begin{equation}\label{HLSecondTerm}
\begin{split}
  &\sum_{m\in\naturaln_0} \frac{\tau^m}{(2m)!!} \frac12
  \int_{\xi\in\realn^+} \dvol\xi \Bigl( 
  \varphi{}^{\langle 2m\rangle}\!(\xi)\psi(\xi) +
  \varphi(\xi)\psi{}^{\langle 2m\rangle}\!(\xi) \Bigr) + \\
  &\qquad + \sum_{m\in\naturaln} \frac{\tau^m}{(2m)!!} 
  \frac12 \sum_{l=0,\dots,2m-1} (-1)^{l} \:
  \varphi{}^{\langle 2m-l-1\rangle}\!(0)\:\psi{}^{\langle l\rangle}\!(0) - \\
  &\qquad - \sum_{m\in\naturaln_0} \frac{\tau^m}{(2m)!!}
  \sum_{k,l\in\naturaln_0}
  \frac{\epsilon^{k+l+1}}{k!\,l!\,(k+l+1)}\:
  \varphi{}^{\langle k\rangle}\!(0)\:\psi{}^{\langle l+2m\rangle}\!(0) + \\
  &\qquad + \exp\Bigl(-\frac12\frac{\epsilon^2}{\tau}\Bigr)
  \order{\tau} \period
\end{split}
\end{equation}

Substituting equations \eqref{HLFirstTerm} and 
\eqref{HLSecondTerm} to equation \eqref{SmHLExpProof} 
and ignoring exponentially suppressed terms 
$\exp\bigl(-\frac12\frac{\epsilon^2}{\tau}\bigr) 
\order{\tau}$ we obtain the desired relation \eqref{SmHLExpLemma}.

Next we will prove another expansion for small $\tau$
\begin{equation}\label{SmHLBndExpLemma}
\begin{split}
  &\frac1{\sqrt{2\pi\isign\tau}}\,
  \omega\Bigl(\frac{\xi\zeta}{\xi+\zeta}\Bigr)\; \exp\Bigl(
  -\frac1{2\isign\tau}(\xi+\zeta)^2\Bigr)\;
  \mu(\xi)\mu(\zeta) = \\
  &\qquad = \sum_{n\in\naturaln_0}
  \Bigl(\frac{\tau\isign}{2}\Bigr)^{\frac{n+1}{2}}
  \frac1{\Gammafc\bigl(\frac{n+1}{2}\bigr)}
  \sum_{\substack{m,k,l\in\naturaln_0\\ k+l+m = n}}
  \frac{\omega{}^{\langle m \rangle}\!(0)}{n+m+1}
  \frac{\binom{m+k}{k} \binom{m+l}{l}}{\binom{n+m}{m}}\;
  \acttol{\etader}{}^{\langle k \rangle}  
  \stint\bnddelta\stint
  \acttor{\etader}{}^{\langle l \rangle}\commae
\end{split}
\end{equation}
where $\omega$ is some smooth function. 
As a corollary for $\omega=1$ we get
\begin{align}
  &\frac1{\sqrt{2\pi\isign\tau}}\; \exp\Bigl(
  -\frac1{2\isign\tau}(\xi+\zeta)^2\Bigr)\;
  \mu(\xi)\mu(\zeta) = \nonumber\\
  &\qquad =
  \sum_{k,l\in\naturaln_0}
  \frac1{\Gammafc\bigl(\frac{k+l+1}{2}+1\bigr)}
  \Bigl(\frac{\tau\isign}{2}\Bigr)^{\frac{k+l+1}{2}} \,\frac12\;
  \acttol{\etader}{}^{\langle k \rangle} \stint\bnddelta\stint
  \acttor{\etader}{}^{\langle l \rangle} = 
  \displaybreak[0]\\
  &\qquad = \sum_{m\in\naturaln} \frac1{m!}
  \Bigl(\frac{\tau\isign}{2}\Bigr)^{m} \;\frac12
  \sum_{\substack{k,l\in\naturaln_0\\k+l+1=2m}}
  \acttol{\etader}{}^{\langle k \rangle}\stint
  \bnddelta\stint\acttor{\etader}{}^{\langle l \rangle} + \nonumber\\
  &\qquad\qquad + \sum_{m\in\naturaln_0}
  \frac1{\Gammafc\bigl(m + \frac32\bigr)}
  \Bigl(\frac{\tau\isign}{2}\Bigr)^{m+\frac12} \;\frac12
  \sum_{\substack{k,l\in\naturaln_0\\k+l=2m}}
  \acttol{\etader}{}^{\langle k \rangle}\stint
  \bnddelta\stint\acttor{\etader}{}^{\langle l \rangle}\period\nonumber
\end{align}

As before, we drop the factor $\isign$ in 
the proof because it can easily be restored from 
the $\tau$ dependence. Again, we smooth the relation with 
test functions $\varphi$ and $\psi$. Thanks to 
the exponential suppression, only small values of 
$\xi$ and $\zeta$ contributes to the integrals. 
Therefore we can rescale $\xi$ and $\zeta$ by 
$\sqrt\tau$, expand $\varphi$, $\psi$, and $\omega$ 
about zero, and get
\begin{align}
  &\frac1{2\pi\tau} \int_{\xi,\zeta\in\realn^+}
  \varphi(\xi)\psi(\zeta)\:\omega\Bigl(\frac{\xi\zeta}{\xi+\zeta}\Bigr)\;
  \exp\Bigl(-\frac1{2\tau} (\xi+\zeta)^2 \Bigr)\;
  \dvol\xi\dvol\zeta =\nonumber\\
\begin{split}
  &\qquad = \frac1{2\pi\tau} \sum_{n\in\naturaln_0}
  \tau^{\frac{n}2+1} \sum_{\substack{k,l,m\in\naturaln_0\\k+l+m=n}}
  \frac{n!}{k!\,l!\,m!}\: \varphi{}^{\langle k\rangle}\!(0)\:
  \psi{}^{\langle l\rangle}\!(0)\: \omega{}^{\langle m\rangle}\!(0)\stimes\\
  &\qquad\qquad\qquad\qquad\stimes \frac1{n!} \int_{\xi,\zeta\realn^+}
  \dvol\xi\dvol\zeta\;\frac{\xi^{m+k}\zeta^{m+l}}{(\xi+\zeta)^m}\;
  \exp\Bigl(-\frac12(\xi+\zeta)^2\Bigr) = 
  \end{split}\displaybreak[0]\nonumber\\
\begin{split}
  & \qquad =\sum_{n\in\naturaln_0}\Bigl(\frac\tau2\Bigr)^{\frac{n+1}2}
  \frac1{\Gammafc\bigl(\frac{n+1}2\bigr)}
  \sum_{\substack{k,l,m\in\naturaln_0\\k+l+m=n}}
  \frac{\omega{}^{\langle m \rangle}\!(0)}{n+m+1}
  \frac{\binom{m+k}{k} \binom{m+l}{l}}{\binom{n+m}{m}}\;
  \varphi{}^{\langle k\rangle}\!(0) \:\psi{}^{\langle l\rangle}\!(0)\commae
\end{split}
\end{align}
what proves the relation \eqref{SmHLBndExpLemma}. 
Here we have used the integral \eqref{IntGenPowExp}.

\subsection{Manifold with boundary --- no reflection contribution}

Now we find an expansion of the contribution to the short 
time amplitude from the trajectories near the geodesic 
without reflection in the domain $\Omega$ with boundary 
$\bound\Omega$. We will prove for small $\tau$ 
\begin{equation}\label{hkrnlwbstamplexp}
\begin{split}
  \frac\stsign{(2\pi\isign\tau)^{\frac\stdim2}}\, 
  \stvol(x) \stvol(z) & \stvvmd(x|z)
  \exp\Bigl( - \frac1{\isign\tau} \stwf(x|z)\Bigr) = \\
  &= \stbivol \,+\,
  \sqrt{\tau} \Bigl(-\frac1\stsign\sqrt{\frac\isign{2\pi}}\Bigr)\bbivol\,-\,
  \tau \frac\isign2\, \acttolr{\bidst{L}} \,+\,
  \order{\tau^{\frac32}} \period
\end{split}
\end{equation}

As usual, we will be proving a smoothed version of this relation 
--- we multiply the expression by test functions $\varphi(x)$ 
and $\psi(z)$, and integrate over $x$ and $z$.  Thanks to 
the exponential suppression the only non-trivial contribution 
is from $x \approx z$. Therefore it is sufficient to prove 
the relation locally. Clearly, for $\varphi$ and $\psi$ 
with support inside of interior of the domain $\Omega$ 
the boundary does not have any influence and the relation 
reduces to the case without boundary. Therefore we will 
investigate only the case when $\varphi$ and $\psi$ are 
localized near the boundary. Thanks to locality we can also, 
without losing generality, assume that the test functions 
are localized on the neighborhood $U \subset\Omega$ of 
the boundary with topology $\realn\times\bound\Omega$ 
on which the geodesics normal to the boundary do not cross. 
In such a neighborhood we can use the method described in 
appendix \ref{apx:GdTh} and change the integration over 
a neighborhood to integration over the boundary 
$\bound\Omega$ and geodesic distance from 
the boundary (see \eqref{wmapDef})
\begin{equation}\label{chnagvarwmap}\begin{gathered}
  x \rightarrow \xonb,\xi \comma x = \wmap(\xonb,\xi)\commae\\
  z \rightarrow \zonb,\zeta \comma y = \wmap(\zonb,\zeta)\period
\end{gathered}\end{equation}
The Jacobian associated with this change of variables is 
\begin{equation}
   \stvol(\xonb) = \dvol\xi \obvol(\xonb,\xi) =
   \dvol\xi \onbnd\jac(\xonb,\xi)\,\svol\period
\end{equation}
Here we use the convention \eqref{wSplittingConv} --- 
we denote the spacetime dependent object $A(x)$ in 
variables $\xonb$, $\xi$ and tensor indices moved 
to the boundary as $\onbnd{A}(\xonb,\xi)$. 
Changing variables we get
\begin{align}
  &\frac\stsign{(2\pi\isign\tau)^{\frac\stdim2}}\,
  \int_{x,z\in\Omega} \stvol(x) \stvol(z) \stvvmd(x|z)
  \exp\Bigl( - \frac1{\isign\tau} \stwf(x|z)\Bigr)
  \varphi(x) \psi(z) = \nonumber\\
\begin{split}
  &\qquad=\frac\stsign{(2\pi\isign\tau)^{\frac\stdim2}}\,
  \int_{\substack{\xonb,\zonb\in\bound\Omega\\\xi,\zeta\in\realn^+}}
  \dvol\xi\dvol\zeta\:
  {\onbnd\jac}(\xonb,\xi)\svol(\xonb)\:
  {\onbnd\jac}(\zonb,\zeta)\svol(z)\:
  \onbnd{\stvvmd}(\xonb,\xi|\zonb,\zeta) \stimes \\
  &\qquad\qquad\qquad\qquad\qquad\qquad\qquad\qquad\stimes
  \exp\Bigl( - \frac1{\isign\tau} \onbnd{\stwf}(\xonb,\xi|\zonb,\zeta)\Bigr)\;
  \onbnd{\varphi}(\xonb,\xi) \onbnd{\psi}(\zonb,\zeta) = 
  \end{split}\nonumber\displaybreak[0] \\
\begin{split}
  &\qquad=\int_{\xonb\in \bound\Omega} \svol(\xonb)\,
  \frac\stsign{(2\pi\isign\tau)^{\frac12}}
  \int_{\xi,\zeta\in\realn^+} \dvol\xi\dvol\zeta\:
  \onbnd{\varphi}(\xonb,\xi) \stimes \\
  &\qquad\qquad\stimes\frac1{(2\pi\isign\tau)^{\frac\stdim2}}
  \int_{\zonb\in\bound\Omega} \svol(\zonb)\: \svvmd(\xonb|\zonb)\:
  \onbnd{\psi}(\zonb,\zeta)  \;
  \exp\Bigl( - \frac1{\isign\tau} \onbnd{\stwf}(\xonb,\xi|\zonb,\zeta)
  + \onbnd l(\xonb,\xi|\zonb,\zeta)\Bigr)
  \period\end{split}
\end{align}
Here we have defined   
\begin{equation}\label{lDef}
  \onbnd l(\xonb,\xi|\zonb,\zeta) = \ln\biggl(
  \onbnd\jac(\xonb,\xi)
  \frac{\onbnd{\stvvmd}(\xonb,\xi|\zonb,\zeta)}{\svvmd(\xonb|\zonb)}
  \onbnd\jac(\zonb,\zeta)\biggr)\commae
\end{equation}
where $\svvmd(\xonb|\zonb)$ is the Van-Vleck Morette 
determinant of the metric $\smtrc$ on the boundary manifold.

Exponential suppression ensures again that 
the only contributions comes from $\xonb\approx\zonb$. 
So we can change variables $\xonb,\zonb \rightarrow \yonb,Y$
\begin{equation}\label{chnagvarbumap}
  \yonb = \xonb \comma
  Y \in\tens_\yonb\bound\Omega\comma
  \zonb = \sgeod_\yonb(\sqrt\tau Y)\commae
\end{equation}
with the exponential map $\sgeod_\yonb(Y)$ defined as in 
\eqref{umapDef} but on the boundary manifold. 
The Jacobian for this change of variables is given 
by an equation similar to \eqref{vvmdetisjac}, 
only with the Van-Vleck Morette determinant 
$\svvmd(\xonb,\yonb)$ defined using the metric $\smtrc$ 
on the boundary manifold. We use covariant expansions
\begin{align}
  \onbnd{\stwf}(\yonb,\xi|\yonb,\sqrt\tau Y,\zeta) &=
  \tau \stsign^2 \frac12 (\xi-\zeta)^2 +
  \sum_{k=2,3,\dots} \tau^{\frac k2} \frac1{k!}\; 
  \stwfonb_{0,k\,\absidx{\mu}_1\dots\absidx{\mu}_k}
  (\yonb;\xi,\zeta) \;
  Y^{\absidx{\mu}_1}\dots Y^{\absidx{\mu}_k} 
  \commae \label{wfbndcovexp}\\
  \onbnd l(\yonb,\xi|\yonb,\sqrt\tau Y,\zeta) &=
  \sum_{k=\naturaln_0} \tau^{\frac k2} \frac1{k!} \;
  {\onbnd l}_{0,k\,\absidx{\mu}_1\dots\absidx{\mu}_k}(\yonb;\xi,\zeta) \; 
  Y^{\absidx{\mu}_1}\dots Y^{\absidx{\mu}_k}
  \commae \label{lbndcovexp}\\
  \psionb(\yonb,\sqrt\tau Y,\zeta) &=
  \sum_{k=\naturaln_0} \tau^{\frac k2} \frac1{k!} \;
  \psionb_{k\,\absidx{\mu}_1\dots\absidx{\mu}_k}(\yonb;\zeta) \;
  Y^{\absidx{\mu}_1}\dots Y^{\absidx{\mu}_k}\commae \label{psibndcovexp}
\end{align}
with coefficients given by expressions 
(\ref{bndcovexpwfzero}--\ref{bndcovexpwffour}), 
(\ref{bndcovexplzero}--\ref{bndcovexpltwo}), 
and \eqref{covexpandsc}. Expanding the exponential, 
gathering all terms up to order $\order{\tau}$, 
and performing a Gaussian integration over $Y$ lead to
\begin{equation}\label{hkrnlwbstamplStep}
  \int_{\yonb\in \bound\Omega} \frac\stsign{(2\pi\isign\tau)^{\frac12}}
  \int_{\xi,\zeta\in\realn^+} \dvol\xi\dvol\zeta\;
  \exp\Bigl(-\frac{\stsign^2}{2\isign\tau}(\zeta-\xi)^2\Bigr)\;
  \tilde \jac(\yonb;\xi,\zeta)\;
  \Bigl( \omega_0(\yonb;\xi,\zeta) + \tau \omega_1(\yonb;\xi,\zeta) 
  + \order{\tau^2} \Bigr)
\end{equation}
with
\begin{align}
\begin{split}
  \tilde\jac(\yonb;\xi\zeta) &= 
    \onbnd\jac(\yonb,\xi)\:
    \onbnd\stvvmd(\yonb,\xi|\yonb,\zeta)\:
    \onbnd\jac(\yonb,\zeta)\:
    \svol(\yonb)\; \Bigl(\Det_\bpar\stwfonb_{0,2}(\yonb;\xi,\zeta)
    \Bigr)^{-\frac12} =\\
  &= \Bigl(\onbnd\jac(\yonb,\xi)\: 
    \onbnd\stvvmd(\yonb,\xi|\yonb,\zeta)\:
    \onbnd\jac(\yonb,\zeta)\Bigr)^{\frac12}\commae
  \end{split} \label{tildejacDef} \displaybreak[0]\\
  \omega_0(\yonb;\xi,\zeta) &=
    \phionb  (\yonb,\xi)\: \psionb(\yonb,\zeta)
    \commae\displaybreak[0]\displaybreak[0]\\
  \omega_1(\yonb;\xi,\zeta) &=
  \isign\: \phionb  (\yonb,\xi)\: \psionb(\yonb,\zeta) \;\Bigl(
  - \frac18 \stwfonb_{0,4\,
    {\absidx{\mu}}{\absidx{\nu}}{\absidx{\kappa}}{\absidx{\lambda}}}\:
    \stwfonb{}_{0,2}^{\dash1\,{\absidx{\mu}}{\absidx{\nu}}}
    \stwfonb{}_{0,2}^{\dash1\,{\absidx{\kappa}}{\absidx{\lambda}}}
  + \frac12 {\onbnd l}_{0,2\,{\absidx{\mu}}{\absidx{\nu}}}\:
    \stwfonb{}_{0,2}^{\dash1\,{\absidx{\mu}}{\absidx{\nu}}}
    +\nonumber\\
  &\qquad + \frac18 \stwfonb_{0,3\,
    {\absidx{\mu}}{\absidx{\nu}}{\absidx{\alpha}}}\:
    \stwfonb_{0,3\,
    {\absidx{\kappa}}{\absidx{\lambda}}{\absidx{\beta}}}\:
    \stwfonb{}_{0,2}^{\dash1\,{\absidx{\mu}}{\absidx{\nu}}}
    \stwfonb{}_{0,2}^{\dash1\,{\absidx{\alpha}}{\absidx{\beta}}}
    \stwfonb{}_{0,2}^{\dash1\,{\absidx{\kappa}}{\absidx{\lambda}}} 
  - \frac12 {\onbnd l}_{0,1\,{\absidx{\mu}}}\:
    \stwfonb_{0,3\,
    {\absidx{\kappa}}{\absidx{\lambda}}{\absidx{\nu}}}\:
    \stwfonb{}_{0,2}^{\dash1\,{\absidx{\mu}}{\absidx{\nu}}}
    \stwfonb{}_{0,2}^{\dash1\,{\absidx{\kappa}}{\absidx{\lambda}}}
    + \nonumber\\
  &\qquad + \frac1{12} \stwfonb_{0,3\,
    {\absidx{\mu}}{\absidx{\kappa}}{\absidx{\alpha}}}\:
    \stwfonb_{0,3\,
    {\absidx{\nu}}{\absidx{\lambda}}{\absidx{\beta}}}\:
    \stwfonb{}_{0,2}^{\dash1\,{\absidx{\mu}}{\absidx{\nu}}}
    \stwfonb{}_{0,2}^{\dash1\,{\absidx{\alpha}}{\absidx{\beta}}}
    \stwfonb{}_{0,2}^{\dash1\,{\absidx{\kappa}}{\absidx{\lambda}}}
  + \frac12 {\onbnd l}_{0,1\,{\absidx{\mu}}}\:
    {\onbnd l}_{0,1\,{\absidx{\nu}}}\:
    \stwfonb{}_{0,2}^{\dash1\,{\absidx{\mu}}{\absidx{\nu}}}
    \Bigr)(\yonb;\xi,\zeta) + \displaybreak[0]\\
  &\quad + \isign\: \phionb  (\yonb;\xi)\:
  \psionb_{1\,{\absidx{\mu}}}(\yonb;\zeta)\;
  \stwfonb{}_{0,2}^{\dash1\,{\absidx{\mu}}{\absidx{\nu}}}\; \Bigl(
  {\onbnd l}_{0,1\,{\absidx{\nu}}} -
  \frac12 \stwfonb_{0,3\,
    {\absidx{\kappa}}{\absidx{\lambda}}{\absidx{\nu}}}\:
    \stwfonb{}_{0,2}^{\dash1\,{\absidx{\kappa}}{\absidx{\lambda}}}
    \Bigr)(\yonb;\xi,\zeta) + \nonumber\\
  &\quad+ \isign\: \phionb  (\yonb;\xi)\:
    \psionb_{2\,{\absidx{\mu}}{\absidx{\nu}}} (\yonb;\zeta)\: 
    \stwfonb{}_{0,2}^{\dash1\,{\absidx{\mu}}{\absidx{\nu}}}
    (\yonb;\xi,\zeta) \period\nonumber
\end{align}
The equality in \eqref{tildejacDef} is proved in 
\eqref{tildejacProof}. Applying the expansion in 
\eqref{SmHLExpLemma} in \eqref{hkrnlwbstamplStep} 
and using \eqref{jacisvvmdetis} we get
\begin{equation}
\begin{split}
  &\int_{y\in\Omega} \stvol(y)\: \varphi(y)\, \psi(y) + \\
  &\qquad+ \sqrt{\tau} \Bigl(-\frac1\stsign\sqrt{\frac\isign{2\pi}}\Bigr)
    \int_{\yonb\in\Omega} \svol(\yonb)\: \varphi(\yonb)\, \psi(\yonb) + \\
  &\qquad+ \tau \int_{\substack{\yonb\in\bound\Omega\\\eta\in\realn^+}}
    \dvol\eta\svol(\yonb)\:\onbnd\jac(\yonb;\eta)\;
    \Bigl(\frac\isign{4\stsign^2}\, {\tilde\jac}^{\dash1}
    \bigl({\tilde\jac}\omega_0\bigr)^{\argl\tdot\,\argl\tdot} +
    \frac\isign{4\stsign^2}\, {\tilde\jac}^{\dash1}
    \bigl({\tilde\jac}\omega_0\bigr)^{\argr\tdot\,\argr\tdot} +
    \omega_1 \Bigr)(\yonb;\eta,\eta) +\\
  &\qquad+ \order{\tau^\frac32}\commae
\end{split}
\end{equation}
where $f^{\argl\tdot}(\xi,\zeta)$ (or $f^{\argr\tdot}(\xi,\zeta)$) 
means derivative of a function $f$ with respect 
of the left (or right) argument.

We see that we have proved already the desired 
expansion up to order $\sqrt\tau$. Now we proceed 
to prove it in the order $\tau$. We split the integrand 
of the last term to pieces and compute each of them. 
First we note that (see \eqref{bndcovexpwftwo}) 
\begin{equation}
  \stCL{\stwfonb_{0,2}} = \obmtrc\commae
\end{equation}
where by a coincidence limit of a function 
$f(\xi,\zeta)$ depending on two real parameters 
we mean $\stCL{f}(\eta) = f(\eta,\eta)$. 
Using \eqref{dotlnjac} and \eqref{dotvvmdcl} we get
\begin{equation}\begin{aligned}
  \frac1\stsign\stCLb{(\ln\tilde\jac)^{\argr\tdot}} &=
    \frac12\stsign\: \extscuronb\commae\\
  \frac1{\stsign^2}\: \stCLb{(\ln\tilde\jac)^{\argr\tdot\argr\tdot}} &=
    \frac1{2\stsign^2}\: (\ln\onbnd \jac)^{\argr\tdot\argr\tdot}
    + \frac1{2\stsign^2} \stCLb{(\ln\onbnd \stvvmd)^{\argr\tdot\argr\tdot}} =
    -\frac{\stsign^2}6 {\extcuronb}{}^2 + 
    \frac13 {\extscuronb}{}^{\tdot}\commae
\end{aligned}\end{equation}
and therefore
\begin{equation}\label{hkrnlwbstamplAStep}
  \frac1{\stsign^2}\:\stCLb{{\tilde\jac}^{\dash1}
  \bigl({\tilde\jac}\omega_0\bigr)^{\argr\tdot\,\argr\tdot}} =
  \frac1{\stsign^2}\: \phionb   \psionb{}^{\tdot\tdot} +
  \extscuronb\: \phionb \psionb{}^{\tdot} +
  \Bigl( \frac13{\extscuronb}{}^{\tdot} - 
  \frac16\stsign^2 {\extcuronb}{}^2 + 
  \frac14 {\extscuronb}{}^2\Bigr)\:
  \phionb  \psionb\period
\end{equation}
Here $\extscuronb$ is a trace of extrinsic curvature 
and $\extcuronb{}^2$ is a square of the extrinsic 
curvature (eq. \eqref{extScQcur}). Next, using 
\eqref{covexpandsc}, \eqref{bndcovexplone}, 
and \eqref{bndcovexpwfthree}, transforming 
the connection $\scnx$ to $\obcnx$ (eq. \eqref{sobcnxrel}) 
and performing an integration by parts gives
\begin{equation}\label{hkrnlwbstamplBStep}\begin{split}
  &\frac\isign2 \int_{\bound\Omega} \svol \onbnd\jac\:
  \phionb  \;\biggl(
  \obmtrc{}^{\dash1\,{\absidx{\mu}}{\absidx{\nu}}}\:
  \psionb_{2\,{\absidx{\mu}}{\absidx{\nu}}} +
  \psionb_{1\,{\absidx{\mu}}}\;
  {\obmtrc}^{\dash1\,{\absidx{\mu}}{\absidx{\nu}}}\;\Bigl(
  \stCLb{{\onbnd l}_{0,1\,{\absidx{\nu}}}} -
  \frac12 \stCLb{\stwfonb_{0,3\,
    {\absidx{\kappa}}{\absidx{\lambda}}{\absidx{\nu}}}}\:
    \obmtrc^{\dash1\,{\absidx{\kappa}}{\absidx{\lambda}}}
    \Bigr) \biggr)= \\
  &\qquad= - \frac\isign2 \int_{\bound\Omega} \obvol \;
  \sgrad_{\absidx{\mu}}\phionb  \:
  \obmtrc{}^{\dash1\,{\absidx{\mu}}{\absidx{\nu}}}\:
  \sgrad_{\absidx{\nu}}\psionb  \period
\end{split}\end{equation}
Substituting for $\stCL{\stwfonb_{0,k}}$ and 
$\stCL{{\onbnd l}_{0,k}}$ in remaining terms of 
$\stCL{\omega_1}$, a straightforward long calculation gives
\begin{equation}\label{hkrnlwbstamplCStep}
\begin{split}
  &-\frac18\stCLb{\stwfonb_{0,4\,
    {\absidx{\mu}}{\absidx{\nu}}{\absidx{\kappa}}{\absidx{\lambda}}}}\:
    \obmtrc^{\dash1\,{\absidx{\mu}}{\absidx{\nu}}}
    \obmtrc^{\dash1\,{\absidx{\kappa}}{\absidx{\lambda}}}
  + \frac12 \stCLb{{\onbnd l}_{0,2\,{\absidx{\mu}}{\absidx{\nu}}}}\:
    \obmtrc^{\dash1\,{\absidx{\mu}}{\absidx{\nu}}} + \\
  &\qquad +\frac18 \stCLb{\stwfonb_{0,3\,
    {\absidx{\mu}}{\absidx{\nu}}{\absidx{\alpha}}}}\:
    \stCLb{\stwfonb_{0,3\,
    {\absidx{\kappa}}{\absidx{\lambda}}{\absidx{\beta}}}}\:
    \obmtrc^{\dash1\,{\absidx{\mu}}{\absidx{\nu}}}
    \obmtrc^{\dash1\,{\absidx{\alpha}}{\absidx{\beta}}}
    \obmtrc^{\dash1\,{\absidx{\kappa}}{\absidx{\lambda}}} 
  + \frac12 \stCLb{{\onbnd l}_{0,1\,{\absidx{\mu}}}}\:
    \stCLb{{\onbnd l}_{0,1\,{\absidx{\nu}}}}\:
    \obmtrc^{\dash1\,{\absidx{\mu}}{\absidx{\nu}}} + \\
  &\qquad  + \frac1{12} \stCLb{\stwfonb_{0,3\,
    {\absidx{\mu}}{\absidx{\kappa}}{\absidx{\alpha}}}}\:
    \stCLb{\stwfonb_{0,3\,
    {\absidx{\nu}}{\absidx{\lambda}}{\absidx{\beta}}}}\:
    \obmtrc^{\dash1\,{\absidx{\mu}}{\absidx{\nu}}}
    \obmtrc^{\dash1\,{\absidx{\alpha}}{\absidx{\beta}}}
    \obmtrc^{\dash1\,{\absidx{\kappa}}{\absidx{\lambda}}} 
   - \frac12 \stCLb{{\onbnd l}_{0,1\,{\absidx{\mu}}}}\:
    \stCLb{\stwfonb_{0,3\,
    {\absidx{\kappa}}{\absidx{\lambda}}{\absidx{\nu}}}}\:
    \obmtrc^{\dash1\,{\absidx{\mu}}{\absidx{\nu}}}
    \obmtrc^{\dash1\,{\absidx{\kappa}}{\absidx{\lambda}}} =\\
  &\qquad = \isign\Bigl( - \frac16 {\extscuronb}{}^\tdot
    + \frac{\stsign^2}{12} {\extcuronb}{}^2
    - \frac{\stsign^2}8 {\extscuronb}{}^2\Bigr)\period
\end{split}\end{equation}

Putting together (\ref{hkrnlwbstamplAStep}--\ref{hkrnlwbstamplCStep}), 
and integrating by parts we find 
\begin{align}
  &\int_{\substack{\yonb\in\bound\Omega\\\eta\in\realn^+}}
    \dvol\eta\svol(\yonb)\:\onbnd\jac(\yonb;\eta)\;
    \Bigl(\frac\isign{4\stsign^2}\, {\tilde\jac}^{\dash1}
    \bigl({\tilde\jac}\omega_0\bigr)^{\argl\tdot\,\argl\tdot} +
    \frac\isign{4\stsign^2}\, {\tilde\jac}^{\dash1}
    \bigl({\tilde\jac}\omega_0\bigr)^{\argr\tdot\,\argr\tdot} +
    \omega_1 \Bigr)(\yonb;\eta,\eta) = \nonumber\\
  &\qquad=\isign\int_{\realn^+}\dvol\eta \int_{\bound\Omega}
    \obvol\biggl( \frac1{4\stsign^2}\, 
    \Bigl( {\phionb}{}^{\tdot\tdot}\psionb
    + {\phionb}\psionb{}^{\tdot\tdot}
    + \stsign^2 \extscuronb\: \bigl(
    {\phionb}{}^{\tdot}\psionb
    + {\phionb}\psionb{}^{\tdot}\bigr)\Bigr)
    -\frac12 \sgrad_{\absidx{\mu}}\phionb  \:
    \obmtrc{}^{\dash1\,{\absidx{\mu}}{\absidx{\nu}}}\:
    \sgrad_{\absidx{\nu}}\psionb\biggr) =
    \nonumber\displaybreak[0]\\
  &\qquad = -\frac\isign2
    \int_{\realn^+}\dvol\eta \int_{\bound\Omega}
    \stsign\obvol\Bigl( \frac1{\stsign^2}\, 
    {\phionb}{}^{\tdot}\psionb{}^{\tdot}
    + \sgrad_{\absidx{\mu}}\phionb  \:
    \obmtrc{}^{\dash1\,{\absidx{\mu}}{\absidx{\nu}}}\:
    \sgrad_{\absidx{\nu}}\psionb\biggr)
    - \frac\isign{4\stsign^2} \int_{\bound\Omega}
    \svol \:\Bigl( {\phionb}{}^{\tdot}\psionb
    + {\phionb}\psionb{}^{\tdot}\Bigr) =
    \nonumber\displaybreak[0]\\
  &\qquad = -\frac\isign2 \int_{\Omega} \stvol\;
    \grad_{\absidx{\mu}}\varphi\:
    \stmtrc{}^{\dash1\,{\absidx{\mu}}{\absidx{\nu}}}\:
    \grad_{\absidx{\nu}}\psi -
    \frac\isign{4\stsign^2}\int_{\bound\Omega}\svol \:\Bigl(
    \psi\:\normv^{\absidx{\mu}}\grad_{\absidx{\mu}}\varphi +
    \varphi\:\normv^{\absidx{\mu}}\grad_{\absidx{\mu}}\psi \Bigr) =
    \nonumber\displaybreak[0]\\
  &\qquad = -\frac\isign2\; \varphi\stint\Bigl(
    \bidst{L}_{\bbcnd{d}} - \frac12 
    \bigl(\acttol{d\bidst{L}}_{\bbcnd{d}} +
    \acttor{d\bidst{L}}_{\bbcnd{d}} \bigr)
    \Bigr)\stint\psi =
  -\frac\isign2\; \varphi\stint
    \acttolr{\bidst{L}} \stint\psi\period
\end{align}
This concludes the proof of the expansion in \eqref{hkrnlwbstamplexp}.

\subsection{Manifold with boundary --- reflection contribution}

Finally we will prove the last expansion 
we have needed in the main text:
\begin{equation}\label{hkrnlwbstamplbndexp}
\begin{split}
  &\frac\stsign{(2\pi\isign\tau)^{\frac\stdim2}}\:
    \stvvmd_\bnd^{\!\!1\dash p}(x|z) \;\beta(\tau,x|z) \;
    \exp\Bigl( - \frac1{\isign\tau} \stwf_\bnd(x|z)
    \Bigr) = \\
  &\qquad= 
  \sqrt{\tau}\; \frac1\stsign\sqrt{\frac\isign{2\pi}} \,\bbivol 
  - \tau \frac\isign2\, \frac12
  (\biFdl_{\bbcnd{d}} + \biFdr_{\bbcnd{d}}) 
  - \tau \frac\isign2\, 
  \Bigl(\frac{1+p}3\extscur + \beta\text{-terms}\Bigr) 
  + \order{\tau^\frac32}\period
\end{split}
\end{equation}

Similarly to the previous section, we smooth 
this expression with test functions, perform 
a change of variables \eqref{chnagvarwmap} 
and consequently \eqref{chnagvarbumap}, 
and use the covariant expansions \eqref{psibndcovexp},
\begin{equation}
\begin{aligned}
  \betaonb(\tau,\xonb,\xi|\zonb,\zeta) &= \betaonb_0(\xonb,\xi|\zonb,\zeta) + 
  \sqrt\tau\betaonb_{\frac12}(\xonb,\xi|\zonb,\zeta) + \order{\tau}\commae\\
  \onbnd \beta_0(\yonb,\xi|\yonb,\sqrt\tau Y,\zeta) &=
  \sum_{k=\naturaln_0} \tau^{\frac k2} \frac1{k!} \;
  {\onbnd \beta}_{0;0,k\,\absidx{\mu}_1\dots\absidx{\mu}_k}(\yonb;\xi,\zeta) \; 
  Y^{\absidx{\mu}_1}\dots Y^{\absidx{\mu}_k} \spce;
\end{aligned}
\end{equation}
similarly for $\beta_{\frac12}(x|z)$; and 
\begin{align}
  \onbnd{\stwf}_\bnd(\yonb,\xi|\yonb,\sqrt\tau Y,\zeta) &=
  \tau \stsign^2 \frac12 (\xi+\zeta)^2 +
  \sum_{k=2,3,\dots} \tau^{\frac k2} \frac1{k!}\; 
  \stwfonb_{\bnd\,0,k\,\absidx{\mu}_1\dots\absidx{\mu}_k}
  (\yonb;\xi,\zeta) \;
  Y^{\absidx{\mu}_1}\dots Y^{\absidx{\mu}_k} \commae\\
  \onbnd l_\bnd(\yonb,\xi|\yonb,\sqrt\tau Y,\zeta) &=
  \sum_{k=\naturaln_0} \tau^{\frac k2} \frac1{k!} \;
  {\onbnd l}_{\bnd\,0,k\,\absidx{\mu}_1\dots\absidx{\mu}_k}(\yonb;\xi,\zeta) \; 
  Y^{\absidx{\mu}_1}\dots Y^{\absidx{\mu}_k} \commae
\end{align}
where
\begin{equation}
  \onbnd l_\bnd(\xonb,\xi|\zonb,\zeta) = \ln\biggl(
  \onbnd\jac(\xonb,\xi)
  \frac{{\onbnd\stvvmd}_\bnd{}^{\!\!1\dash p}(\xonb,\xi|\zonb,\zeta)}
  {\svvmd(\xonb|\zonb)}
  \onbnd\jac(\zonb,\zeta)\biggr)\period
\end{equation}
This leads to a Gaussian integration in 
the variable $Y\in\tens_\yonb \bound\Omega$ 
in which only leading terms in expansions survive and we get
\begin{equation}
\begin{split}
  &\frac\stsign{(2\pi\isign\tau)^{\frac\stdim2}}\:
    \int_{x,z\in\Omega} \stvol(x)\stvol(z)\:\varphi(x)\psi(z)\;
    \stvvmd_\bnd^{\!\!1\dash p}(x|z) \;\beta(\tau,x|z) \;
    \exp\Bigl( - \frac1{\isign\tau} \stwf_\bnd(x|z)
    \Bigr) = \\
  &\qquad= \int_{\yonb\in \bound\Omega}
    \frac\stsign{(2\pi\isign\tau)^{\frac12}}\:\svol(\yonb)
    \int_{\xi,\zeta\in\realn^+} \dvol\xi\dvol\zeta\,
    \exp\Bigl(-\frac{\stsign^2}{2\isign\tau}(\xi+\zeta)^2\Bigr)\;
    \phionb(\yonb,\xi)\,\psionb(\yonb,\zeta)\;
    \stimes\\
  &\qquad\qquad\qquad\qquad\qquad\stimes
    \tilde\jac_\bnd(\yonb;\xi,\tfrac{\xi\zeta}{\xi+\zeta},\zeta)\;
    \Bigl(\tilde\beta_0(\yonb;\xi,\tfrac{\xi\zeta}{\xi+\zeta},\zeta) + \sqrt\tau\,
    \tilde\beta_{\frac12}(\yonb;\xi,\tfrac{\xi\zeta}{\xi+\zeta},\zeta) +
    \order{\tau}\Bigl)\commae
\end{split}\end{equation}
where
\begin{align}
\begin{split}
  \tilde\jac_\bnd(\yonb;\xi,\tfrac{\xi\zeta}{\xi+\zeta},\zeta) &=
    \onbnd\jac(\yonb,\xi)\;{\onbnd\stvvmd}_\bnd{}^{\!\!1\dash p}
    (\yonb;\xi|\yonb,\zeta)\; \onbnd\jac(\yonb,\zeta)\;
    \svol(\yonb)\;\Bigl(\Det_\bpar \stwfonb_{\bnd\,0,2}
    (\yonb;\xi,\zeta)\Bigr)^{-\frac12} = \\
    &= \Bigl(\onbnd\jac(\yonb,\xi)\;{\onbnd\stvvmd}_\bnd{}^{\!\!1\dash 2 p}
    (\yonb;\xi|\yonb,\zeta)\; \onbnd\jac(\yonb,\zeta) \Bigr)^{\frac12}
    \commae
  \end{split}\label{tildajacbndrel}\\
  \tilde\beta_0(\yonb;\xi,\tfrac{\xi\zeta}{\xi+\zeta},\zeta) &=
    \betaonb_0(\yonb,\xi|\yonb,\zeta)\commae
\end{align}
with $\tilde\jac_\bnd$ and $\tilde\beta_0$ 
depending analyticaly on their three real arguments. 
$\tilde\beta_{\frac12}$ is defined in similar 
way as $\tilde\beta_0$. Here we anticipate that 
$\onbnd\stvvmd_\bnd$ and $\betaonb$ can have 
more complicated analytical dependence on $\xi$ and $\zeta$. 
The equality in \eqref{tildajacbndrel} follows from 
\eqref{tildejacBndProof}. Using the expansion 
\eqref{SmHLBndExpLemma} we obtain
\begin{equation}
\begin{split}
  \int_{\yonb\in \bound\Omega}\svol(\yonb) \biggl( &
  \sqrt\tau \frac1\stsign\sqrt{\frac\isign{2\pi}}\:
  \tilde\beta_0(\yonb;0,0,0) \: \varphi(\yonb) \psi(\yonb) +\\
  & + \tau \frac\isign{4\stsign^2}\:\tilde\beta_0(\yonb;0,0,0)\:
  \Bigl( \phionb{}^\tdot(\yonb,0) \psionb(\yonb,0) +
  \phionb(\yonb,0) \psionb{}^\tdot(\yonb,0) \Bigr) +\\
  & + \tau \: \Bigl(
  \frac\isign{4\stsign^2} 
  \bigl(\tilde\jac_\bnd\tilde\beta_0\bigr)^{\argl\tdot} +
  \frac\isign{4\stsign^2} 
  \bigl(\tilde\jac_\bnd\tilde\beta_0\bigr)^{\argr\tdot} + \\
  &\qquad\qquad +\frac\isign{12\stsign^2} 
  \bigl(\tilde\jac_\bnd\tilde\beta_0\bigr)^{\argm\tdot} + 
  \frac1\stsign\sqrt{\frac\isign{2\pi}}\:
  \tilde\beta_{\frac12} \Bigr)\!(\yonb;0,0,0)
  \; \varphi(\yonb) \psi(\yonb) + \\
  &+\order{\tau^{\frac32}} \biggr) \period
\end{split}
\end{equation}
Using the expansion \eqref{jactildeExp} of 
$\tilde\jac_\bnd$ and obvious relations for 
$\tilde\beta_0(\yonb;0,0,0)$ and 
$\tilde\beta_{\frac12}(\yonb;0,0,0)$, 
this expression is equal to
\begin{equation}
\begin{split}
  \int_{\yonb\in \bound\Omega}\svol(\yonb) \biggl( &
  \sqrt\tau \frac1\stsign\sqrt{\frac\isign{2\pi}}\:
  \beta(0,\yonb|\yonb) \: \varphi(\yonb) \psi(\yonb) +\\
  & + \tau \frac\isign{4\stsign^2}\:\beta(0,\yonb|\yonb)\:
  \Bigl( \phionb{}^\tdot(\yonb,0) \psionb(\yonb,0) +
  \phionb(\yonb,0) \psionb{}^\tdot(\yonb,0) \Bigr) +\\
  & + \frac{\tau\isign}2 \: \biggl(
  - \frac{1+6}3\beta(0,\yonb|\yonb) \extscur(\yonb) 
  + \frac2\stsign\sqrt{\frac1{2\pi\isign}}\:\tder\beta(0,\yonb|\yonb)
  +\\&\qquad\qquad
  + \Bigl(\frac1{2\stsign^2}\tilde\beta_0{}^{\argl\tdot}
  + \frac1{2\stsign^2}\tilde\beta_0{}^{\argr\tdot} 
  + \frac1{6\stsign^2}\tilde\beta_0{}^{\argm\tdot}
  \Bigr)\!(\yonb;0,0,0) 
  \biggr)\: \varphi(\yonb) \psi(\yonb) + \\
  &+\order{\tau^{\frac32}} \biggr) \period
\end{split}
\end{equation}
Using the normalization condition \eqref{betazeroCL} 
concludes the proof of the expansion \eqref{hkrnlwbstamplbndexp}. 
By inspection we see that the $\beta$-terms have form
\begin{equation}\label{BetaTerms}
  \beta\text{-terms} = 
  - \frac2\stsign\sqrt{\frac1{2\pi\isign}}\:\tder\beta(0,\yonb|\yonb) -
  \Bigl(
  \frac1{2\stsign^2}\tilde\beta_0{}^{\argl\tdot} +
  \frac1{2\stsign^2}\tilde\beta_0{}^{\argr\tdot} +
  \frac1{6\stsign^2}\tilde\beta_0{}^{\argm\tdot}
  \Bigr)\!(\yonb;0,0,0)\period
\end{equation}


\section{Geodesic theory}
\label{apx:GdTh}


\subsection{Basic definitions}

In this appendix we review some facts from geodesic 
theory and list a number of useful expansion, 
some of which we have used in this work. 
The material related to a manifold without boundary 
is well known --- see for example the classical works 
\cite{DeWittBrehme:1960,DeWitt:book1965} and 
\cite{Christensen:1978}. The theory of expansion near 
a boundary is less known. Some material can be found in 
\cite{McAvityOsborn:1990,McAvityOsborn:1991,McAvityOsborn:1992}. 
The calculations are usually straightforward but cumbersome. 
We will present mostly only results. But all expressions 
presented were computed and checked by the author 
(see \cite{Krtous:PointSplitCalc}).

We start with introducing the \defterm{covariant expansion} 
in a curved manifold. We would like to expand a sufficiently 
smooth tensor field $A_{\absidx{\beta}\dots}^{\absidx{\alpha}\dots}$ 
on a manifold around a point $x$. First we change 
the dependence on a point $z$ in the manifold 
$\stmfld$ to the dependence on a vector $Z$ 
from $\tens_x\, \stmfld$, then we transform vector 
indices from different tangent spaces to one common 
tensor space, and finally we do the usual Taylor 
expansion of a linear-space-valued function on a vector space.

To transform the tensor field on the manifold to a 
linear-space-valued function we need to know how to move 
tensors from one tangent point to another. We assume that 
we have given a metric $\stmtrc$ which define a parallel transport. 
It allows us to transform tensors from the tangent space at 
point $z$ in a normal neighborhood of the point $x$ 
to the space $\tens_x\, \stmfld$ along the geodetic 
joining these two points. In the normal neighborhood of 
$x$ we can parametrize a geodesic by its tangent vector at $x$, 
i.e. we can define an \defterm{exponential map $\stgeod_x$}
\begin{gather}\label{umapDef}
  \stgeod_x : \tens_x\, \stmfld \rightarrow \stmfld\commae\\
  \frac{\stcnx}{d\tau}\frac{\Dmap}{d\tau} \stgeod_x(\tau X) = 0 \comma
  \frac{\Dmap}{d\tau} \stgeod_x(\tau X) |_{\tau = 0} = X \period
\end{gather}
If $f(z)$ is some manifold dependent function, we use notation 
$f(x;Z) = f(\stgeod_x(Z))$. This transformation concludes  
our first step. Next we parallel transform vector indices of 
the tensor field to the space $\tens_x\, \stmfld$ along 
the geodesics starting from $x$. We define the 
\defterm{tensor of geodesic transport $\sttgt(x|z)$} from $z$ to $x$
\begin{equation}
  \sttgt^{\absidx{\mu}}{}_{\absidx{\nu}}(x|z) \in 
  \tens_x\, \stmfld \otimes \tens_z^\dual\, \stmfld \comma
  \frac{\stcnx}{d\tau} \sttgt(x,\stgeod_x(\tau X)) = 0\commae
\end{equation}
and its version with indices up and down
\begin{equation}
  \sttgtu = \sttgt \stctr \stmtrc^{\dash1} 
  \comma \sttgtd = \stmtrc \stctr \sttgt \period
\end{equation}
Using this tensor we can write down the tensor field $A$ 
with transported indices explicitly. We obtain 
the linear-space-valued function on a linear space
\begin{equation}
  A_{\absidx{\nu}\dots}^{\absidx{\mu}\dots}(x;.)\; 
  \sttgt^{\absidx{\alpha}}{}_{\absidx{\mu}}(x|x;.) \dots\;
  \sttgt^{\dash1\,\absidx{\nu}}{}_{\absidx{\beta}}(x|x;.) \dots\quad :\;
  \tens_x\, \stmfld \rightarrow \tens_x{}^k_l\, \stmfld \period
\end{equation}
Finally we can write the \defterm{covariant expansion}
\begin{equation}\label{CovExpGen}
  A_{\absidx{\nu}\dots}^{\absidx{\mu}\dots}(x;Z)\; 
  \sttgt^{\absidx{\alpha}}{}_{\absidx{\mu}}(x|x;Z) \dots\;
  \sttgt^{\dash1\,\absidx{\nu}}{}_{\absidx{\beta}}(x|x;Z) \dots\,  {=}\, 
  \sum_{k\in\naturaln_0} \frac1{k!} A
  _{k}{}_{\absidx{\beta}\dots\,  {\absidx{\mu}}_1\dots{\absidx{\mu}}_k}
  ^{\absidx{\alpha}\dots}(x)\;
  Z^{{\absidx{\mu}}_1}\, \dots Z^{{\absidx{\mu}}_k}\period
\end{equation}
We call $A_k(x)$ the coefficients of the covariant expansion 
of the field $A$ at $x$. They are tensors at $x$ symmetric in indices 
${\scriptstyle{\absidx{\mu}}_1},\,  \dots,\,  {\scriptstyle{\absidx{\mu}}_k}$. 

To compute these coefficients we need to develop 
geodesic theory to greater detail. First we define 
the \defterm{world function $\stwf(x|z)$} of the metric $\stmtrc$. 
It is given by half of the squared geodesic distance 
between points $x$ and $z$ --- see \eqref{stwfDef}. 
For time-like separated points it is negative. 
The \defterm{geodesic distance} is then given by
\begin{equation}
  \stgdist(x|z) = \abs{2\stwf(x|z)}^{\frac12} \period
\end{equation}
We define \defterm{geodesic tangent vectors $\stwfl$, $\stwfr$}
\begin{equation}
  \stwfl(x|z) = \stmtrc^{\dash1}(x) \stctr\grad_\argl \stwf(x|z) \comma 
  \stwfr(x|z) = \grad_\argr \stwf(x|z) \stctr\stmtrc^{\dash1}(z) \period
\end{equation}
Here, as before, $\grad_\argl f$ or $\grad_\argr f$ denote 
the gradient in the left or right argument of a bi-function $f(x|z)$.

The basic properties of the world function 
(see e.g. \cite{DeWitt:book1965}) are that its gradient 
vector $\stwfl(x|z)$ is really tangent to the geodesic between 
$x$ and $z$ and it is normalized to the length of the geodesic. I.e. 
\begin{equation}\label{wfProperty}
  - Z = \stwfl(x|x;Z) \period
\end{equation}
We also introduce a special notation for 
the second derivatives of the world function
\begin{equation}
  \stwfll = \stcnx_\argl\grad_\argl \stwf \comma
  \stwflr = \grad_\argl\grad_\argr \stwf \comma
  \stwfrr = \stcnx_\argr\grad_\argr \stwf \period
\end{equation}
To conclude our definition we introduce also 
determinants of $\sttgt$, $\sttgtd$, $\stwflr$. 
They are well-defined objects --- bi-densities on $\stmfld$
\begin{gather}
  \sttgtdet(x|z) = \frac1{\stsign^2}\Det \sttgt(x|z) =
  \mtrcdet{\stmtrc}^{\dash\frac12}(x) \stvol(z) \comma
  \sttgtddet(x|z) = \frac1{\stsign^2}\Det \sttgtd(x|z) = 
  \stvol(x) \stvol(z) \commae\\
  \stwfdet(x|z) = \frac1{\stsign^2}\Det \bigl(- \stwflr(x|z)\bigr)
  \period
\end{gather} 
Finally we define \defterm{Van-Vleck Morette determinant} 
\begin{equation}\label{stvvmdDef}
  \stvvmd(x|z) =  \stwfdet(x|z)\:\sttgtddet^{\dash1}(x|z) =
  \stwfdet(x|z)\:
  \mtrcdet{\stmtrc}^{\dash\frac12}(x)
  \mtrcdet{\stmtrc}^{\dash\frac12}(z) \period
\end{equation}

For a bi-tensor $F(x|z)$ on the manifold --- 
a tensor object depending on two points in the manifold 
--- we denote the \defterm{coincidence limit}
\begin{equation}
 \stCL{F}(x) = F(x|x) \period
\end{equation}
The generalized Synge's theorem 
(see e.g. \cite{Christensen:1976}) tells us that
\begin{equation}
 \stcnx\stCL{F} = \stCL{\stcnx_\argl F} + \stCL{\stcnx_\argr F} \period
\end{equation}

\subsection{Coincidence limits and covariant expansions}

Equation \eqref{wfProperty} gives\note{nt:DotsConv}
\begin{equation}\label{wfofgeod}
  \stwf = \frac12\: \grad_\argr \stwf \stctr
  \stmtrc^{\dash1} \stctr \grad_\argr \stwf =
  \frac12\: \grad_\argl \stwf \stctr
  \stmtrc^{\dash1} \stctr \grad_\argl \stwf =
  \frac12\: \stwfl \stctr \stmtrc \stctr \stwfl =
  \frac12\: \stwfr \stctr \stmtrc \stctr \stwfr \period
\end{equation}
Taking repeatedly derivatives of this expression in both arguments we can derive
\begin{equation}
  \stwfl \stctr \stmtrc =  \stwfl \stctr \stwfll =
  \stwflr \stctr \stwfr \comma
  \stmtrc \stctr \stwfr   =  \stwfrr \stctr \stwfr =
  \stwfl \stctr \stwflr \commae
\end{equation}
and following identities
\begin{align}
  \stwfl^{\absidx{\mu}} \grad_{\argl\, \absidx{\mu}} \stwf &= 2 \stwf \commae\\
\begin{split}
  \stwfl^{\absidx{\mu}} 
  \stcnx_{\argl\, \absidx{\mu}} \stwfl^{\absidx{\alpha}} &= 
  \stwfl^{\absidx{\alpha}} \commae\\
  \stwfl^{\absidx{\mu}} 
  \stcnx_{\argl\, \absidx{\mu}} \stwfr^{\absidx{\alpha}} &= 
  \stwfr^{\absidx{\alpha}} \commae 
\end{split}\\
\begin{split}
  \stwfl^{\absidx{\mu}} 
  \stcnx_{\argl\, \absidx{\mu}} \stwfll_{\absidx{\alpha\beta}} &= 
  \stwfll_{\absidx{\alpha\beta}} -
  \stwfll_{\absidx{\alpha\mu}} \stwfll_{\absidx{\beta\nu}}
  \stmtrc^{\dash1\, \absidx{\mu\nu}} -
  \stcur_{\absidx{\alpha\mu\beta\nu}}
  \stwfl^{\absidx{\mu}} \stwfl^{\absidx{\nu}}\commae\\
  \stwfl^{\absidx{\mu}} 
  \stcnx_{\argl\, \absidx{\mu}} \stwflr_{\absidx{\alpha\beta}} &= 
  \stwflr_{\absidx{\alpha\beta}} -
  \stwfll_{\absidx{\alpha\mu}} 
  \stmtrc^{\dash1\, \absidx{\mu\nu}} \stwflr_{\absidx{\nu\beta}}\commae\\
  \stwfl^{\absidx{\mu}} 
  \stcnx_{\argl\, \absidx{\mu}} \stwfrr_{\absidx{\alpha\beta}} &= 
  \stwfrr_{\absidx{\alpha\beta}} -
  \stmtrc^{\dash1\, \absidx{\mu\nu}}  \stwflr_{\absidx{\mu\alpha}} 
  \stwflr_{\absidx{\nu\beta}}\commae\\
  \stwfl^{\absidx{\mu}} 
  \stcnx_{\argl\, \absidx{\alpha}} \stcnx_{\argl\, \absidx{\beta}} 
  \stcnx_{\argl\, \absidx{\mu}} \stwf &= 
  \stwfll_{\absidx{\alpha\beta}} -
  \stwfll_{\absidx{\alpha\mu}} \stwfll_{\absidx{\beta\nu}}
  \stmtrc^{\dash1\, \absidx{\mu\nu}} \period
\end{split}
\end{align}
Similar, more complicated relations hold for higher derivatives. 
Using the fact that coincidence limits of the world function 
and tangent geodesic vector are zero, taking coincidence limits 
of relations above and similar relations for higher 
derivatives, and using the Synge's theorem, we get
\begin{gather}
  \stCL{\stwf} = 0 \commae\\
  \stCL{\grad_\argl\stwf} = \stCL{\grad_\argr\stwf} = 0\commae\\
  \stCL{\stcnx_\argl\stcnx_\argl\stwf} = 
  - \stCL{\stcnx_\argl\stcnx_\argr\stwf} = 
  \stCL{\stcnx_\argr\stcnx_\argr\stwf} =
  \stmtrc\commae\\
  \stCL{\stcnx_\argl\stcnx_\argl\stcnx_\argl\stwf} = 
  \stCL{\stcnx_\argl\stcnx_\argl\stcnx_\argr\stwf} = 
  \stCL{\stcnx_\argl\stcnx_\argr\stcnx_\argr\stwf} =
  \stCL{\stcnx_\argr\stcnx_\argr\stcnx_\argr\stwf} =
  0\commae\\
\begin{split}
  &\stCL{\stcnx_{\argl\, \absidx{\alpha}}\stcnx_{\argl\, \absidx{\beta}}
     \stcnx_{\argl\, \absidx{\mu}}\stcnx_{\argl\, \absidx{\nu}}\stwf} = 
  -\stCL{\stcnx_{\argl\, \absidx{\beta}}\stcnx_{\argl\, \absidx{\mu}}
     \stcnx_{\argl\, \absidx{\nu}}\stcnx_{\argr\, \absidx{\alpha}}\stwf} = 
  \stCL{\stcnx_{\argl\, \absidx{\mu}}\stcnx_{\argl\, \absidx{\nu}}
     \stcnx_{\argr\, \absidx{\beta}}\stcnx_{\argr\, \absidx{\alpha}}\stwf} = \\
  &\qquad= -\stCL{\stcnx_{\argl\, \absidx{\nu}}\stcnx_{\argr\, \absidx{\mu}}
     \stcnx_{\argr\, \absidx{\beta}}\stcnx_{\argr\, \absidx{\alpha}}\stwf} =
  \stCL{\stcnx_{\argr\, \absidx{\nu}}\stcnx_{\argr\, \absidx{\mu}}
      \stcnx_{\argr\, \absidx{\beta}}\stcnx_{\argr\, \absidx{\alpha}}\stwf} =
  -\frac13 \bigl(\stcur_{\absidx{\alpha\mu\beta\nu}} +
        \stcur_{\absidx{\alpha\nu\beta\mu}}\bigr)\period 
\end{split}
\end{gather}
Similar relation for the fifth and sixth 
derivative can be found in \cite{Krtous:PointSplitCalc}.

Now we are prepared to compute at least some 
coefficients of covariant expansion. We start with the simplest 
case of the covariant expansion of a function $f$ on the manifold. 
In this case we do not have problems with tensor nature of $f$ 
and we do not have to worry about parallel transport of tensor indices. 
The equation \eqref{CovExpGen} can be rewritten using \eqref{wfProperty} as
\begin{equation}
  f(z) = \sum_{k\in\naturaln_0} \frac{(-1)^k}{k!}
  f_{k\, {\absidx{\mu}}_1\dots{\absidx{\mu}}_k}(x) \;
  \stwfl^{{\absidx{\mu}}_1}(x,z)\dots\stwfl^{{\absidx{\mu}}_k}(x,z)\period
\end{equation}
Taking derivatives of this equation and coincidence 
limits we can find that the coefficients are given by
\begin{equation}\label{covexpandsc}
  f_{k\, {\absidx{\mu}}_1\dots{\absidx{\mu}}_k} = 
  \stcnx_{({\absidx{\mu}}_1}\dots\stcnx_{{\absidx{\mu}}_k)} f\period
\end{equation}

To do a similar calculation for a general tensor field $A$ 
we need to know the coincidence limits of the geodesic 
transport tensor. They can be calculated from the equation
\begin{equation}
  \stwfl = - \sttgt \stctr \stwfr
\end{equation}
by taking derivatives and coincidence limits. 
We give only a list of some of them 
(see \cite{Christensen:1976,Christensen:1978} or \cite{Krtous:PointSplitCalc}).
\begin{gather}
  \stCL{\sttgtd} = \stmtrc
  \displaybreak[0]\commae \\
  \stCL{\stcnx_\argl\sttgtd} = \stCL{\stcnx_\argr\sttgtd} = 0
  \displaybreak[0]\commae \\
  - \stCL{\stcnx_{\argl{\absidx{\beta}}}\stcnx_{\argl{\absidx{\alpha}}}
        \sttgtd_{\absidx{\mu\nu}}} = 
  \stCL{\stcnx_{\argl{\absidx{\alpha}}}\stcnx_{\argr{\absidx{\beta}}}
        \sttgtd_{\absidx{\mu\nu}}} = 
  - \stCL{\stcnx_{\argr{\absidx{\alpha}}}\stcnx_{\argr{\absidx{\beta}}}
        \sttgtd_{\absidx{\mu\nu}}} = 
  \frac12 \stcur_{\absidx{\alpha\beta\mu\nu}}
  \displaybreak[0]\commae\\
  \stCL{\stcnx_{\argl{\absidx{\gamma}}}
        \stcnx_{\argl{\absidx{\beta}}}
        \stcnx_{\argl{\absidx{\alpha}}}
        \sttgtd_{\absidx{\mu\nu}}} = 
  - \frac13 \stcnx_{\absidx{\gamma}}
        \stcur_{\absidx{\alpha\beta\mu\nu}}
  - \frac13 \stcnx_{\absidx{\beta}}
        \stcur_{\absidx{\alpha\gamma\mu\nu}}
  \displaybreak[0]\commae\\
\begin{split}
  &\stCL{\stcnx_{\argl{\absidx{\delta}}}
        \stcnx_{\argl{\absidx{\gamma}}}
        \stcnx_{\argl{\absidx{\beta}}}
        \stcnx_{\argl{\absidx{\alpha}}}
        \sttgtd_{\absidx{\mu\nu}}} = \\
  &\qquad
  - \frac14 \stcnx_{\absidx{\delta}}\stcnx_{\absidx{\gamma}}
    \stcur_{\absidx{\mu\nu\alpha\beta}} 
  - \frac14 \stcnx_{\absidx{\delta}}\stcnx_{\absidx{\beta}}
    \stcur_{\absidx{\mu\nu\alpha\gamma}} 
  - \frac14 \stcnx_{\absidx{\gamma}}\stcnx_{\absidx{\beta}}
    \stcur_{\absidx{\mu\nu\alpha\delta}}
  + \\ &\qquad
  + \frac18\: \stcur_{\absidx{\mu\nu\delta\lambda}}\:
    \stcur_{\absidx{\kappa\alpha\beta\gamma}}\:
    \stmtrc^{\dash1\absidx{\kappa\lambda}}
  + \frac18\: \stcur_{\absidx{\mu\nu\gamma\lambda}}\:
    \stcur_{\absidx{\kappa\alpha\beta\delta}}\:
    \stmtrc^{\dash1\absidx{\kappa\lambda}}
  + \frac18\: \stcur_{\absidx{\mu\nu\beta\lambda}}\:
    \stcur_{\absidx{\kappa\alpha\gamma\delta}}\:
    \stmtrc^{\dash1\absidx{\kappa\lambda}}
  + \\ &\qquad
  + \frac1{24}\: \stcur_{\absidx{\mu\nu\delta\lambda}}\:
    \stcur_{\absidx{\kappa\beta\alpha\gamma}}\:
    \stmtrc^{\dash1\absidx{\kappa\lambda}}
  + \frac1{24}\: \stcur_{\absidx{\mu\nu\gamma\lambda}}\:
    \stcur_{\absidx{\kappa\beta\alpha\delta}}\:
    \stmtrc^{\dash1\absidx{\kappa\lambda}}
  + \frac1{24}\: \stcur_{\absidx{\mu\nu\delta\lambda}}\:
    \stcur_{\absidx{\kappa\gamma\alpha\beta}}\:
    \stmtrc^{\dash1\absidx{\kappa\lambda}}
  + \\ &\qquad
  + \frac18\: \stcur_{\absidx{\nu\lambda\gamma\delta}}\:
    \stcur_{\absidx{\kappa\mu\alpha\beta}}\:
    \stmtrc^{\dash1\absidx{\kappa\lambda}}
  + \frac18\: \stcur_{\absidx{\nu\lambda\beta\delta}}\:
    \stcur_{\absidx{\kappa\mu\alpha\gamma}}\:
    \stmtrc^{\dash1\absidx{\kappa\lambda}}
  + \frac18\: \stcur_{\absidx{\nu\lambda\beta\gamma}}\:
    \stcur_{\absidx{\kappa\mu\alpha\delta}}\:
    \stmtrc^{\dash1\absidx{\kappa\lambda}}
  + \\ &\qquad
  + \frac18\: \stcur_{\absidx{\nu\lambda\alpha\delta}}\:
    \stcur_{\absidx{\kappa\mu\beta\gamma}}\:
    \stmtrc^{\dash1\absidx{\kappa\lambda}}
  + \frac18\: \stcur_{\absidx{\nu\lambda\alpha\gamma}}\:
    \stcur_{\absidx{\kappa\mu\beta\delta}}\:
    \stmtrc^{\dash1\absidx{\kappa\lambda}}
  + \frac18\: \stcur_{\absidx{\nu\lambda\alpha\beta}}\:
    \stcur_{\absidx{\kappa\mu\gamma\delta}}\:
    \stmtrc^{\dash1\absidx{\kappa\lambda}}
  + \\ &\qquad
  + \frac1{24}\: \stcur_{\absidx{\mu\nu\kappa\gamma}}\:
    \stcur_{\absidx{\alpha\beta\delta\lambda}}\:
    \stmtrc^{\dash1\absidx{\kappa\lambda}}
  + \frac1{24}\: \stcur_{\absidx{\mu\nu\kappa\beta}}\:
    \stcur_{\absidx{\alpha\gamma\delta\lambda}}\:
    \stmtrc^{\dash1\absidx{\kappa\lambda}}
  + \frac1{24}\: \stcur_{\absidx{\mu\nu\kappa\beta}}\:
    \stcur_{\absidx{\alpha\delta\gamma\lambda}}\:
    \stmtrc^{\dash1\absidx{\kappa\lambda}}
  + \\ &\qquad
  + \frac1{24}\: \stcur_{\absidx{\mu\nu\kappa\alpha}}\:
    \stcur_{\absidx{\beta\gamma\delta\lambda}}\:
    \stmtrc^{\dash1\absidx{\kappa\lambda}}
  + \frac1{24}\: \stcur_{\absidx{\mu\nu\kappa\alpha}}\:
    \stcur_{\absidx{\beta\delta\gamma\lambda}}\:
    \stmtrc^{\dash1\absidx{\kappa\lambda}}
  + \frac18\: \stcur_{\absidx{\nu\mu\kappa\alpha}}\:
    \stcur_{\absidx{\beta\lambda\gamma\delta}}\:
    \stmtrc^{\dash1\absidx{\kappa\lambda}}
\end{split}
\end{gather}
Derivatives in other argument can be obtained using 
Synge's theorem and commuting covariant derivatives.

Now it is straightforward to compute the coefficients 
in a covariant expansion of a general field. It can 
be done by taking covariant derivatives and coincidence 
limits of the rewritten equation \eqref{CovExpGen}
\begin{equation}
  A_{\absidx{\nu}\dots}^{\absidx{\mu}\dots}(z)\; 
  \sttgt_{\absidx{\mu}}^{\absidx{\alpha}}(x|z) \dots\;
  \sttgt^{\dash1}{}_{\absidx{\beta}}^{\absidx{\nu}}(x|z) \dots\;  {=}\, 
  \sum_{k\in\naturaln_0} \frac{(-1)^k}{k!} A
  _{k}{}_{\absidx{\beta}\dots\,  {\absidx{\mu}}_1\dots{\absidx{\mu}}_k}
  ^{\absidx{\alpha}\dots}(x)\;
  \stwfl^{{\absidx{\mu}}_1}(x|z)\, \dots \stwfl^{{\absidx{\mu}}_k}(x|z)\period
\end{equation}
We will not list explicit results.

We can also expand a bi-tensor $A(x|z)$ in both its 
arguments around some point $y$. We denote coefficients 
of such expansion $A_{k,l}(y)$. I.e.
\begin{equation}
  \sttgt^{\diffind}(y|y;X) \sttgt^{\diffind}(y|y;Z) A(y;X|y;Z) =
  \sum_{k,l\in\naturaln}
  A_{k,l\, {\absidx{\mu}}_1\dots{\absidx{\mu}}_k
     {\absidx{\nu}}_1\dots{\absidx{\nu}}_l}(y)\;
  X^{{\absidx{\mu}}_1}\dots X^{{\absidx{\mu}}_k}\;
  Z^{{\absidx{\nu}}_1}\dots Z^{{\absidx{\nu}}_k}\commae
\end{equation}
where by $\sttgt^{\diffind}(y|z)A(z)$ we mean a parallel 
transport of all indices from $z$ to $y$. In the case of 
a bi-scalar $f(x|z)$, similarly to \eqref{covexpandsc} 
we can derive that 
\begin{equation}\label{CovExpCoeffBiGen}
  f_{k,l\,{\absidx{\mu}}_1\dots{\absidx{\mu}}_k
     {\absidx{\nu}}_1\dots{\absidx{\nu}}_l} = \stCL{
  \stcnx_{(\argl{\absidx{\mu}}_1} \dots
  \stcnx_{\argl{\absidx{\mu}}_k)}
  \stcnx_{(\argr{\absidx{\nu}}_1} \dots
  \stcnx_{\argr{\absidx{\nu}}_l)} f }\period
\end{equation}

For calculations in appendix \ref{apx:AEHK} we need 
the covariant expansion of the world function $\stwf(x|z)$. 
When we expand both its arguments at point $y$ using 
the method described above we obtain
\begin{equation}
  \stwf(y;X|y;Z) = \frac12
  \bigl(X-Z\bigr)^{\absidx{\mu}}\,
  \stmtrc_{\absidx{\mu\nu}}(y)\,
  \bigl(X-Z\bigr)^{\absidx{\nu}} 
  - \frac16  X^{\absidx{\mu}} X^{\absidx{\nu}}
  Z^{\absidx{\kappa}} Z^{\absidx{\lambda}}\;
  \stcur_{\absidx{\mu\kappa\nu\lambda}}(y) + \dots \period
\end{equation}
Clearly, the expansion of the world function at one of its 
arguments is given by equation \eqref{wfofgeod}.

Similarly, it is possible to derive 
(see \cite{Christensen:1976,Krtous:PointSplitCalc}) 
that coincidence limits of derivatives of 
the Van-Vleck Morette determinant are
\begin{gather}
  \stCL{\stvvmd} = 1\commae\\
  \stCL{\grad_\argl\stvvmd} = \stCL{\grad_\argr\stvvmd} = 0\commae\\
  \stCL{\stcnx_{\argl\absidx{\mu}}
    \stcnx_{\argl\absidx{\nu}}\stvvmd} =
  - \stCL{\stcnx_{\argl\absidx{\mu}}
    \stcnx_{\argr\absidx{\nu}}\stvvmd} =
  \stCL{\stcnx_{\argr\absidx{\mu}}
    \stcnx_{\argr\absidx{\nu}}\stvvmd} =
  \frac13 \stric_{\absidx{\mu\nu}}\period
\end{gather}
and the covariant expansion 
\begin{equation}
  \stvvmd(y;X|y;Z) = 1 +
  \frac16\: (X-Z) \stctr \stric \stctr (X-Z) + \dots \period
\end{equation}

Finally let us note that the Jacobian associated with a map
\begin{equation}
  \stgeod_x^{\dash1} : z \rightarrow Z = - \stwfl(x|z)
\end{equation}
is given by
\begin{equation}\label{vvmdetisjac}
  \abs{\Det \Dmap \stgeod_x^{\dash1}(z)} = 
  \abs{\Det \bigl(\stmtrc^{\dash1}(x)\stctr\stwflr(x|z)\bigr)} =
  \mtrcdet{\stmtrc}^{\dash1}(x)\; \stwfdet(x|z) =
  \sttgtdet(x|z)\; \stvvmd(x|z) \period
\end{equation}

\subsection{$\boldsymbol{(\stdim-1)+1}$ splitting near a boundary}

Now we turn to investigate the domain $\Omega$ with a boundary. 
We will study this situation locally --- i.e. we will 
work on a neighborhood of the boundary with topology 
$\realn\times \Sigma$ where $\Sigma$ is part of boundary manifold. 
In such neighborhood we can perform a $(\stdim-1)+1$ 
splitting which is discussed for example in \cite{MTW}.
It is given by a \defterm{time function $t$} and 
\defterm{time flow vector $\vec{t}$} 
such that $\vec{t}\stctr\grad t = 1$. We use notation of usual 
$3+1$ splitting of spacetime even if we do not necessarily assume 
that $t$ plays the role of a time coordinate. We assume that 
the condition $t=0$ defines the boundary and that $t>0$ inside 
of the domain $\Omega$. We denote $\Sigma_t$ hypersurfaces defined 
by conditions $t = \text{const}$. We denote $\normf$ and $\normv$ 
inside oriented normalized normal form and vector, $\smtrc$ 
orthogonal projection of the metric $\stmtrc$ on the hypersurfaces 
$\Sigma_t$, and $\sdelta$ orthogonal projector 
to hypersurfaces $\Sigma_t$. I.e.
\begin{equation}
  \stmtrc = \stsign^2\, \normf \normf + \smtrc \comma
  \stmtrc^{\dash1} = \stsign^{\dash2}\, \normv \normv + \smtrc^{\dash1} \comma 
  \delta = \normv \normf + \sdelta \commae
\end{equation}
where $\smtrc^{\dash1}$ is inverse of $\stmtrc$ in the tangent 
space to the hypersurfaces. The phase factor $\stsign$ governs 
the signature of the metric $\stmtrc$ and the character of 
the hypersurfaces. We will use shorthands
\begin{equation}
  A^{\dots\bpar{\absidx{\alpha}}\dots\bper\dots}
   _{\dots\bper\dots} =
  A^{\dots{\absidx{\beta}}\dots{\absidx{\mu}}\dots}
   _{\dots{\absidx{\nu}}\dots}\; 
  \sdelta^{\absidx{\alpha}}_{{\absidx{\beta}}}\;
  \normv^{\absidx{\nu}}\,\normf_{{\absidx{\mu}}}\period
\end{equation}
We also use
\begin{equation}
  \stwfll_\bpar \defeq \stwfll_{\bpar\,\bpar}\comma
  \stwflr_\bpar \defeq \stwflr_{\bpar\,\bpar}\comma
  \stwfrr_\bpar \defeq \stwfrr_{\bpar\,\bpar}\period
\end{equation}

Decomposition of the time flow vector $\vec{t}$ 
defines \defterm{lapse $N$} and \defterm{shift $\vec{N}$}
\begin{equation}
  \vec{t} = N \normv + \vec{N} \comma \grad t = N \normf \period
\end{equation}

We denote $\sgrad$ the \defterm{hypersurface gradient} --- 
an orthogonal projection of a spacetime gradient to 
the hypersurfaces $\Sigma_t$ 
\begin{equation}
  \sgrad f = \sdelta \stctr \grad f \commae
\end{equation}
and $\scnx$ the \defterm{hypersurface connection} of 
the metric $\smtrc$. It is related to the spacetime connection as
\begin{equation}
  \scnx A = \sdelta^\diffind \stcnx A 
  \qquad\text{for $A$ such that}\qquad
  A = \sdelta^\diffind A \period
\end{equation}
where by $\sdelta^\diffind A$ we mean projection of 
all tensor indices to the spaces tangent to the boundary. 
We denote by $\scur$, $\sric$, $\sscur$, and $\slaplace$ 
the Riemann curvature tensor, Ricci tensor, scalar curvature 
and Laplace operator of the metric $\smtrc$.

The \defterm{extrinsic curvature $\extcur$} is given 
by covariant derivative of the normal form
\begin{equation}\label{extcurDef}
  \extcur = \sdelta \stctr \stcnx \normf \commae
\end{equation}
and we use shorthands
\begin{equation}\label{extScQcur}
\begin{gathered}
  \extscur=
  \extcur_{\absidx{\mu\nu}} \stmtrc^{\dash1\,\mu\nu}\commae\\
  \extcur^2 =
  \extcur_{\absidx{\kappa\mu}} \extcur_{\absidx{\lambda\nu}}
  \stmtrc^{\dash1\,\kappa\lambda}  \stmtrc^{\dash1\,\mu\nu}\period
\end{gathered}\end{equation}

We define the \defterm{time derivative} of 
a tensor field $A$ tangent to the hypersurfaces:
\begin{equation}
  A^\tdot = \sdelta^\diffind \lieder_{\vec{t}} A
  \qquad\text{for $A$ such that}\qquad
  A = \sdelta^\diffind A \period
\end{equation}

Now we list a number of useful relations between 
spacetime quantities and \vague{space} quantities, 
derivations of which are straightforward and for 
the case $\stsign^2 = -1$ can be mostly found, 
for example, in \cite{MTW}.
\begin{gather}
\begin{gathered}
  \normv \stctr \stcnx \normf = - \sgrad \ln N \commae\\
  \stcnx\stctr\normv = \stsign^2 \extscur \commae
\end{gathered}\displaybreak[0]\\
\begin{gathered}
  \normv\stctr \stcnx\smtrc = \stsign \Bigl(
  \normf\, (\sgrad \ln N) + (\sgrad\ln N)\,\normf \Bigr) \commae\\
  \sdelta^{\absidx{\mu}}_{{\absidx{\gamma}}}\,
  \stcnx_{\absidx{\mu}} \smtrc_{\absidx{\alpha\beta}} =
  - \normf_{\absidx{\alpha}} \extcur_{\absidx{\beta\gamma}}
  - \normf_{\absidx{\beta}} \extcur_{\absidx{\alpha\gamma}} \commae
\end{gathered}\displaybreak[0]\\
\begin{gathered}
  \stcnx\stctr\sdelta = - \stsign^2 \normf \extscur +
  \sgrad\ln N \commae\\
  \sdelta \stctr ( \stcnx \sdelta ) \stctr \sdelta =
  - \extcur \,\normv
\end{gathered}\displaybreak[0]\\
\begin{gathered}
  \smtrc^\tdot = 2 \stsign^2 N \extcur +
  \lieder_{\vec{N}} \smtrc \commae\\
  \extcur^\tdot = N\,
  \bigl(\normv \stctr \stcnx
  \extcur\bigr)_{\bpar\,\bpar} +
  2 \stsign^2 N \,
  \extcur \stctr \smtrc^{\dash1} \stctr \extcur +
  \lieder_{\vec{N}} \extcur\commae\\
  \extscur^\tdot = N\, \normv\stctr\grad \extscur +
  \vec{N} \extscur \period
\end{gathered}
\end{gather}
The curvature tensors of the spacetime metric $\stmtrc$ 
and of the space metric $\smtrc$ are related by
\begin{gather}
\begin{gathered}
\begin{aligned}
  \stcur_{\bpar{\absidx{\alpha}}\,\bpar{\absidx{\beta}}\,
          \bpar{\absidx{\gamma}}\,\bpar{\absidx{\delta}}} &=
    \scur_{\absidx{\alpha\beta\gamma\delta}} +
    \stsign^2 \bigl(
    \extcur_{\absidx{\alpha\delta}}
    \extcur_{\absidx{\beta\gamma}}
    - \extcur_{\absidx{\alpha\gamma}}
    \extcur_{\absidx{\beta\delta}}\bigr) \commae\\
  \stcur_{\bpar{\absidx{\alpha}}\bpar{\absidx{\beta}}
          {\absidx{\gamma}}\bper} &=
    \stsign^2 \bigl(
    \scnx_{\absidx{\alpha}}
    \extcur_{\absidx{\beta\gamma}}
    - \scnx_{\absidx{\beta}}
    \extcur_{\absidx{\alpha\gamma}}\bigr) \commae
  \end{aligned}\\
  \begin{aligned}
  \stcur_{\bpar\bper\bpar\bper} &=
    \stsign^4 \bigl(
    \extcur \extscur
    - \extcur \stctr \smtrc^{\dash1} \stctr \extcur \bigr)
    - \stsign^2 \Bigl(
    \scnx\scnx \ln N + (\sgrad\ln N) \, (\sgrad\ln N) +
    \bigl(\stcnx\stctr(\normv\extcur)\bigr)_{\bpar\,\bpar}\Bigr) = \\
    &= \stsign^4 
    \extcur\stctr \smtrc^{\dash1} \stctr  \extcur
    - \stsign^2 \Bigl(
    \scnx\scnx \ln N
    + (\sgrad\ln N) \, (\sgrad\ln N) +
    \frac1N \extcur^\tdot -
    \frac1N \lieder_{\vec{N}} \extcur
    \Bigr) \commae 
\end{aligned}\end{gathered}\displaybreak[0]\\
\begin{aligned}
  \stric_{\bpar\,\bpar}&=
    \sric - \scnx\scnx \ln N - (\sgrad\ln N) \, (\sgrad\ln N)
    - \bigl(\stcnx\stctr(\normv
    \extcur)\bigr)_{\bpar\,\bpar} \commae\\
  \stric_{\bpar\bper} &=
    \stsign^2 \bigl(\scnx\stctr\smtrc^{\dash1}\stctr\extcur
    -\sgrad k \bigr) \commae\\
  \stric_{\bper\bper} &=
    \stsign^4(\extscur^2 - \extcur^2) -
    \stsign^2 \stcnx \stctr \bigl(\normv \,\extscur +
    \smtrc^{\dash1}\stctr(\sgrad\ln N) \bigr) = \\
    &= - \stsign^4 \extcur^2 -
    \stsign^2 \Bigl(\slaplace \ln N
    + (\sgrad\ln N) \stctr\smtrc^{\dash1}\stctr (\sgrad\ln N)
    + \normv\stctr\grad \extscur \Bigr)
\end{aligned}\displaybreak[0]\\
\begin{aligned}
  \stscur &= \sscur + \stsign^2(\extscur^2 - \extcur^2) -
    2 \stcnx \stctr \bigl(\normv\, \extscur +
    \smtrc^{\dash1}\stctr(\sgrad\ln N) \bigr) =\\
  &= \sscur - \stsign^2(\extscur^2 + \extcur^2) -
  2 \Bigl(\normv\stctr\grad \extscur + \slaplace\ln N
    + (\sgrad\ln N) \stctr\smtrc^{\dash1}\stctr (\sgrad\ln N )\Bigr)
\end{aligned}
\end{gather}

\subsection{Geodesic theory near a boundary}

We can develop geodesic theory on a hypersurface $\Sigma$ 
similarly to what we did for the spacetime $\stmfld$. 
On the boundary we denote the exponential map $\sgeod_x$, 
the tensor of geodesic transform and its determinant $\stgt$ 
and $\stgtdet$, the world function, its derivatives 
and its determinant $\swf$, $\swfl$, $\swfr$, $\swfll$, 
$\swflr$,  $\swfrr$ and $\swfdet$, and 
Van-Vleck Morette determinant $\svvmd$. Finally, 
we denote $\sCL{.}$ the coincidence limit on the boundary.

In the neighborhood of a part of the boundary $\Sigma$ 
of the domain $\Omega$ in which geodesics orthogonal to 
the boundary do not cross we can also define the map $\wmap$
\begin{equation}\label{wmapDef}
\begin{gathered}
  \wmap : \Sigma \times \realn \rightarrow \stmfld \commae\\
  \wmap(x,\eta) \quad\text{is geodesic} \comma
  \wmap(x,0) = x \comma
  \wmap^\tdot(x,0) = \normv \period
\end{gathered}\end{equation}
It maps point $x$ on the boundary \vague{orthogonally} 
to the domain $\Omega$ by the distance $\eta$. 
We denote $\Sigma_\eta$ the hypersurface which we obtain 
by shifting $\Sigma = \Sigma_0$ by the distance $\eta$. 
We also use the notation 
\begin{equation}\begin{aligned}
  \wmap_\eta :&\; \Sigma \rightarrow \Sigma_\eta \comma
  \wmap_\eta(x) = \wmap(x,\eta) \commae\\ 
  \wmap_{\xi,\zeta} :&\; \Sigma_\xi \rightarrow \Sigma_\zeta \comma
  \wmap_{\xi,\zeta} = \wmap_\xi(\wmap_\zeta^{-1}) \period 
\end{aligned}\end{equation}
This foliation is special case of the foliation discussed above. 
We obtain it for the choice of lapse and shift $N = 1$ and $\vec{N} = 0$.

For a tensor field $A(x)$ on spacetime we denote by $A(\xonb,\xi)$ 
its dependence on $\xonb$ and $\xi$, and $\onbnd A(\xonb,\xi)$ 
the tensor field on the boundary manifold obtained by transformation of 
tensor indices of $A(\xonb,\xi)$ to the boundary tangent bundle 
using $\wmap_{-\xi}^\diffind$
\begin{equation}\label{wSplittingConv}
  A(\xonb,\xi) = A\bigl(\wmap(\xonb,\xi)\bigr)\comma
  \onbnd A(\xonb,\xi) = \wmap_{-\xi}^\diffind
  A\bigl(\wmap_\xi(\xonb)\bigr) \commae
\end{equation}
where $\wmap_{\xi}^\diffind$ is the induced transformation on 
tangent bundles. For a bi-tensor $A(x|z)$ we mean by the 
\defterm{boundary coincidence limit} 
\begin{equation}
  \sCL{A}(\yonb;\xi,\zeta) = A(\yonb,\xi|\yonb,\zeta)\period
\end{equation}

Specially, we have a metric $\obmtrc(\yonb,\eta)$ 
(generally different from $\smtrc(\yonb)$) on 
the boundary manifold, volume element 
$\obvol(\yonb,\eta)$, and associated connection $\obcnx$. 
It is related to the connection $\scnx$ by 
\begin{equation}\label{sobcnxrel}
  \obcnx = \scnx \oplus \sobChS\period
\end{equation}
The relation of corresponding curvature tensors is (see e.g. \cite{MTW})
\begin{equation}\begin{aligned}
  \onbndcr[0.7ex]\scur{}_{\absidx{\gamma\alpha}}
    {}^{\absidx{\delta}}{}_{\absidx{\beta}} &=
  \scur{}_{\absidx{\gamma\alpha}}
    {}^{\absidx{\delta}}{}_{\absidx{\beta}}
  + \obcnx_{\absidx{\gamma}}
    \sobChS_{\absidx{\alpha\beta}}^{\absidx{\delta}} 
  - \obcnx_{\absidx{\alpha}}
    \sobChS_{\absidx{\gamma\beta}}^{\absidx{\delta}} 
  + \sobChS_{\absidx{\alpha\mu}}^{\absidx{\delta}}
    \sobChS_{\absidx{\gamma\beta}}^{\absidx{\mu}} 
  - \sobChS_{\absidx{\gamma\mu}}^{\absidx{\delta}}
    \sobChS_{\absidx{\alpha\beta}}^{\absidx{\mu}} \commae\\
  \onbndcr[0.7ex]\sric_{\absidx{\alpha\beta}} &=
  \sric_{\absidx{\alpha\beta}}
  + \obcnx_{\absidx{\mu}}
    \sobChS_{\absidx{\alpha\beta}}^{\absidx{\mu}} 
  - \obcnx_{\absidx{\alpha}}
    \sobChS_{\absidx{\beta\mu}}^{\absidx{\mu}} 
  + \sobChS_{\absidx{\alpha\mu}}^{\absidx{\nu}}
    \sobChS_{\absidx{\beta\nu}}^{\absidx{\mu}} 
  - \sobChS_{\absidx{\mu\nu}}^{\absidx{\nu}}
    \sobChS_{\absidx{\alpha\beta}}^{\absidx{\mu}} \period
\end{aligned}\end{equation}

From the definition of the map $\wmap$ we have
\begin{gather}
  \sCL{\stwfl}(\yonb;\xi,\zeta) = 
  (\xi-\zeta)\; \normv(\yonb,\xi) \commae\\
  \sCL{\sgrad_\argl \stwf} = 0 \period\label{orthggeod}
\end{gather}
Differentiating this equation we obtain 
the differential map $\Dmap\wmap$
\begin{equation}
\begin{gathered}
  \Dmap\wmap_{\xi,\zeta}(x) :
  \tens_x\,\Sigma_\xi \rightarrow \tens_z\,\Sigma_\zeta 
  \comma z = \wmap_{\xi,\zeta}(x) \commae\\
  \Dmap^{\absidx{\nu}}_{\absidx{\mu}}\wmap_{\xi,\zeta}(x) =
  - \bigl(\scnx_{\argl\absidx{\mu}}\scnx_{\argl\absidx{\kappa}}
  \stwf\bigr)(x|z)\; \stwflr_\bpar^{\dash1\,\absidx{\kappa\nu}}(x|z) \period
\end{gathered}
\end{equation}
In the special case $\xi = 0$ we get
\begin{equation}\label{Dwmap}
  \Dmap^{\absidx{\nu}}_{\absidx{\mu}}\wmap_{\eta}(\yonb) =
  - \bigl(\eta\stsign^2 \extcur_{\absidx{\mu\kappa}}(\yonb) +
  \stwfll_{\bpar\,{\absidx{\mu\kappa}}}(\yonb|\yonb,\eta)\bigr)\;
  \stwflr_\bpar^{\dash1\,\absidx{\kappa\nu}}(\yonb|\yonb,\eta) \period
\end{equation}
Here $\stwflr_\bpar^{\dash1}$ is the inverse of $\stwflr_\bpar$ 
in spaces tangent to hypersurfaces $\Sigma_\eta$. 
Because $\Dmap\wmap_{\xi,\zeta} = \Dmap\wmap^{\dash1}_{\zeta,\xi}$ we have 
\begin{equation}
  \sCL{\stwflr_{\bpar\,\absidx{\mu\nu}}} =
  \sCL{\bigl(\scnx_{\argl\absidx{\mu}}
  \scnx_{\argl\absidx{\kappa}} \stwf\bigr) 
  \bigl(\scnx_{\argr\absidx{\nu}}
  \scnx_{\argr\absidx{\lambda}} \stwf\bigr)
  \stwflr_\bpar^{\dash1\,\absidx{\kappa\lambda}}} \period
\end{equation}

Using this relation, the definition of 
the Van-Vleck Morette determinant and 
\begin{equation}
  \normv(x)\stctr\stwflr(x|z) =
  - \normv(z)\stctr\stmtrc(z) \qquad\text{for}\qquad
  z = \wmap_{\xi,\zeta}(x)\commae
\end{equation}
we get an expression for the Jacobian 
associated with the map $\wmap_{\xi,\zeta}$,
\begin{equation}\label{wSplittingGenJac}
\begin{split}
  \onbnd\jac(\yonb;\xi,\zeta) &=
  \abs{\Det_\bpar\Dmap\wmap_{\xi,\zeta}}(\yonb,\xi) = \\
  &= \sCLb{\stvvmd^{\dash1} \; \mtrcdet{\smtrc}^{\dash1}
  \bigl(\Det_\bpar \scnx_\argl \scnx_\argl \stwf\bigr)}(\yonb;\xi,\zeta) =
  \sCLb{\stvvmd \; \mtrcdet{\smtrc}
  \bigl(\Det_\bpar \scnx_\argr
  \scnx_\argr \stwf\bigr)^{\dash1}}(\yonb;\xi,\zeta) \period
\end{split}
\end{equation}
As special cases we have
\begin{equation}\label{wSplittingJac}
  \onbnd\jac(\yonb,\eta) = \onbnd\jac(\yonb;0,\eta) \comma
  \onbnd\jac(\yonb;\xi,\zeta) =
  \onbnd\jac{}^{\dash1}(\yonb,\xi)
  \onbnd\jac(\yonb,\zeta) \period
\end{equation}
This also gives the expression for 
the Van-Vleck Morette determinant, 
\begin{equation}
  \stvvmd(\yonb,\xi|\yonb,\zeta) = 
  \mtrcdet{\smtrc}^{\dash\frac12}(\yonb,\xi)\,
  \mtrcdet{\smtrc}^{\dash\frac12}(\yonb,\zeta)\;
  \Bigl( 
  \bigl(\Det_\bpar \scnx_\argl \scnx_\argl \stwf\bigr)\,
  \bigl(\Det_\bpar \scnx_\argr \scnx_\argr \stwf\bigr)
  \Bigr)^{\frac12}\!\!(\yonb,\xi|\yonb,\zeta) \period
\end{equation}

Finally we can prove that
\begin{equation}\begin{split}
  &\onbnd\jac(\yonb,\xi)\, \onbnd\stvvmd(\yonb,\xi|\yonb,\zeta)\,
  \onbnd\jac(\yonb,\zeta) =\\
  &\quad\quad=\mtrcdet{\smtrc}^{\dash1}\!(\yonb)\;
  \bigl(\Det_\bpar \scnx_\argl
  \scnx_\argl\stwfonb\bigr)\!(\yonb,\xi|\yonb,\zeta)
  = \mtrcdet{\smtrc}^{\dash1}\!(\yonb)\;
  \bigl(\Det_\bpar \scnx_\argr
  \scnx_\argr \stwfonb\bigr)\!(\yonb,\xi|\yonb,\zeta)
   \period
\end{split}\end{equation}
Thanks to \eqref{orthggeod}, \eqref{wSplittingGenJac}, 
and \eqref{wSplittingJac}, the function $\tilde j$ 
defined in \eqref{tildejacDef} is
\begin{equation}\label{tildejacProof}
\begin{split}
  \tilde\jac(\yonb;\xi,\zeta)
  &=
  \onbnd\jac(\yonb,\xi)\, \onbnd\stvvmd(\yonb,\xi|\yonb,\zeta)\;
  \onbnd\jac(\yonb,\zeta)\;
  \svol(\yonb)\;
  \Bigl(\Det_\bpar \scnx_\argr
  \scnx_\argr\stwfonb\Bigr)^{\dash\frac12}\!(\yonb,\xi|\yonb,\zeta)
  = \\  &= 
  \onbnd\jac(\yonb,\xi)\, \onbnd\stvvmd(\yonb,\xi|\yonb,\zeta)\;
  \obvol(\yonb,\zeta)\;
  \Bigl(\Det_\bpar \obcnx_\argr
  \obcnx_\argr\stwfonb\Bigr)^{\dash\frac12}\!(\yonb,\xi|\yonb,\zeta)
  = \\ &= 
  \jac(x) \stvvmd(x|z) \;\svol(z) \;
  \bigl(\Det_\bpar \scnx_\argr
  \scnx_\argr\stwf\bigr)^{\dash\frac12}(x|z) =
  \jac(x)\: \stvvmd^{\!\frac12}(x|z)\:
  \onbnd\jac(\yonb;\xi,\zeta)^{\frac12} 
  = \\ &=
  \Bigl(\onbnd\jac(\yonb,\xi)\,
  \onbnd\stvvmd(\yonb,\xi|\yonb,\zeta)\,
  \onbnd\jac(\yonb,\zeta)\Bigr)^{\frac12}\period
\end{split}
\end{equation}
The equation previous to this one is a straightforward consequence.

\subsection{Reflection on the boundary}

In section \ref{sc:SOH-partwbnd} we have worked with 
the \defterm{geodesic reflected on the boundary}. 
We recall its definition here and list some useful 
properties which allows us to prove the relation 
\eqref{tildajacbndrel}.

We will study the geodesic $\extr{\traj{x}}_\bnd(x|z)$ 
between points $x$ and $z$ which is reflecting on 
the boundary at point $\rflpt(x|z)$ --- an extreme 
trajectory of the functional given by the half of 
squared length, with the condition that it has to 
touch the boundary. We use the convention, that for 
any quantity depending on two spacetime points 
$f(x|z)$ we denote
\begin{equation}\label{lrIdxConvention}
  f_\bdl(x|z) = f(x|\rflpt(x|z)) \comma 
  f_\bdr(x|z) = f(\rflpt(x|z)|z) \period
\end{equation}
If we denote the parameter at which the geodesic 
reflects on the boundary  $\rfltmr(x|z)$ and its 
complement $\rfltml(x|z)$ 
\begin{equation}\begin{gathered}
  \rflpt(x|z) = \extr{\traj{x}}_\bnd(x|z)|_{\rfltmr(x|z)}
  \in \bound\Omega\commae\\
  1 = \rfltml(x|z) + \rfltmr(x|z)\period
\end{gathered}\end{equation}
we can write the reflected geodesic as 
joining of two geodesics 
\begin{equation}
  [\tau,\extr{\traj{x}}_\bnd] =
  [\rfltml\tau,\extr{\traj{x}}_\bdl] \join
  [\rfltmr\tau,\extr{\traj{x}}_\bdr]\period 
\end{equation}
The extremum conditions on position of 
the reflection point and reflection parameter are
\begin{equation}
  \frac{(\sgrad\stwf)_\bdl}{\rfltml} +
  \frac{(\sgrad\stwf)_\bdr}{\rfltmr} = 0 \comma
  \frac{\stwf_\bdl}{\rfltml^2} = \frac{\stwf_\bdr}{\rfltmr^2}\commae
\end{equation}
where $\sgrad$ acts in the argument on the boundary.

We define \defterm{reflection world function $\stwf_\bnd$} as
\begin{equation}
  \stwf_\bnd = \frac{\stwf_\bdl}{\rfltml} + \frac{\stwf_\bdr}{\rfltmr} =
  \frac{\stwf_\bdl}{\rfltml^2} = \frac{\stwf_\bdr}{\rfltmr^2}\period
\end{equation}
Clearly
\begin{equation}
 \begin{gathered}
 \rfltml = \sqrt{\frac{\stwf_\bdl}{\stwf_\bnd}} \comma
  \rfltmr = \sqrt{\frac{\stwf_\bdr}{\stwf_\bnd}} \commae\\
  \sqrt{\stwf_\bnd} = \sqrt{\stwf_\bdl} + \sqrt{\stwf_\bdr} \comma
  0 = (\sgrad\sqrt{\stwf})_\bdl + (\sgrad\sqrt{\stwf})_\bdr \comma
\end{gathered}
\end{equation}
Using the last equation we can get
\begin{equation}
  \grad_\argl\sqrt{\stwf_\bnd} = (\grad_\argl \sqrt\stwf)_\bdl \comma
  \grad_\argr\sqrt{\stwf_\bnd} = (\grad_\argr \sqrt\stwf)_\bdr \period
\end{equation}

Similarly to the case without boundary we define
\begin{gather}
  \stgdist_\bnd = \abs{\sqrt{2\stwf_\bnd}} 
  = \stgdist_\bdl + \stgdist_\bdr\\
  \stwfl_\bnd = \stmtrc^{\dash1}\stctr\grad_\argl\stwf_\bnd
  = \frac{\stwfl_\bdl}{\rfltml}\comma
  \stwfr_\bnd = \grad_\argr\stwf_\bnd \stctr\stmtrc^{\dash1}
  = \frac{\stwfr_\bdr}{\rfltmr}\commae\\
  \stwfll_\bnd = \stcnx_\argl\stcnx_\argl\stwf_\bnd \comma
  \stwflr_\bnd = \grad_\argl\grad_\argr\stwf_\bnd \comma
  \stwfrr_\bnd = \stcnx_\argr\stcnx_\argr\stwf_\bnd \period
\end{gather}
Additionally we define
\begin{equation}
  \stgdist_\bper 
  = - \frac1\rfltml\: \normv(\rflpt)\stctr(\grad_\argr\stwf)_\bdl
  = - \frac1\rfltmr\: \normv(\rflpt)\stctr(\grad_\argl\stwf)_\bdr 
  = \normv \stctr (\grad_\argl\stwf_\bnd)
  = \normv \stctr (\grad_\argr\stwf_\bnd) \commae
\end{equation}
and we denote differentials of maps $x \rightarrow \rflpt(x|z)$ 
and $z \rightarrow \rflpt(x|z)$ as
\begin{equation}
  \rflptl = \Dmap_\argl\rflpt \comma 
  \rflptr = \Dmap_\argr\rflpt \comma 
\end{equation}
i.e. if we displace points $x$ and $z$ in 
directions $X$ and $Z$, the reflection point 
moves in direction $X\stctr\rflptl(x|z) + \rflptr(x|z)\stctr Z$.
Finally we define the \defterm{reflection 
Van-Vleck Morette determinant $\stvvmd_\bnd$}
\begin{equation}\label{stvvmdBndDef}
  \stvvmd_\bnd = \abs{\Det \stwflr_\bnd} \;\sttgtddet^{\dash1} \period 
\end{equation}

Some long algebra gives
\begin{equation}\begin{aligned}
  \stwfll_\bnd &= - \rflptl \stctr \rflB \stctr \rflptl -
  \frac1{2\stwf_\bnd} \frac{\rfltmr}{\rfltml}
  (\grad_\argl\stwf_\bnd)(\grad_\argl\stwf_\bnd)
  + \frac1{\rfltml} \stwfll_\bdl\commae\\
  \stwflr_\bnd &= - \rflptl \stctr \rflB \stctr \rflptr +
  \frac1{2\stwf_\bnd}
  (\grad_\argl\stwf_\bnd)(\grad_\argr\stwf_\bnd)\commae\\
  \stwfrr_\bnd &= - \rflptr \stctr \rflB \stctr \rflptr -
  \frac1{2\stwf_\bnd} \frac{\rfltml}{\rfltmr}
  (\grad_\argr\stwf_\bnd)(\grad_\argr\stwf_\bnd)
  + \frac1{\rfltmr} \stwfrr_\bdl\commae
\end{aligned}\end{equation}
where
\begin{equation}
  \rflB = 2\sqrt{\stwf_\bnd}\,
  \Bigl(\stcnx_\argr\stcnx_\argr\sqrt\stwf)_\bdl +
  \stcnx_\argl\stcnx_\argl\sqrt\stwf)_\bdr\Bigr) =
  \frac{\stwfrr_\bpar\,\bdl}{\rfltml} +
  \frac{\stwfll_\bpar\,\bdr}{\rfltmr} +
  \frac12 \frac{\stwf_\bnd}{\stwf_\bdl\stwf_\bdr}
  (\sgrad_\argl\stwf)_\bdr (\sgrad_\argr\stwf)_\bdl
  + 2 \stgdist_\bper \extcur(\rflpt) \period
\end{equation}
Using these relations, a more intricate 
calculation gives the space coincidence limits
\begin{gather}
  \sCLb{\stwflr_{\bnd\bpar}} = 
  - \sCLb{\rflptl\stctr\rflB\stctr\rflptr}\commae\label{sCLofbBb}\\
  \sCLb{\frac{\stwflr_{\bpar\,\bdl}}{\rfltml}} =
  - \sCLb{\rflptl\stctr\rflB} \comma
  \sCLb{\frac{\stwflr_{\bpar\,\bdr}}{\rfltmr}} =
  - \sCLb{\rflB\stctr\rflptr} \commae \label{sCLofbBandBb}\\
\begin{split}
  &\sCLb{(\scnx_\argl\scnx_\argl\stwf_\bnd)^{\dash1} }=
  \sCLb{\stwflr_{\bnd\bpar}^{\dash1}\stctr
  (\scnx_\argr\scnx_\argr\stwf_\bnd)^{\dash1} 
  \stctr \stwflr_{\bnd\bpar}^{\dash1}} =\\
  &\qquad =\sCLB{\stwflr_{\bpar\,\bdl}^{\dash1}\stctr
  (\scnx_\argr\scnx_\argr\stwf)_\bdl \stctr
  \Bigl( \rfltml (\scnx_\argr\scnx_\argr\stwf)_\bdl^{\dash1} +
  \rfltmr (\scnx_\argl\scnx_\argl\stwf)_\bdr^{\dash1} \Bigr)
  \stctr (\scnx_\argr\scnx_\argr\stwf)_\bdl
  \stctr \stwflr_{\bpar\,\bdl}^{\dash1}}\commae
\end{split}\\
  \sCLb{(\scnx_\argl\scnx_\argl\stwf_\bnd)^{\dash1} \stctr
  \stwflr_{\bnd\bpar}} =
  \sCLb{\stwflr_{\bnd\bpar}^{\dash1} \stctr
  (\scnx_\argr\scnx_\argr\stwf_\bnd)} \period
\end{gather}
Here inverses are taken in the spaces tangent 
to the boundary. Taking the determinant 
of the last equation, we find
\begin{equation}\begin{gathered}
  \sCLb{(\Det_\bpar \stwflr_{\bnd\bpar})^2} =
  \sCLb{(\Det_\bpar\scnx_\argl\scnx_\argl\stwf_\bnd)\,
  (\Det_\bpar\scnx_\argr\scnx_\argr\stwf_\bnd)} \commae\\
  \stvvmd_\bnd(\yonb,\xi|\yonb,\zeta) = \Bigl(
  \mtrcdet{\smtrc}^{\dash\frac12}(\yonb,\xi)\,
  \mtrcdet{\smtrc}^{\dash\frac12}(\yonb,\zeta)\;
  (\Det_\bpar\scnx_\argl\scnx_\argl\stwf_\bnd)\,
  (\Det_\bpar\scnx_\argr\scnx_\argr\stwf_\bnd)
  \Bigr)^{\frac12}\!(\yonb,\xi|\yonb,\zeta)\period
\end{gathered}\end{equation}
and
\begin{equation}
\begin{split}
  &\frac{\jac^2(x)}{\mtrcdet{\smtrc}(x)}\:
  \bigl(\Det_\bpar \scnx_\argl\scnx_\argl\stwf_\bnd\bigr)(x|z) =
  \frac{\jac^2(z)}{\mtrcdet{\smtrc}(z)}\:
  \bigl(\Det_\bpar \scnx_\argr\scnx_\argr\stwf_\bnd\bigr)(x|z) = \\
  &\qquad = \frac{\jac^2(\yonb)}{\mtrcdet{\smtrc}(\yonb)}\:
  \Bigl( \rfltml (\scnx_\argr\scnx_\argr\stwf)_\bdl^{\dash1} +
  \rfltmr (\scnx_\argl\scnx_\argl\stwf)_\bdr^{\dash1} \Bigr)(x|z) 
\end{split}
\end{equation}
for $x = \wmap(\yonb,\xi)$ and $z = \wmap(\yonb,\zeta)$. 
Putting these relation together we obtain
\begin{equation}\label{tildejacBndProof}
  \jac(x) \, \stvvmd_\bnd(x|z) \jac(z) = 
  \frac{\jac^2(x)}{\mtrcdet{\smtrc}(x)}\:
  \bigl(\Det_\bpar \scnx_\argl\scnx_\argl\stwf_\bnd\bigr)(x|z) =
  \frac{\jac^2(z)}{\mtrcdet{\smtrc}(z)}\:
  \bigl(\Det_\bpar \scnx_\argr\scnx_\argr\stwf_\bnd\bigr)(x|z)
\end{equation}
for $x = \wmap(\yonb,\xi)$ and $z = \wmap(\yonb,\zeta)$. 
The equality in \eqref{tildajacbndrel} is 
a straightforward consequence of this relation.

\subsection{Covariant expansions near boundary}

Finally we will write down coefficients in 
covariant expansions \eqref{wfbndcovexp} and 
\eqref{lbndcovexp} of the world function $\stwf$ 
and function $l$ defined in \eqref{lDef}. 
These are expansions \emph{inside} of the boundary 
manifold of $\wmap$-mapped functions 
$\stwfonb(\xonb,\xi|\zonb,\zeta)$ and 
$\onbnd l(\xonb,\xi|\zonb,\zeta)$ around point 
$\xonb$. The derivation is long and technical. 
It uses the general method discussed above and a 
transformation of the connection $\scnx$ to 
the connection $\obcnx$. Fortunately, we need only 
the spacetime coincidence limit of the coefficients 
(i.e. $\stwfonb_{k,l}(\yonb;\eta,\eta)$), which 
simplifies the calculations significantly. 
But even then the calculations is too long and 
uninteresting to be included it here. We list 
only the results. See also 
\cite{McAvityOsborn:1990,McAvityOsborn:1991,McAvityOsborn:1992} 
for similar calculations.

The coefficients of the boundary covariant 
expansion of the spacetime world function 
$\stwfonb$ at some general point $\yonb$ 
(slight generalization of eq. \eqref{wfbndcovexp}) are
\begin{gather}
  \stwfonb_{0,0}(\yonb;\xi,\zeta) = \frac12 \stsign^2\,(\xi-\zeta)^2  
  \commae\label{bndcovexpwfzero}\displaybreak[0]\\
  \stwfonb_{0,1}(\yonb;\xi,\zeta) = \stwfonb_{1,0}(\yonb;\xi,\zeta) = 0
  \commae\label{bndcovexpwfone}\displaybreak[0]\\
  \stCL{\stwfonb_{2,0}} = - \stCL{\stwfonb_{1,1}} =
  \stCL{\stwfonb_{0,2}} = \obmtrc
  \commae\label{bndcovexpwftwo}\displaybreak[0]\\
    \stCL{\stwfonb_{3,0\,{\absidx{\alpha\beta\gamma}}}} =
    \stCL{\stwfonb_{0,3\,{\absidx{\alpha\beta\gamma}}}} =
    3 \sobChS^{\,\:\;\absidx{\mu}}_{(\absidx{\alpha\beta}}
    \obmtrc_{\absidx{\gamma}){\absidx{\mu}}}\comma
    \stCL{\stwfonb_{2,1\,{\absidx{\alpha\beta\,\kappa}}}} =
    \stCL{\stwfonb_{1,2\,{\absidx{\kappa\,\alpha\beta}}}} =
    - \sobChS^{\,\,\absidx{\mu}}_{\absidx{\alpha\beta}}
    \obmtrc_{\absidx{\kappa\mu}}\commae
  \label{bndcovexpwfthree}\displaybreak[0]\\
  \begin{aligned}
    {}&\stCL{\stwfonb_{4,0\,{\absidx{\alpha\beta\gamma\delta}}}} =
    \stCL{\stwfonb_{0,4\,{\absidx{\alpha\beta\gamma\delta}}}} = 
    \\{}&\qquad = 
    - \stsign^2 \extcuronb_{(\absidx{\alpha\beta}}
    \extcuronb_{\absidx{\gamma\delta})}
    + 4 (\obcnx_{(\absidx{\alpha}} 
    \sobChS^{\,\,\absidx{\mu}}_{\absidx{\beta\gamma}})
    \obmtrc_{\absidx{\delta}){\absidx{\mu}}}
    + 8 \sobChS^{\;\;\absidx{\nu}}_{\absidx{\mu}(\absidx{\alpha}}
    \sobChS^{\,\,\absidx{\mu}}_{\absidx{\beta\gamma}}
    \obmtrc_{\absidx{\delta}){\absidx{\nu}}}
    + 3 \sobChS^{\,\:\;\absidx{\mu}}_{(\absidx{\alpha\beta}}
    \sobChS^{\,\,\absidx{\nu}}_{\absidx{\gamma\delta})}
    \obmtrc_{\absidx{\mu\nu}}\commae\\
    {}&\stCL{\stwfonb_{3,1\,{\absidx{\alpha\beta\gamma\,\kappa}}}} =
    \stCL{\stwfonb_{1,3\,{\absidx{\kappa\,\alpha\beta\gamma}}}} =
    \\{}&\qquad = 
    \stsign^2 \extcuronb_{(\absidx{\alpha\beta}}
    \extcuronb_{\absidx{\gamma}){\absidx{\kappa}}}
    - (\obcnx_{(\absidx{\alpha}} 
    \sobChS^{\,\,\absidx{\mu}}_{\absidx{\beta\gamma})})
    \obmtrc_{\absidx{\mu\kappa}}
    - \sobChS^{\,\:\;\absidx{\mu}}_{(\absidx{\alpha\beta}}
    \sobChS^{\;\;\absidx{\nu}}_{\absidx{\gamma}){\absidx{\mu}}}
    \obmtrc_{\absidx{\nu\kappa}}\commae\\
    {}&\stCL{\stwfonb_{2,2\,{\absidx{\alpha\beta\,\kappa\lambda}}}} =
    -\frac13 \bigl( \scur_{\absidx{\alpha\kappa\beta\lambda}}
    + \scur_{\absidx{\alpha\lambda\beta\kappa}}\bigr)
    -\\{}&\qquad  \qquad
    -\frac13 \stsign^2
    \extcuronb_{\absidx{\alpha\kappa}}
    \extcuronb_{\absidx{\beta\lambda}}
    -\frac13 \stsign^2
    \extcuronb_{\absidx{\alpha\lambda}}
    \extcuronb_{\absidx{\beta\kappa}}
    -\stsign^2
    \extcuronb_{\absidx{\alpha\beta}}
    \extcuronb_{\absidx{\kappa\lambda}}
    -\sobChS^{\,\,\absidx{\mu}}_{\absidx{\alpha\beta}}
    \sobChS^{\,\,\absidx{\nu}}_{\absidx{\kappa\lambda}}
    \obmtrc_{\absidx{\mu\nu}}\period
  \end{aligned}\label{bndcovexpwffour}
\end{gather}

The coefficients of the boundary covariant expansion 
of the function $l$ at some general point $\yonb$ 
(slight generalization of eq. \eqref{lbndcovexp}) are
\begin{gather}
  \stCL{\onbnd l_{0,0}} = 2 \ln \onbnd\jac 
  \commae\label{bndcovexplzero}\\
  \stCL{\onbnd l_{1,0\,{\absidx{\alpha}}}} =
  \stCL{\onbnd l_{0,1\,{\absidx{\alpha}}}} =
  \sobChS^{\,\,\absidx{\mu}}_{\absidx{\alpha\mu}}
  \commae\label{bndcovexplone}\displaybreak[0]\\
  \begin{aligned}
    {}&\stCL{\onbnd l_{2,0\,{\absidx{\alpha\beta}}}} =
    \stCL{\onbnd l_{0,2\,{\absidx{\alpha\beta}}}} =
    \\ {}& \qquad =
    \frac13 \Bigl(
    - \extcuronb{}^\tdot_{\absidx{\alpha\beta}}
    + 2 \stsign^2 \extcuronb_{\absidx{\alpha\mu}}
    \extcuronb_{\absidx{\beta\nu}}
    \obmtrc^{\dash1\,\absidx{\mu\nu}}
    - \stsign^2 \extcuronb_{\absidx{\alpha\beta}}
    \extscuronb 
    + \\ {}& \qquad\qquad\qquad 
    + 3 \obcnx_{(\absidx{\mu}} 
    \sobChS^{\,\,\absidx{\mu}}_{\absidx{\alpha\beta})}
    + \sobChS^{\,\,\absidx{\mu}}_{\absidx{\alpha\nu}}
    \sobChS^{\,\,\absidx{\nu}}_{\absidx{\beta\mu}}
    + 2 \sobChS^{\,\,\absidx{\mu}}_{\absidx{\alpha\beta}}
    \sobChS^{\,\,\absidx{\nu}}_{\absidx{\mu\nu}}
    \Bigr)\commae\\
  {}&\stCL{\onbnd l_{1,1\,{\absidx{\alpha\,\beta}}}} =
    \frac13 \Bigl(
    \extcuronb{}^\tdot_{\absidx{\alpha\beta}}
    - 2 \stsign^2 \extcuronb_{\absidx{\alpha\mu}}
    \extcuronb_{\absidx{\beta\nu}}
    \obmtrc^{\dash1\,\absidx{\mu\nu}}
    + \stsign^2 \extcuronb_{\absidx{\alpha\beta}}
    \extscuronb \Bigr)
  \end{aligned}\label{bndcovexpltwo}
\end{gather}

Computing normal derivatives we also get
\begin{equation}\label{dotlnjac}
\begin{gathered}
  \frac1\stsign(\ln \onbnd\jac)^\tdot = \stsign \extscuronb\commae\\
  \frac1{\stsign^2}(\ln \onbnd\jac)^{\tdot\tdot} = \extscuronb{}^\tdot\commae
\end{gathered}\end{equation}
and
\begin{equation}\label{dotvvmdcl}
\begin{gathered}
  \frac1\stsign \stCL{(\ln \onbnd\jac)^{\argr\tdot}} = 0\commae\\
  \frac1{\stsign^2} \stCL{(\ln \onbnd\jac)^{\argr\tdot\argr\tdot}} =
  - \frac13(\extscuronb{}^\tdot
  + \stsign^2 \extcuronb{}^2 \bigr)\period
\end{gathered}\end{equation}

Finally, we have boundary coincidence limits
\begin{gather}
  \sCL{\stwf_\bnd}(\yonb;\xi,\zeta) =
  \frac12\stsign^2 (\xi+\zeta)^2 \comma
  \sCL{\stgdist_\bper}(\yonb;\xi,\zeta) =
  \stsign^2 (\xi+\zeta) \comma\\
  \sCL{\stwfl_\bnd}(\yonb;\xi,\zeta) =
  (\xi+\zeta)\normv(\yonb,\xi) \comma
  \sCL{\stwfr_\bnd}(\yonb;\xi,\zeta) =
  (\xi+\zeta)\normv(\yonb,\zeta) \commae\\
  \sCL{\rfltml}(\yonb;\xi,\zeta) =
  \frac\xi{\xi+\zeta} \comma
  \sCL{\rfltmr}(\yonb;\xi,\zeta) =
  \frac\zeta{\xi+\zeta} \period
\end{gather}
Using these relations, equations \eqref{sCLofbBb}, 
\eqref{sCLofbBandBb}, with help of \eqref{Dwmap} and 
\begin{equation}
  \sCL{\onbnd\stwfll_{\bpar\,\bdr}}(\yonb;\xi,\zeta) =
  \smtrc(\yonb) + \order{\zeta^2} \comma
  \sCL{\onbnd\stwfrr_{\bpar\,\bdl}}(\yonb;\xi,\zeta) =
  \smtrc(\yonb) + \order{\xi^2}
\end{equation}
we can derive 
\begin{equation}
  -\sCL{\onbnd\stwflr_{\bnd\bpar}}(\yonb;\xi,\zeta) =
  \smtrc(\yonb) +
  \Bigl( \xi + \zeta - 2 \frac{\xi\zeta}{\xi+\zeta} \Bigr)
  \stsign^2 \extcur(\yonb) + \order{(\xi+\zeta)^2}\period
\end{equation}
From this follows
\begin{equation}
  \sCL{\onbnd\stvvmd_\bnd}(\yonb;\xi,\zeta) =
  1 + \Bigl( \xi + \zeta - 2 \frac{\xi\zeta}{\xi+\zeta} \Bigr)
  \stsign^2 \extscur(\yonb) + \order{(\xi+\zeta)^2}\period
\end{equation}
Together with
\begin{equation}
  \onbnd\jac(\yonb,\eta) = 1 + \stsign^2 \extscur(\yonb) \eta +
  \order{\eta^2}
\end{equation}
and the definition \eqref{tildajacbndrel}, 
it finally gives the expansion for $\tilde\jac_\bnd$,
\begin{equation}\label{jactildeExp}
  \tilde\jac_\bnd(\yonb;\xi,\lambda,\zeta) = 
  1 + \Bigl( \frac{\xi+\zeta}2
  - (1-2p)\lambda\Bigr) \stsign^2\extscur(\yonb)
  + \order{(\xi+\zeta+\lambda)^2} \period
\end{equation}


\section{Special functions $\boldsymbol{\rfc_\nu}$}
\label{apx:SFcR}


In appendix \ref{apx:AEHK} we have used various integrals 
of exponentials of quadratic exponent and integrals 
of such integrals. Here we summarize properties of 
such integrals. We will introduce a special function 
$\rfc_\nu$ closely related to error functions $\erfc(x)$, 
the definition and properties of which can be found, 
for example, in \cite{GradshteinRyzhik:book}. 
The derivation of the properties below 
are not all simple, and we do not include them here.

We define the function $\rfc_\nu(x)$ for positive $\nu$ as
\begin{equation}\label{rfcDef}
  \rfc_\nu(z) = \frac1{\Gammafc(\nu)}
  \int_{\realn^+} \dvol{x} x^{\nu-1}
  \exp\Bigl(-\frac12(x-z)^2\Bigr) \period
\end{equation}
It is a solution of the differential equation
\begin{equation}
  \rfc_\nu'(z) = \nu \rfc_{\nu+1}(z) - z \rfc_\nu(z)\comma
  \rfc_\nu \xrightarrow{z\rightarrow-\infty} 0 \period
\end{equation}
In the limit $\nu\rightarrow0$ and for $\nu=1$ we have
\begin{gather}
  \rfc_0(z) = \exp\Bigl(-\frac12z^2\Bigr) \commae\\
  \rfc_1(z) = \sqrt{2\pi} -
  \sqrt{\frac\pi2}\;\erfc\Bigl(\frac{z}{\sqrt2}\Bigr)\period
\end{gather}
We also have the recurrence relation
\begin{equation}\label{RRecRel}
  \rfc_{\nu+2}(z) = \frac1{\nu+1}
  \bigl( z\, \rfc_{\nu+1}(z) + \rfc_\nu(z) \bigr) \period
\end{equation}

For $\nu\in\naturaln$ these functions are combinations 
of $\rfc_0$ and $\rfc_1$ with polynomial coefficients
\begin{equation}
  \rfc_{n+1} = \pfc_n \rfc_1 + \qfc_{n-1}\rfc_0
  \qquad\text{for}\qquad n\in\naturaln \quad\commae
\end{equation}
where
\begin{equation}\begin{gathered}
  \pfc_{n+1}(z) = \frac1{n+1}
  \bigl(z\, \pfc_n(z) + \pfc_{n-1}(z)\bigr)
  \comma \pfc_0 = 1 \comma \pfc_1 = z \commae\\
  \qfc_{n+1}(z) = \frac1{n+2}
  \bigl(z\, \qfc_n(z) + \qfc_{n-1}(z)\bigr)
  \comma \qfc_0 = 1 \comma \qfc_1 = \frac12 z \period
\end{gathered}\end{equation}
These polynomials satisfy
\begin{equation}
  \pfc_{n}' = \pfc_{n-1} \comma 
  \qfc_n' = (n+2)\,\qfc_{n+1} - \pfc_{n+1}  \commae
\end{equation}
and
\begin{equation}
  \sqrt{2\pi}\,\pfc_n(z) = 
  \rfc_{n+1}(z) + (-1)^n \rfc_{n+1}(-z) = 
  \frac{\sqrt{2\pi}}{i^n 2^{\frac n2}n!}
  \Hermpol_n\bigl(i\tfrac{z}{\sqrt2}\bigr)\commae
\end{equation}
where $\Hermpol_n$ are the Hermite polynomials 
(see \cite{GradshteinRyzhik:book}).
  
Values at zero are
\begin{alignat}{2}
  \rfc_\nu(0) &=
  2^{\frac \nu2} \frac{\Gammafc(\frac\nu2+1)}{\Gammafc(\nu+1)} =
  \frac12 \frac{\sqrt{2\pi}}{2^{\frac{\nu-1}2}\Gammafc(\frac{\nu-1}2+1)}
  \hspace{-0.9em}&&=
  \begin{cases}
    \frac1{(\nu-1)!!} & \text{for $\nu$ natural and even}\\
    \frac1{\nu!!}\sqrt{\frac\pi2} & \text{for $\nu$ natural and odd}
  \end{cases}\commae\label{ZeroOfrfc}\\
  \pfc_n(0) &=
  \begin{cases}
    \frac1{n!!}     &\text{for $n$ even} \\
    0               &\text{for $n$ odd}
  \end{cases}\commae &
  \pfc_n' &=
  \begin{cases}
    0               &\text{for $n$ even} \\
    \frac1{(n-1)!!} &\text{for $n$ odd}
  \end{cases}\commae\label{ZeroOfpfc}\\
  \qfc_n(0) &= 
  \begin{cases}
    \frac1{(n+1)!!} &\text{for $n$ even} \\
    0               &\text{for $n$ odd}
  \end{cases}\commae &
  \qfc_n' &=
  \begin{cases}
    0               &\text{for $n$ even} \\
    \tfrac1{n!!} - \tfrac1{(n+1)!!} &\text{for $n$ odd}
  \end{cases}\period
\end{alignat}
The behavior for small $z$ can be found for 
natural $\nu$ with help of relations \eqref{RRecRel} and
\begin{equation}\begin{aligned}
  \rfc_0(z) &= \sum_{k\in\naturaln_0}
    \frac{(-1)^k}{(2k)!!} z^{2k}\commae\\
  \rfc_1(z) &= \sqrt{\frac\pi2} + 
  \sum_{k\in\naturaln_0}
    \frac{(-1)^k}{(2k+1)\,(2k)!!} z^{2k+1}\period
\end{aligned}\end{equation}
The behavior for $\abs{z}\gg1$ and $n\in\naturaln_0$ is
\begin{equation}\label{rfcAsymptB}
  \rfc_{n+1}(z) = \sqrt{2\pi}\, \pfc_n(z)\, \thetafc(z) +
  \exp\Bigl(-\frac12 z^2\Bigr)\; \orderB{\frac1{z^{n+1}}} \commae
\end{equation}
where $\thetafc(z)$ is the step function.

Now we can write down results of some integrals 
in terms of these functions. For $n\in\naturaln_0$ we have
\begin{equation}\label{pfcIntDef}
  \frac1{n!} \int_\realn \dvol{x} x^n \exp\Bigl(-\frac12(x-z)^2\Bigr) 
  = \sqrt{2\pi}\,\pfc_n(z)\period
\end{equation}
 Further for $k,l\in\naturaln_0$ we have
\begin{gather}
\begin{split}
  &\frac1{l!} \int_{ \langle - \infty,x \rangle }
  \dvol{\xi} \xi^l \rfc_k(\xi) =
  \sum_{m=0,\dots,l} \frac{(-1)^{l+m}}{m!} x^m
  \rfc_{\scriptscriptstyle k+l-m+1}(x) = \\
  &\qquad = \sqrt{2\pi} (-1)^l \pfc_{k+l}(0) +
  \frac{\sqrt{2\pi}}{l!\,(k-1)!}
  \sum_{\substack{m\in\naturaln\\2m\leq k-1}}
  \frac{(2m-1)!!}{k+l-2m} \binom{k-1}{2m} x^{k+l-2m} 
  -\\&\hspace{20em}
  - \sum_{m=0,\dots,l} \frac1{m!} x^m 
  \rfc_{\scriptscriptstyle k+l-m+1}(-x) \commae
\end{split}\label{intInftyXofrfc}\displaybreak[0]\\
\begin{split}
  &\frac1{l!} \int_{ \langle 0,x \rangle }
  \dvol{\xi} \xi^l \rfc_k(\xi) = \\
  &\qquad =  (-1)^k \rfc_{k+l+1}(0) +
  \frac{\sqrt{2\pi}}{l!\,(k-1)!}
  \sum_{\substack{m\in\naturaln\\2m\leq k-1}}
  \frac{(2m-1)!!}{k+l-2m} \binom{k-1}{2m} x^{k+l-2m} 
  -\\&\hspace{20em}
  - \sum_{m=0,\dots,l} \frac1{m!} x^m 
  \rfc_{\scriptscriptstyle k+l-m+1}(-x) \period
\end{split}\label{intZeroXofrfc}
\end{gather}
Finally for $m,k,l\in\realn^+$ and $n= m+k+l$ we have
\begin{equation}\label{IntGenPowExp}
  \frac1{n!} \int_{\xi,\zeta\in\realn^+} \dvol{\xi}\dvol{\zeta}
  \frac{\xi^{m+k}\zeta^{m+l}}{(\xi+\zeta)^m} 
  \exp\Bigl(-\frac12(\xi+\zeta)^2\Bigr) =
  \sqrt{2\pi} \;2^{-\frac{n+1}2}\;
  \frac{\Gammafc(m+k+1)\,\Gammafc(m+l+1)}
  {\Gammafc(b+m+1)\,\Gammafc(\frac{n+1}2)} \period
\end{equation}

  

\begin{notes}
\noteitem{nt:Conventions}{
We use the MTW sign convention ([\citen{MTW}]) 
for spacetime quantities generalized 
to accomodate both Euclidian and Lorentzian versions 
of the theory. The version is determined by the constant factor
$\stsign$ which is equal to $1$ in the Euclidian version or
to $i$ in the Lorentzian version. Specifically, geometric 
quantities defined in spacetime $\stmfld$ with an embeded 
hypersurface $\Sigma$ are related in the following way:
\begin{equation*}\begin{aligned}
  {}&\text{metric:}&\quad&
  \stmtrc_{\absidx{\alpha\beta}} =
  \stsign^2\; \normf_ {\absidx{\alpha}}\normf_ {\absidx{\beta}}
  + \smtrc_{\absidx{\alpha\beta}}\comma
  \text{$\smtrc$ positive definite}\commae\\
  {}&\text{volume element:}&\quad&
  \stvol = \frac{1}{\stsign}(\Det\stmtrc)^{\frac12}\commae\\
  {}&\text{curvature:}&\quad&
  \stcur_{\absidx{\alpha\beta}}{}^{\absidx{\mu}}{}_{\absidx{\nu}} 
  a^{\absidx{\nu}} = \stcnx_{[\absidx{\alpha}}\stcnx_{\absidx{\beta}]}
  a^{\absidx{\mu}}\comma
  \stric_{\absidx{\alpha\beta}} =
  \stcur_{\absidx{\mu\alpha}}{}^{\absidx{\mu}}{}_{\absidx{\beta}}
  \comma
  \stscur = \stric_{\absidx{\alpha\beta}}\;
  \stmtrc^{\dash 1\,\absidx{\alpha\beta}} \commae\\
  {}&\text{extrinsic curvature:}&\quad&
  \extcur_{\absidx{\alpha\beta}} =
  \sdelta^{\absidx{\mu}}_{\absidx{\alpha}}
  \sdelta^{\absidx{\nu}}_{\absidx{\beta}}\;
  \stcnx_{\absidx{\mu}}\normf_{\absidx{\nu}}\comma
  \extscur = \extcur_{\absidx{\alpha\beta}}\;
  \stmtrc^{\dash 1\,\absidx{\alpha\beta}}\commae\\
  {}&&\quad&\text{$\normf$ has the orientation 
  of the hypersurface $\Sigma$}\commae\\
  {}&\text{projector on $\Sigma$:}&\quad&
  \sdelta^{\absidx{\beta}}_{\absidx{\alpha}} =
  \delta^{\absidx{\beta}}_{\absidx{\alpha}}
   - \normv^{\absidx{\beta}}\, \normf_{\absidx{\alpha}}\comma
   \normv^{\absidx{\alpha}} = \stsign^2\; \normf_{\absidx{\mu}}
   \stmtrc^{\dash 1\,\absidx{\mu\alpha}}\period 
\end{aligned}\end{equation*}
  }
\noteitem{nt:WhyEuclidian}{
  Euclidian and physical actions are related by
  \begin{equation*}
  -I(\hist{h}) = \isign S(\hist{h}) \period
  \end{equation*}
  The version of the theory is determined by 
  actual value of the factor $\isign$ --- 
  whether it is real or imaginary.}
\noteitem{nt:stsignAndProp}{
  The factor $\stsign$, which governs the signature 
  of the spacetime metric, is chosen here for convenience 
  and reflects that we are using Lorentzian convention 
  for volume element\note{nt:Conventions}. 
  I.e., the physical amplitude 
  is $\frac1\stsign\hkrnl$, but the quantity $\hkrnl$ will 
  have nicer properties in language of Lorentzian quantities. 
  Similarly, for the Green function the physical amplitude 
  is $\frac1\stsign \GFF$, but we will use often the quantity 
  $\GFF$ to express properties of the amplitude.} 
\noteitem{nt:DotsConv}{
  To abbreviate formulas we often use different dots to
  indicate contraction in different vector spaces. 
  We use ``$\stctr$'' for contractions of tangent tensor 
  indices, i.e. 
  $\stmtrc_{\absidx{\alpha\beta}}\, a^{\absidx{\beta}} = 
  \stmtrc\stctr a$. We use ``$\stint$'' for contraction
  in functional vector spaces of function and densities on
  spacetime, i.e. the bullet means an integration 
  over spacetime (or an action of a distribution on a test 
  function --- a formal integration):
  \begin{equation*}
  \phi\stint\alpha = \int_\stmfld \phi\, \alpha \period
  \end{equation*}
  Similarly we use ``$\sint$'' for integration over 
  a hypersurface $\Sigma$ or a boundary
  $\bound\Omega$ of a spacetime domain.
  }
\noteitem{nt:DistrNot}{
  A volume element $\stvol$ on spacetime defines a 
  bi-distribution $\stbivol = \stvol\deltadst$ --- 
  a delta function normalized to the volume element,
  i.e. $\phi\stint\stbivol\stint\psi =
  \int \phi\,\psi\;\stvol$. Clearly for a smooth function 
  $f$ we can define a distribution $f\stbivol$.
  
  It is convenient to represent differential operators 
  on the manifold $\stmfld$ as bi-dis\-tri\-bu\-tions. 
  We use arrows $\acttol{}$ and $\acttor{}$ to indicate 
  direction of derivatives. So, for example,
  \begin{equation*}
  \psi \stint \bigl(\bigradl_{\absidx{\alpha}}
  a^{\absidx{\alpha}}\bigr) \stint\omega =
  \omega\stint \bigl(a^{\absidx{\alpha}}
  \bigradr_{\absidx{\alpha}}\bigr) \stint \psi =
  \int \omega\; a^{\absidx{\alpha}}
  \grad_{\absidx{\alpha}} \psi
  \end{equation*}
  for a test function $\psi$, a test density $\omega$, 
  and a vector field $a$.}
\noteitem{nt:RevIndx}{
  A meaning of the tilde in the index of the quadratic form
  $\biF_{\revidx\bbcnd{k}}$ will not be discussed in this paper.
  It will be explained in the following paper and 
  can be safely ignored here.}
\noteitem{nt:stsignAndSrc}{
  Again, we factorize out the prefactor $\stsign$ motivated 
  by the fact that $J$ is a density, i.e. proportional 
  to volume element $\stvol$.}
\end{notes}

    

\end{document}